\documentclass[11pt]{article}

\usepackage{amsfonts}
\usepackage{amssymb}
\usepackage{amstext}
\usepackage{amsmath}
\usepackage{enumerate}
\usepackage{algorithm}
\usepackage{algorithmic}

\usepackage{graphicx}

\usepackage{bbm}

\usepackage{amsthm}

\usepackage{thmtools}

\usepackage{verbatim} 

\usepackage[usenames,dvipsnames,svgnames,table]{xcolor}
\definecolor{dartmouthgreen}{rgb}{0.05, 0.5, 0.06}
\definecolor{ceruleanblue}{rgb}{0.16, 0.32, 0.75}

\usepackage[colorlinks=true,pdfpagemode=UseNone,citecolor=dartmouthgreen,linkcolor=ceruleanblue,urlcolor=BrickRed,pdfstartview=FitH]{hyperref}

\usepackage{framed}
\usepackage{enumitem}


\newtheorem{theorem}{Theorem}[section]
\newtheorem{fact}[theorem]{Fact}
\newtheorem{lemma}[theorem]{Lemma}
\newtheorem{definition}[theorem]{Definition}

\newtheorem{corollary}[theorem]{Corollary}
\newtheorem{proposition}[theorem]{Proposition}
\newtheorem{claim}[theorem]{Claim}

\newtheorem*{problem*}{Problem}
\newtheorem{remark}[theorem]{Remark}
\newtheorem*{remark*}{Remark}

\newcommand{\st}{\mbox{\rm subject to }}

\numberwithin{equation}{section}
\numberwithin{table}{section}

\renewcommand{\preceq}{\preccurlyeq}
\renewcommand{\succeq}{\succcurlyeq}

\renewcommand{\tilde}{\widetilde}

\newcommand{\R}{\ensuremath{\mathbb R}}

\newcommand{\N}{\ensuremath{\mathbb N}}
\newcommand{\F}{\ensuremath{\mathcal F}}

\newcommand{\E}[1]{{\mathbb{E}}\left[#1\right]}

\newcommand{\ceil}[1]{\ensuremath{\left\lceil#1\right\rceil}}

\newcommand{\poly}{\operatorname{poly}}

\newcommand{\junk}[1]{}

\renewcommand{\l}{\lambda}

\newcommand{\edgeexp}{\ensuremath{\vec{\phi}_\pi}}
\newcommand{\primal}{\ensuremath{\lambda_{\pi}^{\triangle}}}
\newcommand{\vertexexp}{\ensuremath{\vec{\psi}_\pi}}

\newcommand{\norm}[1]{\left\lVert#1\right\rVert}

\newcommand{\vertiii}[1]{{\left\vert\kern-0.25ex\left\vert\kern-0.25ex\left\vert #1 \right\vert\kern-0.25ex\right\vert\kern-0.25ex\right\vert}}

\newcommand{\one}{\ensuremath{\mathbbm{1}}}

\newenvironment{proofof}[1]{{\medbreak\noindent \em Proof of #1.  }}{\hfill\qed\medbreak}

\def\b1{{\bf 1}}
\def\eps{{\epsilon}}

\def\R{\mathbb{R}}

\newcommand{\M}{\mathcal{M}}
\newcommand{\p}{\mathcal{P}}
\newcommand{\D}{\mathcal{D}}

\def\diag{\operatorname{diag}} 

\def\tr{\operatorname{tr}}

\global\long\def\E{\mathbb{E}}

\global\long\def\R{\mathbb{R}}

\newcommand{\inner}[2]{\langle #1, #2 \rangle} 
\newcommand{\biginner}[2]{\Big\langle #1, #2 \Big\rangle} 

\DeclareMathOperator{\supp}{supp}

\makeatletter
\DeclareRobustCommand{\cev}[1]{%
  {\mathpalette\do@cev{#1}}%
}
\newcommand{\do@cev}[2]{%
  \vbox{\offinterlineskip
    \sbox\z@{$\m@th#1 x$}%
    \ialign{##\cr
      \hidewidth\reflectbox{$\m@th#1\vec{}\mkern4mu$}\hidewidth\cr
      \noalign{\kern-\ht\z@}
      $\m@th#1#2$\cr
    }%
  }%
}
\makeatother

\DeclareMathOperator{\diam}{diam}

\newcommand{\Fil}{\mathcal{F}} 

\title{Fast Algorithms for Directed Graph Partitioning\\ Using Flows and Reweighted Eigenvalues}

\author{
Lap Chi Lau\footnote{Cheriton School of Computer Science, University of Waterloo. Supported by NSERC Discovery Grant. 
},~~~~~
Kam Chuen Tung\footnote{Cheriton School of Computer Science, University of Waterloo. Supported by NSERC Discovery Grant. 
},~~~~~
Robert Wang\footnote{Cheriton School of Computer Science, University of Waterloo. Supported by NSERC Discovery Grant and Canada Graduate Scholarship. 
}}

\date{}

\begin{document}

\begin{titlepage}
\def\thepage{}
\thispagestyle{empty}

\maketitle

\begin{abstract}
We consider a new semidefinite programming relaxation for directed edge expansion, which is obtained by adding triangle inequalities to the reweighted eigenvalue formulation.
Applying the matrix multiplicative weight update method on this relaxation,
we derive almost linear-time algorithms to achieve $O(\sqrt{\log n})$-approximation and Cheeger-type guarantee for directed edge expansion, as well as an improved cut-matching game for directed graphs.
This provides a primal-dual flow-based framework to obtain the best known algorithms for directed graph partitioning.
The same approach also works for vertex expansion and for hypergraphs, providing a simple and unified approach to achieve the best known results for different expansion problems and different algorithmic techniques.
\end{abstract}

\end{titlepage}

\thispagestyle{empty}


\newpage

\section{Introduction}

The main combinatorial quantity that we study in this work is the directed edge expansion with arbitrary vertex weights.

\begin{definition}[$\pi$-Weighted Directed Edge Expansion]
\label{def:phi-pi}
Let $G = (V, E, w)$ be a directed graph with edge weights $w: E \rightarrow \R^+$, 
equipped with vertex weights $\pi: V \rightarrow \mathbb{R}^+$.
For $S \subseteq V$, let $\delta^+(S) := \{ij\in E:i\in S,\;j\notin S\}$ be the set of edges going out of $S$, and let $\delta^-(S) = \delta^+(\overline{S})$. Let $\pi(S) := \sum_{i\in S}\pi(i)$ be the $\pi$-weight of $S$. 
The $\pi$-weighted edge expansion of $S \subseteq V$ and of the graph $G$ are defined as
\begin{align*}
    \edgeexp(S) := \frac{\min\{w(\delta^+(S)), w(\delta^-(S))\}}{\min\{\pi(S),\pi(\overline{S})\}}
    \quad {\rm and} \quad
    \edgeexp(G) := \min_{\emptyset \neq S \subset V}\edgeexp(S).
\end{align*}
\end{definition}

This is a general problem that encompasses various expansion problems studied in the literature.
The directed edge expansion problem is when $\pi(i) = 1$ for all $i \in V$, and this is equivalent (up to a factor of $\Theta(n)$ where $n$ is the number of vertices) to the directed sparsest cut of $G$
\[
  \min_{\emptyset \neq S \subset V}
  \frac{\min\{w(\delta^+(S)), w(\delta^+(\overline{S}))\}}
  {|S|\cdot |V\backslash S|}
\]
studied in \cite{ACMM05,AK07,Kal07}.
The directed edge conductance problem studied in~\cite{Yos19,LTW23} is when $\pi(i) = w(\delta^+(i)) + w(\delta^-(i))$, the weighted total degree of vertex $i$.  Clearly, the corresponding problems in undirected graphs as studied in~\cite{ARV09,KRV06,AK07} can be reduced to \autoref{def:phi-pi} by bidirecting the edges in the undirected graph.
Also, the undirected vertex expansion problem studied in~\cite{FHL08,LRV13} 
and the directed vertex expansion problem studied in~\cite{LTW23} can be reduced to \autoref{def:phi-pi} through a standard reduction of splitting each vertex into two.
Furthermore, the corresponding problems in undirected and directed hypergraphs can be reduced to \autoref{def:phi-pi} through a reduction of replacing each hyperedge by a vertex as shown in~\cite{CS18}.
Therefore, the main goal of this work is to design fast algorithms for approximating $\edgeexp$.

\subsection{Previous Work}

Before presenting our results, we first review previous work on approximating various graph expansion problem to provide the context of our work.
We let $n$ be the number of vertices and $m$ be the number of edges unless otherwise specified.

\subsubsection{Undirected Graphs}

The edge expansion, sparsest cut, and the edge conductance problems in undirected graphs are central problems in approximation algorithms.
These problems have a rich literature with various techniques developed.

{\em Spectral Method:} Cheeger's inequality~\cite{AM85,Alo86} provides a near-linear time algorithm to return a set $S$ with conductance $\lambda_2 \lesssim \phi(S) \lesssim \sqrt{\lambda_2}$ where $\lambda_2$ is the second smallest eigenvalue of the normalized Laplacian matrix of the graph.

{\em Linear Programming:}
Leighton and Rao~\cite{LR99} gave an $O(\log n)$-approximation algorithm for sparsest cut based on linear programming.
The dual problem of their linear program is to embed a complete graph into the original graph using flows. 

{\em Semidefinite Programming:}
Arora, Rao, and Vazirani~\cite{ARV09} gave a celebrated semidefinite programming $O(\sqrt{\log n})$-approximation algorithm for sparsest cut.
They introduced novel geometric ideas in analyzing the triangle inequalities of the Goemans-Linial SDP relaxation.
The dual problem of their SDP is to embed an expander graph into the original graph using flows.

{\em Cut-Matching Game:}
Developing the idea of expander flows in~\cite{ARV09}, 
Khandekar, Rao, and Vazirani~\cite{KRV06} introduced the cut-matching game as a combinatorial approach to obtain fast approximation algorithm for sparsest cut.
Orecchia, Schulman, Vazirani, Vishnoi~\cite{OSVV08} improved the analysis of cut-matching game to give an $O(\log n)$-approximation algorithm for sparsest cut using $O(\log^3 n)$ undirected approximate max-flow computations.
Since then, the cut-matching game has become a useful algorithmic tool on its own, with interesting applications in different problems~\cite{And10,Chu12,CE13,CL16,CGLNPS20,BGS20}.

{\em Primal-Dual Algorithms:}
Arora and Kale~\cite{AK07} developed a general primal-dual combinatorial approach to solve SDPs based on the matrix multiplicative weight update (MMWU) method.
Using this, they gave an $\tilde{O}(n^2)$-time $O(\sqrt{\log n})$-approximation algorithm for sparsest cut using $O(\log^2 n)$ multi-commodity flow computations.
Notably, the cut-matching game in~\cite{OSVV08} can be interpreted as an instantiation of the matrix multiplicative weight update method.

{\em Almost Linear-Time Algorithm:}
Sherman~\cite{She09} pushed the approach in~\cite{AK07} further to get the best of the semidefinite programming approach and the combinatorial approach.  
He gave an $O\big(\sqrt{\frac{1}{\eps} \log n}\big)$-approximation algorithm for sparsest cut using $n^{O(\eps)}$ approximate max-flow computations,
which implies an almost linear-time $O(\sqrt{\log n})$-approximation algorithm for the problem.

\subsubsection{Directed Graphs}

The corresponding problems for directed graphs are not as well-understood, particularly in relation to fast algorithms.

{\em Spectral Method:}
There was no known analog of Cheeger's inequality for directed graphs until recently.
Lau, Tung, Wang~\cite{LTW23} defined a ``spectral'' quantity (using semidefinite programming) called the reweighted eigenvalue $\lambda_2^*$, and showed that there is a polynomial-time algorithm to return a set $S \subseteq V$ with $\lambda_2^* \lesssim \vec{\phi}(S) \lesssim \sqrt{\lambda_2^* \log(1/\lambda_2^*)}$.
It is left as an open question to design a fast algorithm to return such a set.

{\em Semidefinite Programming:}
Agarwal, Charikar, Makarychev and Makarychev \cite{ACMM05} formulated an SDP using directed semi-metrics, and extended the analysis in~\cite{ARV09} to obtain $O(\sqrt{\log{n}})$-approximation algorithms for directed sparsest cut, directed balanced separator and other related problems.

{\em Cut-Matching Game:}
Louis~\cite{Lou10} defined an analog of the cut-matching game in~\cite{KRV06} for directed graphs, and used it to obtain an $O(\log^2 n)$-approximation algorithm for directed sparsest cut using $O(\log^3 n)$ max-flow computations.
This directed cut-matching game has found applications in dynamic algorithms~\cite{BGS20}. 

{\em Primal-Dual Algorithms:}
Using the matrix multiplicative weight update method on the SDP formulation in~\cite{ACMM05}, Arora and Kale~\cite{AK07,Kal07} claimed an $O(\sqrt{\log n})$-approximation algorithm for directed sparsest cut with time complexity $O(n^{2+o(1)})$ plus $O(\log^3 n)$ maximum flow computations.
Chan and Sun~\cite{CS18} pointed out an issue (which was acknowledged by Kale) in the trace bound in the analysis in~\cite{AK07}, and consequently the number of iterations is only bounded by $\tilde{O}(n^2)$ instead of $\tilde{O}(1)$, and so the time complexity should be $O(n^{4+o(1)})$ plus $\tilde{O}(n^2)$ maximum flow computations.
Therefore, even with the recent breakthrough~\cite{CKLPPS22} in maximum flow computations in directed graphs, the time complexity of Arora-Kale's algorithm remains $\Omega(n^4)$ for directed sparsest cut.
Moreover, unlike for undirected graphs, the connection between the cut-matching game in~\cite{Lou10} and the matrix multiplicative weight update method is not known.

\subsection{Our Results} \label{sec:results}

We consider a new semidefinite program for directed edge expansion based on the reweighted eigenvalue formulation.
Using the MMWU method on this new SDP, we improve the algorithmic results for directed edge expansion, matching the corresponding results for undirected edge expansion.

\subsubsection{Primal Formulation}

We consider a new SDP relaxation for directed edge expansion in \autoref{def:phi-pi}.
For undirected graphs, the SDP formulation in~\cite{ARV09} can be understood as the spectral formulation for second smallest Laplacian eigenvalue plus the $\ell_2^2$ triangle inequalities (see~\cite{Tre16}).
For directed graphs, our SDP formulation is to use the spectral formulation for reweighted eigenvalue in~\cite{LTW23} plus the $\ell_2^2$ triangle inequalities.

\begin{definition}[Reweighted Eigenvalue with Triangle Inequalities]
\label{def:main-sdp}
Given an edge-capacitated directed graph $G = (V,E,w)$,
$F: E \to \R_{\geq 0}$ is called a circulation\footnote{In~\cite{LTW23}, $F$ is called an Eulerian reweighting of $G$. In this paper, network flows is a unifying theme, and so we find it more suitable to call $F$ a circulation.} on $G$ if $\sum_{j:ij \in E} F(i,j) = \sum_{j:ji \in E} F(j,i)$ for all $i \in V$.  
Let $\mathcal{F}(G)$ be the set of all circulations on $G$ that also satisfy the capacity constraints $F(e) \leq w(e)$ for all $e \in E$.
Given also vertex weights $\pi: V \to \R_{\geq 0}$, the $\primal(G)$ program for directed edge expansion is
  \begin{align}
  \label{eqn:phi-mu-sdp}
  \begin{split}
    \primal(G) :=
     \min_{v_1, \ldots, v_n \in \mathbb{R}^n}\max_{F\in \mathcal{F}(G)} 
     &~~~ \sum_{i<j} \frac12\big(F(i,j) + F(j,i)\big) \cdot \norm{v_i - v_j}^2 \\
     \st 
     &~~~\sum_{i=1}^n \pi(i) \cdot v_i = \vec{0}\\
     &~~~\sum_{i=1}^n \pi(i) \cdot \norm{v_i}^2 = 1\\
     &~~~\norm{v_i - v_k}^2 + \norm{v_k - v_j}^2 \geq \norm{v_i-v_k}^2\;\qquad \forall i,j,k\in V.
  \end{split}
  \end{align}
Note that we use the convention that $F(i, j) = 0$ if $ij \not\in E$.
\end{definition}

Note that the formulation in \autoref{def:main-sdp} without the $\ell_2^2$ triangle inequalities in the last line is exactly the formulation for reweighted eigenvalues in~\cite[Proposition 3.4]{LTW23}.
Just as the addition of triangle inequalities to the spectral formulation reduces the integrality gap of undirected edge expansion to $O(\sqrt{\log n})$ in~\cite{ARV09}, we show the exact analog for directed edge expansion by using the spectral formulation for reweighted eigenvalues.

\begin{theorem}[$O(\sqrt{\log n})$-Approximation for Directed Vertex Expansion]
\label{thm:primal-rounding}
For any edge-capacitated directed graph $G = (V, E, w)$  with vertex weights $\pi: V \rightarrow \R_+$,
\begin{align*}
    \primal(G) \lesssim \edgeexp(G) \lesssim \sqrt{\log{n}}\cdot \primal(G).
\end{align*}
\end{theorem}

The proof is a simple adaptation of that in~\cite{ARV09}.
We will compare our formulation with that in~\cite{ACMM05} in \autoref{sec:ACMM}, and we will compare the two dual formulations in~\autoref{sec:Arora-Kale-directed}.
We note that the same approach of adding $\ell_2^2$ triangle inequalities to reweighted eigenvalues provides considerably simpler formulations and proofs for undirected vertex expansion and hypergraph edge expansion than that in~\cite{FHL08} and in \cite{LM14}, while having the same integrality gap $O(\sqrt{\log n})$;
see \autoref{sec:others} for more details.


\subsubsection{Dual Formulation}

As in~\cite{ARV09}, the dual program of $\primal$ in \autoref{def:main-sdp} can be interpreted as embedding a directed expander flow into the original directed graph.
Since our formulation in \autoref{def:main-sdp} requires the flow to be a circulation, we obtain a new structural result about the existence of a circulation of high edge expansion as a dual certificate, which may be of independent interest.

\begin{proposition}[$O(\sqrt{\log n})$ Dual Certificate]
  \label{prop:reweighted-phi}
  Given an edge-capacitated directed graph $G = (V, E, w)$ with vertex weights $\pi: V \rightarrow \R_+$, there exists a circulation $F\in \mathcal{F}(G)$ satisfying edge capacity constraints with 
  \begin{align*}
    \phi_\pi(F)
    :=
    \min_{\emptyset \neq S \subset V} 
    \frac
    {\sum_{i\in S, j \notin S} F(i,j)}
    {\min\{\pi(S),\pi(\overline{S})\}}
    \gtrsim \frac{\edgeexp(G)}{\sqrt{\log{n}}}.
  \end{align*}
\end{proposition}


\subsubsection{Primal-Dual Algorithms}

Using the matrix multiplicative weight update method in~\cite{AK07,Kal07} on $\primal$, 
combining with the chaining techniques in~\cite{She09}, we extend Sherman's result to directed graphs.

\begin{theorem}[Fast $O(\sqrt{\log{n}})$-Approximation to Directed Edge Expansion]
    \label{thm:directed-Sherman}
    For small enough $\epsilon > 0$, there is a randomized algorithm that, given any edge-capacitated directed graph $G = (V, E, w)$ with vertex weights $\pi: V \rightarrow \R_+$,
    uses $\Tilde{O}(n^{3\epsilon})$ directed max-flow computations to compute a cut $S \subseteq V$ with $\edgeexp(S) \lesssim \sqrt{\frac{\log{n}}{\epsilon}}\cdot \primal(G)$ with constant probability.
\end{theorem}

Using the recent breakthrough~\cite{CKLPPS22} on directed maximum flow, 
\autoref{thm:directed-Sherman} implies an $O(m^{1+O(\eps)})$-time $O(\sqrt{(\log n)/\eps})$-approximation algorithm for directed edge expansion.
This is a significant improvement over the previous $O(n^{4+o(1)})$-time $O(\sqrt{\log n})$-approximation algorithm for directed sparsest cut by Arora and Kale~\cite{AK07,Kal07}.

Since undirected vertex expansion can be reduced to directed edge expansion, this is also a significant improvement over the previous results~\cite{CK19,CS21} in fast approximation algorithms for undirected vertex expansion, where the best known result is a $O(\log^2 n)$-approximation using $O(\log^3 n)$ vertex-capacitated max-flow computations.

We remark that our algorithm is simpler than Sherman's when restricted to undirected graphs, bypassing the use of multi-commodity flows. 
See \autoref{sec:expander-flow-dual} and \autoref{sec:Arora-Kale-directed} for more discussions.


\subsubsection{Cheeger-Type Guarantee}

We show that the matrix multiplicative weight update method can also be used to obtain a fast algorithm to output a set with the Cheeger-type guarantee in~\cite{LTW23}.

\begin{theorem}[Fast Cheeger-type Approximation] \label{thm:fast-Cheeger}
Given an edge-capacitated directed graph $G=(V,E,w)$, there is an almost linear time algorithm for approximating the directed edge conductance $\vec{\phi}(G)$ that returns a set $S$ with
$\vec{\phi}(S) \lesssim
     \sqrt{\vec{\phi}(G)\cdot \log{\frac{1}{\vec{\phi}(G)}}}.
$
 \end{theorem}

This answers an open question in~\cite{LTW23} and provides a fast ``spectral'' algorithm for directed graph partitioning.


\subsubsection{Cut-Matching Game}

The cut-matching game is an interesting and useful way to construct an expander graph; see \autoref{sec:cut-matching-game} for an introduction.
Using the matrix multiplicative weight update method, we also obtain a cut-player strategy that matches the cut-matching game result in~\cite{OSVV08} for undirected graphs.

\begin{theorem}[Cut-Matching Game for Directed Edge Expansion]
\label{thm:cut-matching-game}
In the cut-matching game for directed graphs (see \autoref{sec:Louis} for definition), there is a cut player strategy so that, in $O(\log^2{n})$ iterations, the union of the matchings played by the matching player is an Eulerian graph with edge expansion $\Omega(\log{n})$.
\end{theorem}

This is an improvement over the cut-matching game by Louis~\cite{Lou10}, which only had an expansion lower bound of $\Omega(1)$.
A corollary of \autoref{thm:cut-matching-game} is a simple almost linear-time $O(\log n)$-approximation algorithm for directed edge expansion.


\subsubsection{Unifying Framework}

The reweighted eigenvalue formulations in~\cite{KLT22,LTW23} provide a unifying framework to obtain Cheeger-type inequalities for vertex expansion, directed graph expansions, and hypergraph expansions.
In this study, we show that in all these cases, adding $\ell_2^2$ triangle inequality constraints to the reweighted eigenvalue formulations gives $O(\sqrt{\log{n}})$-approximation algorithms for estimating these quantities, as well as fast algorithms for computing such approximations using expander flows and the chaining techniques~\cite{ARV09,AK07,Kal07,She09}.
Our results bring the more general expansion problems closer to the basic undirected edge expansion problem, since both the formulations and the proofs are close analogs of the corresponding results for undirected edge expansion.
Moreover, our proofs show that the MMWU method and the max-flow min-cut theorem can also be used to recover the Cheeger-type inequality and the cut-matching game, providing a common framework to analyze these different algorithmic techniques for graph expansion problems.
Overall, we believe that our results simplify and unify the state-of-the-art of various problems and approaches studied in the literature.

\section{Technical Review and Overview}

Since our work revisits and extends several previous works~\cite{ARV09,KRV06,AK07,Kal07,She09,ACMM05,Lou10,LTW23}, we review these previous techniques and mention some of our ideas for improvements along the way in the corresponding subsections, and we conclude with the common themes in~\autoref{sec:techniques}.

{\bf Notations:}
We introduce some notation that we will use throughout the paper. We use $\R_+$ to denote the set of positive real numbers and $\R_{\ge 0}$ to denote the set of non-negative real numbers.
Given two functions $f, g: X \rightarrow \R_{\ge 0}$, we use $f \lesssim g$ to denote the existence of a positive constant $c > 0$, such that $f \le c \cdot g$ always holds.
We use $f \sim g$ to denote $f \lesssim g$ and $g \lesssim f$.


\subsection{Semidefinite Program with Triangle Inequalities}

The seminal work of Arora, Rao and Vazirani~\cite{ARV09} proved that the following Goemans-Linial SDP relaxation for the undirected sparsest cut problem has an integrality gap of $O(\sqrt{\log n})$.
\begin{align}
  \label{eqn:ARV}
  \begin{split}
     \min_{v_1, \ldots, v_n \in \mathbb{R}^n} 
     &~~~ \sum_{ij \in E} \norm{v_i - v_j}^2 \\
     \st 
     &~~~\sum_{i<j} \norm{v_i - v_j}^2 = 1\\
     &~~~\norm{v_i - v_k}^2 + \norm{v_k - v_j}^2 \geq \norm{v_i-v_k}^2\;\qquad \forall i,j,k\in V.
  \end{split}
\end{align}
Note that this formulation without the triangle inequalities in the last line is equivalent to the second smallest eigenvalue of the normalized Laplacian matrix when the graph is regular (see e.g.~\cite{Tre16}).

A major contribution in~\cite{ARV09} is a structure theorem on vectors satisfying the $\ell_2^2$ triangle inequalities.
It asserts that, given a ``well-spread'' set of vectors satisfying the $\ell_2^2$ triangle inequalities, there are two large subsets $L$ and $R$, such that all vectors in $L$ are far away from all vectors in $R$.

\begin{definition}[Well-Spread Vectors] \label{def:well-spread}
Let $\{v_i\}_{i=1}^n$ be a set of vectors that satisfy $\sum_{i<j} \norm{v_i - v_j}_2^2 = n^2$.
Let $B(i, \delta) := \{j \in V: \norm{v_j - v_i}_2 \leq \delta\}$ denote the closed $\delta$-ball centered at $v_i$. 
We say that $\{v_i\}_{i=1}^n$ is well-spread if $\Big|B\Big(i, \frac{1}{\sqrt{10}}\Big)\Big| \le \frac{n}{10}$ for all $i \in V$.
\end{definition}

\begin{theorem}[$\ell_2^2$ Structure Theorem {\cite[Theorem 1]{ARV09}}]
  \label{thm:arv-structure}
  Let $\{v_i\}_{i=1}^n$ be a set of vectors\footnote{In~\cite{ARV09}, the vectors $v_i$ are assumed to be of unit length.  We note that the structure theorem holds without this assumption as well; see for example \cite{Rot16} for a writeup.} 
  that satisfy the $\ell_2^2$ triangle inequalities and $\sum_{i, j \in V} \norm{v_i - v_j}^2 = n^2$.
  If $\{v_i\}_{i=1}^n$ is well-spread, then there exist two sets $L, R \subseteq V$ such that $|L|, |R| \ge \Omega(n)$ and
  \[
    d(L, R) := \min_{i \in L, j \in R} \norm{v_i - v_j}^2 \gtrsim 1/\sqrt{\log n}.
  \]
  Moreover, there is a randomized polynomial-time algorithm that finds such sets with high probability.
\end{theorem}

The proof consists of novel geometric arguments involving measure concentration and chaining.
We will use \autoref{thm:arv-structure} straightforwardly to prove that the new SDP formulation in \autoref{def:main-sdp} has integrality gap $O(\sqrt{\log n})$.
We will also use a refined version of the chaining result by Sherman~\cite{She09} for our fast algorithm in \autoref{thm:directed-Sherman}.

\subsection{Expander Flows} \label{sec:expander-flows-review}

Another important contribution of~\cite{ARV09} is the concept of expander flows.
The idea of using multi-commodity flow to certify edge expansion was first introduced by Leighton and Rao~\cite{LR99}.

\begin{definition}[Multi-Commodity Flow and Demand Graph]
\label{def:demand}
Let $G=(V,E,w)$ be an edge-capacitated undirected graph.
Given demands $d_{ij}$ for each $i,j \in V$,
a multicommodity flow $f$ assigns a value $f_p \ge 0$ to each path $p$ in $G$ such that
(i) $\sum_{p \ni e} f_p \leq w_e$ for all $e \in E$ and
(ii) $\sum_{p \in \mathcal{P}_{ij}} f_p = d_{ij}$ for all $i,j \in V$, where $\mathcal{P}_{ij}$ denotes the set of paths from $i$ to $j$.
The demand graph $D$ is defined on the same vertex set $V$, with the weight of each edge $ij$ being $d_{ij}$.
\end{definition}

For an edge-capacitated undirected graph $G=(V,E,w)$,
let
\[\Phi(G) := \frac{\min_{S \subseteq V: |S| \leq |V|/2} w(\delta(S)) }{|S||\overline{S}|}\]

be the value of the sparsest cut of $G$.
If there is a multi-commodity flow in $G$ with demand graph $D$, 
then it is not difficult to check that $\Phi(G) \geq \Phi(D)$.
Leighton and Rao~\cite{LR99} used linear programming with the demand graph $D=K_n$, the complete graph on $n$ vertices, to approximate the sparsest cut of $G$ up to an approximation ratio $O(\log n)$.

The new idea in~\cite{ARV09} was to use semidefinite programming to search for a demand graph $D$ with a feasible multi-commodity flow on $G$, and to lower bound  the sparsest cut of $G$ using the second eigenvalue of the Laplacian matrix of $D$ through Cheeger's inequality. 
This approach can be summarized as
\begin{align} \label{eqn:expander-flow}
    \max_{D,f}
    & \quad
    \lambda_2(L(D))
    \\
    \st
    & \quad
    \text{$f$ is a multi-commodity flow on $G$ with demand graph $D$}
    \nonumber
\end{align}

\subsubsection{Expander Flows vs Dual Program} \label{sec:expander-flow-dual}

Indeed, the above approach of lower bounding $\Phi(G)$ can be understood as lower bounding the objective value of the dual of the Goemans-Linial SDP in (\ref{eqn:ARV}).
To see this, we first express the triangle inequalities as
\begin{align*}
   \norm{v_{i_1}-v_{i_2}}^2 + 
    \norm{v_{i_2} - v_{i_3}}^2 +
    \dots + 
    \norm{v_{i_{\ell-1}} - v_{i_{\ell}}}^2
    \geq
    \norm{v_{i_1} - v_{i_\ell}}^2
    \quad \forall p = (i_1,\dots, i_\ell)\in \mathcal{P}(K_n),
\end{align*}

where $\mathcal{P}(K_n)$ denotes the set of paths in the complete graph $K_n$ on the same vertex set $V$.
We write the primal program in matrix form.
Let $U$ be the matrix with the $i$-th column being $v_i$ for $1 \leq i \leq n$ and let $X = U^T U$.
Let $L_{i,j}$ be the Laplacian of the edge $ij$ and 
\begin{equation} \label{eqn:Tp}
    T_p := \sum_{k=1}^{\ell-1} L_{i_k,i_{k+1}} - L_{i_1,i_l}.
\end{equation}
Then the Goemans-Linial SDP in (\ref{eqn:ARV}) can be written as
\begin{align} \label{eqn:ARV-matrix}
    \min_{X \succeq 0}
    & \quad
    \inner{L(G)}{X} \nonumber
    \\
    \st
    & \quad
    \inner{L(K_n)}{X} = 1 
    \\
    & \quad
    \inner{T_p}{X} \geq 0\quad\forall p\in \mathcal{P}(K_n). \nonumber
\end{align}
One can check that strong duality holds, and the dual program can be written as
\begin{align*}
     \max_{f_p\geq 0 : p\in \mathcal{P}(K_n)} & \quad  \lambda 
    \\
    \st& \quad
    \lambda \cdot L(K_n) \preceq L(G) - \sum_p f_p T_p.
\end{align*}
Therefore, the dual program of the Goemans-Linial SDP can be succinctly written as 
\begin{equation} \label{eqn:ARV-dual}
\max_{f}~~ \lambda_2\Big(L(G) - \sum_{p} f_p T_p\Big).
\end{equation}

The expander flow formulation in~\eqref{eqn:expander-flow} is weaker than this dual program.

\begin{claim}
The objective value of (\ref{eqn:expander-flow}) is a lower bound on the objective value of (\ref{eqn:ARV-dual}).
\end{claim}
\begin{proof}
Let $f$ be a multi-commodity flow on $G$ with demand graph $D$, 
and $F$ be the $n \times n$ matrix with $F(i,j) = \sum_{p \ni ij} f_p$.
Then, check that $\sum_p f_p T_p = L(F) - L(D)$, and hence
\[
L(D) = L(F) - \sum_p f_p T_p \preceq L(G) - \sum_p f_p T_p 
\quad \implies \quad
\lambda_2(L(D)) \leq \lambda_2\Big( L(G) - \sum_p f_p T_p  \Big),
\]
where the inequality $L(F) \preceq L(G)$ is because $F(i,j) \leq w_{ij}$ for all $(i,j) \in V \times V$.
\end{proof}

We remark that all previous works on undirected graphs~\cite{ARV09,KRV06,AK07,Kal07,She09} use the expander flow formulation in (\ref{eqn:expander-flow}) to approximate sparsest cut.
This can be understood as the dual program in (\ref{eqn:ARV-dual}) with the additional constraint that $\sum_{p \ni ij} f_p \leq w_{ij}$ for all $i,j \in V \times V$, which in particular implies that only the path variables $f_p$ when $p$ is a path in $G$ are used. Since we will discuss several variations of the program (\ref{eqn:ARV}) and take their duals, we will refer to dual programs with additional capacity constraints on the $f_p$ variables such as (\ref{eqn:expander-flow}) as the ``constrained dual programs'' and the original dual programs such as (\ref{eqn:ARV-dual}) as the ``unmodified dual programs.''

In proving \autoref{thm:directed-Sherman}, we will use the unmodified dual program of $\lambda^\Delta_\pi$.
As we will explain later, this will allow us to design a simpler primal-dual algorithm using the MMWU method, bypassing the use of multi-commodity flow as in~\cite{AK07,Kal07,She09}.

\subsection{Cut-Matching Game} \label{sec:cut-matching-game}

The cut-matching game was first introduced by Khandekar, Rao and Vazirani~\cite{KRV06} 
as a fast combinatorial method for approximating sparsest cut in undirected graphs using flows. 
In this game, there is a cut player and a matching player who try to build an expander from the empty graph as follows.
In each round, the cut player chooses a bisection $(S,\overline{S})$ of the vertices, and the matching player chooses a perfect matching between $(S,\overline{S})$. 
The goal of the cut player is to minimize the number of rounds so that the union of the matchings is guaranteed to be a good expander.
Khandekar, Rao and Vazirani~\cite{KRV06} gave a cut player strategy that builds a graph with $\Omega(1)$ edge expansion in $O(\log^2 n)$ rounds.
Orecchia, Schulman, Vazirani, and Vishnoi~\cite{OSVV08} gave an improved cut player strategy that builds a graph with $\Omega(\log n)$ edge expansion in $O(\log^2 n)$ rounds.
The proofs of these results are based on ad-hoc potential functions, although in hindsight the algorithm in~\cite{OSVV08} is very similar to the one using MMWU method in~\cite{AK07}.

The original motivation of the cut-matching game is to build an expander flow to approximate sparsest cut.
In each round, we aim to send a flow between the cut $(S,\overline{S})$ provided by the cut player.
On the one hand, if such a flow cannot be sent, then we obtain a sparse cut by the max-flow min-cut theorem and the algorithm stops.
On the other hand, if such a flow can be sent, then the demand pairs routed by this flow
form a perfect matching between $S$ and $\overline{S}$.
Therefore, if we successfully send such a flow in each round, then the average of the flows is a multicommodity flow in the original graph, with the demand graph being the average of the perfect matchings, which is guaranteed to be an expander by the cut-matching game. 
In this case, we can prove a lower bound on the sparsest cut by the expander flow formulation in~(\ref{eqn:expander-flow}), with the approximation ratio depending on the parameters in the cut-matching game.
The cut player strategy in~\cite{KRV06} gave an $O(\log^2 n)$-approximation for undirected sparsest cut using $O(\log^3 n)$ max-flow computations,
while the one in~\cite{OSVV08} gave an $O(\log n)$-approximation using $O(\log^3 n)$ max-flow computations.

We remark that the cut-matching game has become a useful algorithmic tool on its own, with interesting applications in other important problems such as edge-disjoint paths~\cite{And10,Chu12,CL16} and dynamic graph problems~\cite{CGLNPS20,BGS20}.

\subsection{Matrix Multiplicative Weight Update Method}
\label{sec:mmw}

Arora and Kale~\cite{AK07,Kal07} developed a general primal-dual framework to solve SDPs using the matrix multiplicative weight update method. 
For our purpose, it would be better to understand this method from the viewpoint of regret minimization, which is the setting in online optimization.
In each iteration $t$, the player chooses a density matrix $X_t$, which represents a probability distribution over the set of unit vectors.
The player then observes a feedback matrix $M_t$ with bounded spectral norm and incurs a loss of $\inner{X_t}{M_t}$.
The objective of the player is to minimize the total loss.
In hindsight, if the player had knowledge of all the feedback matrices $M_t$ from the start, then the best strategy would be to choose the density matrix $vv^T$ where $v$ is a unit-length minimum eigenvector of $\sum_{t} M_t$, with total loss $\lambda_{\min}(\sum_t M_t)$.
The regret of the player is thus defined as $\sum_{t} \inner{M_t}{X_t} - \lambda_{\min}(\sum_{t} M_t)$, the difference of the player's loss to this offline loss.
Arora and Kale~\cite{AK07,Kal07} analyzed the following algorithm that sets $X_t$ to be the matrix exponential of the feedback matrices.

\begin{algorithm}[ht]\caption{Matrix Multiplicative Weight Update Algorithm}
 \label{alg:mmw}
    \textbf{Initialization:}
    $X_0 = \frac{1}{n} I_n$, $\eta\in (0,1)$ as a step size\\
    \textbf{For} $t=0, \dots, T-1$
    \begin{enumerate}
        \item Observe feedback matrix $M_t$ such that $\norm{M_t} \leq \rho$. Incur a loss of $\inner{M_t}{X_t}$.
        \item Compute $X_{t+1}' := \exp(-\eta \sum_{i=0}^t\frac1\rho M_i)$ and update $X_{t+1} := X_{t+1}'/\tr(X_{t+1}')$.
    \end{enumerate}
\end{algorithm}

The requirement that $M_t$ has bounded spectral norm, or $\norm{M_t}\leq \rho$, is to control the regret bound. 
The $\rho$ parameter is called the ``width'' and is the key parameter in analyzing the matrix multiplicative weight update method in many applications.

\begin{theorem}[Regret Bound {\cite[Theorem 10]{Kal07}}]
    \label{thm:regret-bound}
    After $T$ iterations of \autoref{alg:mmw}, 
    let $\overline{M} := \frac1T\sum_{t=0}^{T-1} M_t$, then
    \begin{align}
        \label{eqn:general-regret-bound}
        \lambda_{\min}(\overline{M}) \gtrsim \frac{1}{T}\sum_{t=0}^{T-1} \inner{M_t}{X_t} - \eta\rho- \frac{\rho\log{n}}{\eta T}.
    \end{align}
    If, in addition, each $M_t$ satisfies $M_t \succeq 0$, then we have the stronger bound that
    \begin{align}
        \label{eqn:psd-regret-bound}
        \lambda_{\min}(\overline{M}) \gtrsim \frac{1}{T}\sum_{t=0}^{T-1} \inner{M_t}{X_t}(1 - \eta) - \frac{\rho\log{n}}{\eta T}.
    \end{align}
\end{theorem}

\autoref{thm:regret-bound} is a key result that we will use to design fast algorithms.

%

\subsection{Primal-Dual Algorithms for Sparsest Cut} \label{sec:primal-dual-sparsest}

Arora and Kale~\cite{AK07} uses the regret bound in \autoref{thm:regret-bound} to design a primal-dual algorithm for approximating the sparsest cut problem.
The setup is to either certify that the optimal value is at least $\Omega(\alpha)$ by building an expander flow solution to (\ref{eqn:expander-flow}), or to find a cut of sparsity at most $O(\sqrt{\log n} \cdot \alpha)$.
In each iteration, the algorithm uses the density matrix $X_t$ given by the matrix multiplicative weight update algorithm as a candidate primal solution to (\ref{eqn:ARV-matrix}).
To build a dual solution to (\ref{eqn:expander-flow}), the idea is to use the regret minimization framework to reduce to the simpler task of finding a multi-commodity flow $f_t$ whose demand graph $D_t$ satisfies $\inner{L(D_t)}{X_t} \geq \alpha$.
If such a multi-commodity flow with demand graph $D_t$ can be found in each iteration $t$ for $O(\log n)$ iterations, then the regret bound in \autoref{thm:regret-bound} would imply that $\lambda_2(L(\frac{1}{T}\sum_t D_t)) \gtrsim \alpha$, and thus the average of the flows $f_t$ is an expander flow solution to (\ref{eqn:expander-flow}) with objective value at least $\Omega(\alpha)$.

The remaining task is that, given a density matrix $X_t$, either to find a multi-commodity flow $f_t$ whose demand graph $D_t$ satisfies $\inner{L(D_t)}{X_t} \geq \alpha$ and $\norm{L(D_t)} \leq \rho$, or to find a cut with sparsity at most $O(\sqrt{\log n} \cdot \alpha)$.
This task is usually called implementing the ``oracle'' for the MMWU method.
To do so, consider the Gram decomposition $v_1,\ldots, v_n$ of $X$ and note that $\inner{L(D_t)}{X_t} = \sum_{i,j} D_t(i,j) \norm{v_i - v_j}^2$.
To ensure that the width $\rho$ is small, the algorithm only searches for demand graphs with bounded maximum degree.
To ensure that the inner product $\inner{L(D_t)}{X_t}$ is large, the algorithm only routes flow between pairs of vertices $(i, j)$ with $\norm{v_i-v_j} = \Omega(1)$.
If such a multi-commodity flow can be sent, then the oracle succeeds and the primal-dual algorithm proceeds to the next iteration.
If not, using the dual solution to the multi-commodity flow problem, along with the geometric chaining arguments used in~\cite{ARV09}, they showed how to find a cut with sparsity at most $O(\sqrt{\log n} \cdot \alpha)$ (see \cite[Lemma 6.6 and Theorem 6.7]{AK07}).
The time complexity of their algorithm is $\tilde{O}(n^2)$, where the bottleneck is in the multi-commodity flow computation in the implementation of the oracle.

To achieve $O(\log n)$-approximation, there is a much easier way to implement the oracle using only max-flow computations.
The algorithm is to project the vectors $v_1,\ldots,v_n$ along a random direction, and set up a single-commodity flow between the $\Omega(n)$ vertices with the lowest projection values and the $\Omega(n)$ vertices with the highest projection values.
This algorithm is very similar to the cut-matching game in~\cite{OSVV08} that uses matrix exponentials to define a cut-player strategy. 

\subsection{Almost Linear-Time Primal-Dual Algorithm}

Sherman~\cite{She09} pushed the approach in~\cite{AK07} further to almost get the best of the semidefinite programming approach ($O(\sqrt{\log n})$-approximation) and the combinatorial cut-matching game approach (near linear-time algorithms).  

The approach in~\cite{She09} is to use an inner multiplicative weight update algorithm to compute the multicommodity flow in the oracle implementation, rather than doing it in a black-box manner as in~\cite{AK07}.
Specifically, each iteration of this inner multiplicative weight update algorithm consists of chaining together matchings corresponding to 
flow paths of single-commodity flows.
The single-commodity flows are set up using the random projection method as in the $O(\log n)$-approximation in~\cite{AK07}, but the random directions for these flows are correlated and the distribution of the random directions is explicit and can be sampled efficiently.
The main contribution of \cite{She09} was to show that, after chaining together $\Theta(\sqrt{\log{n}})$ of these correlated random matchings, one can find not just one (as in \cite{ARV09}), but many flow paths between pairs $(i, j)$ such that $\norm{v_i-v_j}$ is $\Omega(1)$. 
Using this chaining method as a subroutine, one can either find a good multicommodity flow whose demand graph satisfies $\sum_{i,j}D(i,j) \norm{v_i-v_j}^2\geq \alpha$ in $O(n^{1+\eps})$ time by running the inner multiplicative weight update algorithm,
or find some direction along which the single commodity flow cannot be sent and an associated min-cut $S$ with $\phi(S)\lesssim \alpha \cdot \sqrt{\frac{\log{n}}{\eps}}$.

Sherman's algorithm and its analysis are rather technical and we will provide more details in \autoref{sec:well-spread-case}.
We will use his main chaining result as a black-box in our algorithm for \autoref{thm:directed-Sherman}.


\subsection{Directed Semi-Metric for Directed Sparsest Cut} \label{sec:ACMM}

Agarwal, Charikar, Macharychev and Macharychev~\cite{ACMM05} introduced an SDP for approximating directed sparsest cut using a directed semi-metric. 
The idea was to introduce an extra vector $v_0$ to the embedding, and to define the semi-metric as $d(i,j) := \norm{v_i - v_j}^2 - \norm{v_i - v_0}^2 + \norm{v_j - v_0}^2 \geq 0$. 
The program is formulated as follows:
\begin{align}
  \label{eqn:acmm}
  \begin{split}
    \min_{v: V \cup \{0\} \rightarrow \R^n} &~~~  
    \sum_{ij \in E} w(i,j) \left(\norm{v_i - v_j}^2 - \norm{v_i - v_0}^2 + \norm{v_j - v_0}^2 \right)
    \\
    \st&~~~
    \norm{v_i - v_j}^2 + \norm{v_j - v_k}^2 \geq \norm{v_i - v_k}^2  \quad \quad \forall i,j,k \in V \cup \{0\}
    \\
    &~~~
    \sum_{i \in V} \pi(i) \cdot v_i = \vec{0}
    \\
    &~~~ \sum_{i \in V} \pi(i) \cdot \norm{v_i}^2 = 1.
  \end{split}
\end{align}
The $\primal$ program that we introduce in \autoref{def:main-sdp} is less constrained than this program. 
We can see this by taking the linear programming dual of the inner maximization problem with respect to the $F(i,j)$ variables (see \cite[Lemma 3.21 and Lemma 3.22]{LTW23}):
\begin{align*}
    \max_{F\in \mathcal{F}(G)} \sum_{i<j} \big(F(i,j) + F(j,i)\big) \cdot \norm{v_i - v_j}^2
    =
    \min_{r: V \rightarrow \R} \sum_{ij\in E}
    w(i,j) \cdot \max \Big\{0, \norm{v_i - v_j}^2 - r(i) +r(j) \Big\}.
\end{align*}
Thus, we see that every feasible solution to \eqref{eqn:acmm} corresponds to a feasible solution to the $\primal(G)$ program with the same objective value by taking $r(i) = \norm{v_i - v_0}^2$.
The reason we present $\primal$ throughout the paper in the min-max form is that all our analyses make use of this min-max formulation of the problem, as it can be naturally captured by flows.

\subsection{Primal-Dual Algorithm for Directed Sparsest Cut} \label{sec:Arora-Kale-directed}

Arora and Kale~\cite{AK07,Kal07} used the matrix multiplicative update method on the SDP~(\ref{eqn:acmm}) in~\cite{ACMM05} to obtain a primal-dual $O(\sqrt{\log{n}})$-approximation algorithm for directed sparsest cut.

One important difference with the algorithm for undirected sparsest cut is that they used the unmodified dual program of (\ref{eqn:acmm}), which can be expressed as $\max_f \lambda_2\big(\vec{L}(D) - \sum_p f_pT_p\big)$, where $D$ is the demand graph of a flow on $G$ (see \autoref{sec:expander-flow-dual} for a discussion about these two dual programs).
Recall from our previous discussion that simply using $\vec{L}(D)$ instead of $\vec{L}(D) - \sum_p f_p T_p$ as in the undirected case (i.e.~using the constrained dual program instead of the unmodified dual program)  would correspond to only enforcing $\ell_2^2$ triangle inequalities along paths in the directed graph.
Since paths in the directed graph are restricted by the orientation of the edges, it seems arbitrarily restrictive to only enforce triangle inequalities along directed paths. 

Using the dual program $\max_f \lambda_2\big(\vec{L}(D) - \sum_p f_pT_p\big)$,
there remains an important difference between the primal-dual algorithm here with that for the undirected sparsest cut.
Unlike in the algorithm for undirected sparsest cut, 
the algorithm for directed sparsest cut does not involve the use of multicommodity flows.
Instead, it tries to find a single-commodity flow $f$ with demand graph $D$ that pushes a lot of flow between pairs of vertices $(i,j)$ such that $\norm{v_i-v_j}_2$ is large, and to then use $\vec{L}(D)$ as the feedback matrix.
If it fails to do so, then it finds many paths that violate the $\ell_2^2$ triangle inequality, and it then uses $-y \sum_p T_p$ as the feedback matrix, where the sum is over the violating paths $p$ and $y$ is an appropriate scaling factor.
The procedure for finding violating paths is implemented in time $O(n^{2+o(1)})$ using a special data structure about dynamic decremental spanners.
This is the bottleneck and thus the runtime per matrix multiplicative weight update iteration is $O(n^{2+o(1)})$.

The original claim in~\cite{Kal07} was that $O(\log n)$ iterations suffice, but Chan and Sun~\cite{CS18} found that their analysis should only yield the weaker bound of $\tilde{O}(n^2)$ iterations and thus a total runtime of $O(n^{4+o(1)})$. This is because of a technical issue in bounding the trace of feasible solutions in the primal program (see footnotes 1 and 2 in~\cite{CS18}, with Kale's acknowledgement).
Chan and Sun simplified their approach and obtained an $O(n^{4})$-algorithm with the same approximation ratio $O(\sqrt{\log n})$, that also works for directed hypergraphs.

As mentioned in \autoref{sec:expander-flows-review},
we will use the unmodified dual program of $\lambda_\pi^\Delta$ similar to how Arora-Kale's uses the unmodified dual program of (\ref{eqn:acmm}).
We also use their ``flows or violating paths'' oracle for this dual program,
thus bypassing the multicommodity flow computation in~\cite{AK07,She09}.
We observe that Sherman's chaining result can be used to find many violating paths efficiently, without using any special data structures.
This gives us an almost linear-time $O(\sqrt{\log n})$-approximation algorithm for directed sparsest cut, which also simplifies the corresponding algorithm for undirected sparsest cut.

We end this subsection with the following technical remark about the primal-dual algorithm for directed sparsest cut using the SDP in~\cite{ACMM05}.

\begin{remark} \label{rem:directed-technical}
Because of the directed semi-metric with the special vector $v_0$,
Arora and Kale needed to work with a non-PSD Laplacian $\vec{L}(G)$ with vertex set $V \cup \{0\}$ and with both positive and negative edge weights (specifically, edges $(i,j)$ and $(j,0)$ have weight $1$ while edge $(i,0)$ has weight -1).
the Laplacian of the demand graph of a flow is used as a feedback matrix in each iteration.
However, the newly introduced vertex $0$ has large degree in any demand graph, thus making it difficult to bound the spectral norm of the feedback matrix, i.e. the width of the oracle.
To address this, Arora and Kale duplicated the vertex $0$ into $n$ copies, and considered a graph on $2n$ vertices in order to have a better bound on the width.

One advantage of our formulation $\primal$ in \autoref{def:main-sdp} is that it is defined on the original graph, and this simplifies the primal-dual algorithm and the analysis for \autoref{thm:directed-Sherman} considerably.
\end{remark}

\subsection{Cut-Matching Game for Directed Graphs} \label{sec:Louis}

Louis~\cite{Lou10} developed a cut-matching game for directed graphs,
where the matching player plays a directed matching, which is defined as an Eulerian graph where each vertex has indegree and outdegree exactly one. 
He analyzed a cut-player strategy that is similar to the one in~\cite{KRV06} and proved that, in $O(\log^2 n)$ iterations, the union of the directed matchings is an Eulerian graph with edge expansion $\Omega(1)$.

For undirected graphs, the matrix multiplicative update method can be used to give an improved cut-player strategy~\cite{AK07,OSVV08}.
For directed graphs, however, the primal-dual algorithm is more complicated because of the directed semi-metric formulation as discussed in \autoref{rem:directed-technical}, and it does not directly translate to a cut-matching game.
Using the simpler $\primal$ formulation in \autoref{def:main-sdp}, which also has a natural correspondence with Eulerian subgraphs, we obtain an improved cut-player strategy as stated in \autoref{thm:cut-matching-game} using the matrix multiplicative weight update method on $\primal$.


\subsection{Reweighted Eigenvalues for Directed Graphs} \label{sec:LTW23}
 
Lau, Tung, and Wang~\cite{LTW23} defined the reweighted eigenvalue for directed edge expansion and use it to prove a Cheeger-type inequality for directed graphs. 
Given a directed graph $G = (V, E, w)$ with edge weights $w:E \to \R_+$, 
the maximum reweighted second eigenvalue problem seeks to find a circulation $F$ satisfying edge capacity constraints (see \autoref{def:main-sdp}) such that the second smallest eigenvalue of the symmetric Laplacian of $F$ is maximized. 

\begin{definition}[Maximum Reweighted Second Eigenvalue for $\pi$-Weighted Edge Expansion]
\label{def:reweighted-eigenvalue}
Given an edge-capacitated directed graph $G = (V, E, w)$ and vertex weights $\pi: V \to R_+$, define the maximum reweighted second eigenvalue as
\[
    \lambda_2^*(G) := \max_{F\in \mathcal{F}(G)}
    \lambda_2\bigg(
      \Pi^{-1/2} \bigg( D_F - \frac{F+F^\top}{2} \bigg) \Pi^{-1/2}
    \bigg)
\]
where $\Pi = \diag(\pi)$, $F$ is the $n \times n$ adjacency matrix of the circulation, and $D_F$ is the diagonal degree matrix of $(F+F^T)/2$ with $D_F(i,i) = \sum_{j \in V} \frac12 (F(i,j)+F(j,i))$ for $1 \leq i \leq n$. 
\end{definition}

Using the semidefinite programming formulation for the second eigenvalue and von-Neumann min-max theorem, 
$\lambda_2^*(G)$ can be rewritten as the form in \autoref{def:main-sdp} without the triangle inequalities.

The directed edge conductance $\vec{\phi}$ studied in~\cite{LTW23} is a special case of the directed edge expansion $\edgeexp$ in \autoref{def:phi-pi} when $\pi(i) = w(\delta^+(i)) + w(\delta^-(i))$ for all $i \in V$.
The directed Cheeger inequality in~\cite{LTW23} states that
\begin{equation} \label{eqn:Cheeger-directed}
\lambda_2^*(G) \lesssim \vec{\phi}(G) 
\lesssim \sqrt{ \lambda_2^*(G) \cdot \log \frac{1}{\vec{\phi}(G)} }. 
\lesssim \sqrt{ \lambda_2^*(G) \cdot \log \frac{1}{\lambda_2^*(G)} }. 
\end{equation}

In \autoref{thm:fast-Cheeger}, we provide an almost linear-time algorithm to return a set $S$ with $\vec{\phi}(S) \leq \sqrt{ \lambda_2^*(G) \cdot \log \frac{1}{\lambda_2^*(G)} }$.
The idea is to use the regret minimization framework to construct an optimal circulation iteratively, and the observation is that this converges quickly when $\lambda_2^*(G)$ is large.
This combines with our almost linear-time $O(\sqrt{\log n})$-approximation algorithm in \autoref{thm:directed-Sherman} gives \autoref{thm:fast-Cheeger}.

\subsection{Our Techniques} \label{sec:techniques}

We have already discussed the ideas of our main results in \autoref{sec:results} in the corresponding subsections above when we reviewed the previous techniques.
Here we highlight two common themes in our techniques.

One common theme is called the ``metric rounding lemma'' that we prove in \autoref{sec:metric-rounding}, which is to use the max-flow min-cut theorem to find a sparse cut in a geometric embedding of the graph.
All the algorithms in this paper use this lemma to find sparse cuts, including the almost linear-time $O(\sqrt{\log n})$-approximation in \autoref{thm:directed-Sherman}, the improved cut-matching game in \autoref{thm:cut-matching-game}, and interestingly even the Cheeger-type result whose original proof in~\cite{LTW23} is based on a threshold rounding algorithm.

Another common theme is the matrix multiplicative weight update method developed in~\cite{AK07}.
All the algorithms in this paper use this method to construct the dual objects, including the expander flows in the $O(\sqrt{\log n})$-approximation in \autoref{thm:directed-Sherman} and the cut-matching game in \autoref{thm:cut-matching-game}, as well as the circulation in reweighted eigenvalues in \autoref{thm:fast-Cheeger} and in the dual certificate in \autoref{prop:reweighted-phi}.
The cut-matching game was considered original when it was introduced, but now we see that it can be derived systematically from the matrix multiplicative weight update method.


An important element in all our results is the reweighted eigenvalue formulation from~\cite{LTW23}.
We believe that it is the right formulation, as it allows us to extend all known results for undirected graphs to directed graphs, in a way that is consistent with the formulations and the proofs for undirected graphs.
As we discuss in \autoref{sec:others}, our technique of adding $\ell_2^2$ triangle inequalities to reweighted eigenvalue formulations can be extended to directed vertex expansion and hypergraph edge expansion as well, providing a unifying method to extend the results for undirected graphs to more general settings. 




\subsection{Organization}
In \autoref{sec:rounding-algorithms},
we present the metric rounding lemma, and use it to prove \autoref{thm:primal-rounding} and to provide an alternative proof of the directed Cheeger inequality. 
In \autoref{sec:dual-rounding}, we extend Sherman's result to directed graphs and prove \autoref{thm:directed-Sherman}.
In \autoref{sec:primal-dual}, we also use the matrix multiplicative weight update method to compute reweighted eigenvalues, proving \autoref{thm:fast-Cheeger} and to design cut-matching game, proving \autoref{thm:cut-matching-game}.
Finally, in \autoref{sec:others}, we outline how these results can be extended easily to vertex expansion and to hypergraphs.

\section{Rounding Algorithms} \label{sec:rounding-algorithms}

In this section, we first present the metric rounding lemma in \autoref{sec:metric-rounding}. 
Then, we will use it to prove that $\primal$ in \autoref{def:main-sdp} has integrality gap $O(\sqrt{\log n})$ in \autoref{sec:ARV-approx}, and also to provide an alternative proof of the Cheeger-type inequality in~\cite{LTW23} in \autoref{sec:Cheeger}.

\subsection{Metric Rounding Lemma}
\label{sec:metric-rounding}

The following metric rounding lemma will be used to find sparse cuts in all algorithms in this paper.

\begin{lemma}[Metric Rounding Lemma]
    \label{lemma:metric-rounding-lemma}
    Let $G = (V, E, w)$ be an edge-capacitated directed graph.
    Let $d(\cdot, \cdot)$ be a metric on $V$, and let
    $\pi: V \rightarrow \mathbb{R}^+$ be an arbitrary weight function over $V$.
    Suppose we are given disjoint vertex subsets $L, R\subseteq V$ as input to the algorithm. 
    Let $r := \pi(R) / \pi(L)$ and $r' := \max\{1,r\}$. Then there is an algorithm using $O(\log n)$ maximum flow computations to output a set $S$ with 
    \begin{align*}
        \edgeexp(S) &
        \lesssim
        \frac{
          r' \cdot \max_{F\in \mathcal{F}(G)} \sum_{i, j \in V} F(i, j) \cdot d(i, j)
        }{\sum_{i \in R} \pi(i) \cdot d(i, L)}.
    \end{align*}
\end{lemma}

Our proof of the lemma is constructive. 
\autoref{alg:bidirectional-max-flow}, Bidirectional Max-Flow, finds a maximum flow $\vec{f}$ from $L$ to $R$ and also a flow $\cev{f}$ from $R$ to $L$ with a prescribed target amount of flow.
If either of the flow is not ``saturating'', then we find a sparse cut $S$ using the max-flow min-cut theorem.
Otherwise, we combine $\vec{f}$ and $\cev{f}$ to form a circulation $F$, which helps upper bound the expansion of the graph through the flow parameter $\beta$.

\begin{algorithm}
 \caption{Bidirectional Max-Flow}
 \label{alg:bidirectional-max-flow}
 \textbf{Input:}
 Graph $G$, semi-metric $d(\cdot, \cdot)$, vertex weights $\pi:V \to \R^+$ as given in \autoref{lemma:metric-rounding-lemma};
 $L, R\subseteq V$ such that $L\cap R = \emptyset$, flow value parameter $\beta\in \mathbb{R}^+$, and congestion parameter $\kappa \in \mathbb{R}^+$
\begin{enumerate}
    \item
    Let $r := \pi(R)/\pi(L)$.
    Construct flow network $\vec{G}$ from $G$ as follows: add vertices $s$ and $t$ to $G$.
    Connect $s$ to each vertex $i \in L$ with an arc $(s, i)$ of capacity $r \cdot \beta \cdot \pi(i)$.
    Connect each vertex $j \in R$ to $t$ with an arc $(j, t)$ of capacity $\beta \cdot \pi(j)$.
    Multiply the capacities of the edges in $G$ by $\kappa$.
    
    \item
    Construct $\cev{G}$ in the same way as $\vec{G}$, but with arcs directed from $L$ to $s$ and from $t$ to $R$ instead.
    
    \item
    Compute $s$-$t$ maximum flow $\vec{f}$ on $\vec{G}$ and $t$-$s$ maximum flow $\cev{f}$ on $\cev{G}$.
    If one of $\vec{f}$ or $\cev{f}$ does not saturate all source and sink edges (i.e. if maximum flow value is less than $\beta \cdot \pi(R)$), output the minimum cut $S$ associated with the non-saturating flow. 
    Otherwise, output the circulation
    $F = \frac{1}{2}(\vec{f} +\cev{f})$.
\end{enumerate}
\end{algorithm}

In the case where the flows $\vec{f}$ and $\cev{f}$ are saturated, we upper bound the flow value parameter $\beta$. 

\begin{lemma}[Saturated Case]
    \label{lemma:metric-rounding-saturated}
    Suppose $d(\cdot,\cdot)$ is a metric and \autoref{alg:bidirectional-max-flow} outputs a circulation $F$. Then,
    \[
        \beta \leq
        \frac{\sum_{ij\in E} F(i, j) \cdot d(i, j)}
        {\sum_{i\in R} \pi(i) \cdot d(i, L)},
    \]
    where $F(i,j) = \frac{1}{2}\sum_{p\ni (i,j)}\big(\vec{f}(p) + \cev{f}(p)\big)$ defines the flow graph of the circulation returned in step 3.
\end{lemma}

\begin{proof}
Each flow can be decomposed into a set of (weighted) flow paths from source to sink.
For each $j \in R$, let $\vec{\mathcal{P}}(j)$ be the set of $s$-$t$ flow paths in $\vec{f}$ entering $t$ through $j$, and let $\cev{\mathcal{P}}(j)$ be the set of $t$-$s$ flow paths in $\cev{f}$ leaving $t$ through $j$.
For a particular flow path $p = (s, i_1, i_2, \dots, i_k, t)$ or $p = (t, i_k, i_{k-1}, \dots, i_1, s)$, let $d(p) = \sum_{\ell=1}^{k-1} d(i_\ell, i_{\ell+1})$ be its length.
Note that for any path $p \in \vec{\mathcal{P}}(j)$ or $p \in \cev{\mathcal{P}}(j)$, by triangle inequality, 
\[
  d(p) = \sum_{\ell=1}^{k-1} d(i_\ell, i_{\ell+1}) \geq d(i_1, i_k) \geq d(j, L),
\]
where the last inequality is because $i_k = j$ and $i_1 \in L$. Then,
 \begin{align*}
     \sum_{p \in \vec{f}} \vec{f}(p) \cdot d(p)
     + \sum_{p \in \cev{f}} \cev{f}(p) \cdot d(p)
     &=
     \sum_{j \in R} \left[
     \sum_{p \in \vec{\mathcal{P}}(j)} \vec{f}(p) \cdot d(p) + 
     \sum_{p' \in \cev{\mathcal{P}}(j)} \cev{f}(p') \cdot d(p')
     \right]
     \\
     & \geq
     \sum_{j \in R} d(j, L) \left[
     \sum_{p \in \vec{\mathcal{P}}(j)} \vec{f}(p)
     + \sum_{p'\in \cev{\mathcal{P}}(j)}\cev{f}(p') \right]
     \\
     &= 2 \beta \sum_{j \in R} \pi(j) \cdot d(j, L),
 \end{align*}
where the last equality is due to both $\vec{f}$ and $\cev{f}$ being saturating. Thus, we have
 \begin{align*}
     2 \sum_{ij \in E} F(i, j) \cdot d(i, j)
     &=
       \sum_{p \in \vec{f}} \vec{f}(p) \cdot d(p)
       + \sum_{p\in \cev{f}} \cev{f}(p) \cdot d(p)
    \geq
    2\beta \sum_{i \in R}\pi(i) \cdot d(i, L).
 \end{align*}
Rearranging gives the desired result.
\end{proof}

On the other hand, if either of the flows $\vec{f}$ or $\cev{f}$ is unsaturated, we extract from it a cut with bounded expansion. 
This is a slight extension of \cite[Lemma 3.7]{KRV06} to the $\pi$-weighted and vertex-capacitated settings, and so we include a proof here.

\begin{lemma}[Unsaturated Case]
    \label{lemma:flow-cut-lemma}
    Suppose \autoref{alg:bidirectional-max-flow} outputs a cut $S$. 
    Then $\vec{\phi}_\pi(S) \leq \beta r'/\kappa$, where $r' := \max\{1, r\}$.
\end{lemma}

\begin{proof}
Suppose $\vec{f}$ is the non-saturating flow; the other case is similar (we would look at $\delta^-(S)$ for $S$ defined below).
  We obtain from it a cut, which is a set of edges whose removal would make it impossible to go from $s$ to $t$.
  Let $S \subseteq V$ be the set of vertices reachable from $s$ after removing the cut edges.
  Let $V_s \subseteq L$ be the set of vertices connected by a cut edge from $s$, and let $V_t \subseteq R$ be the set of vertices connected by a cut edge to $t$. 
  We claim that
  \[
    w(\delta_G^+(S)) \leq
    \frac{\beta}{\kappa} \big( \pi(R) - r \cdot \pi(V_s) - \pi(V_t) \big)
    , \quad
    \pi(S) \geq \pi(L) - \pi(V_s), \quad
    \text{ and }
    \pi(V-S) \geq \pi(R) - \pi(V_t).
  \]
  The first inequality comes from the fact that $\{si \mid i \in V_s\} \cup \delta_G^+(S) \cup \{jt \mid j \in V_t\}$ is the minimum cut obtained, with total weight equal to $r \cdot \beta \cdot \pi(V_s) + \beta \cdot \pi(V_t) + \kappa \cdot w(\delta_G^+(S))$ by our construction of $\vec{G}$, which is at most $\beta \cdot \pi(R)$.
  The second and third inequalities follow from the facts that $L \setminus V_s \subseteq S$ and $R \setminus V_t \subseteq V-S$.
  Since $\pi(R) = r \cdot \pi(L)$, it follows that
  \[
     \vec{\phi}_\pi(S) = 
     \frac{w(\delta^+(S))}{\min\{\pi(S), \pi(V-S)\}}
     \leq
     \frac{\beta}{\kappa}
     \max \left\{\frac{\pi(R)-\pi(V_t)}{\pi(R)-\pi(V_t)},
     \frac{r(\pi(L)-\pi(V_s))}{\pi(L) - \pi(V_s)} \right\} 
     =
     \frac{\beta \cdot r'}{\kappa}.
  \]
\end{proof}

Now we are ready to prove the metric rounding lemma.

\begin{proof}[Proof of \autoref{lemma:metric-rounding-lemma}]
  In \autoref{alg:bidirectional-max-flow}, choose $\kappa = 2r'$.
  Let $\alpha$ be such that the algorithm outputs a circular flow $f$ when $\beta = \alpha$ and outputs a cut $S$ when $\beta = 2 \alpha$.
  When a cut $S$ is output at $\beta = 2\alpha$,
  by \autoref{lemma:flow-cut-lemma} (unsaturated case), 
  the vertex or edge expansion of $S$ is at most $\beta \cdot r' / \kappa = \alpha$.
  When a circulation $F$ is output at $\beta = \alpha$, then by construction $F' = F/\kappa$ is a circulation satisfying the edge or vertex capacity constraints of $G$, i.e. $F' \in \mathcal{F}(G)$.
  Therefore, by \autoref{lemma:metric-rounding-saturated} (saturated case), 
  \[
  \edgeexp(S) \leq
  \alpha \leq \kappa \cdot 
  \frac{\sum_{ij\in E} F'(i, j) \cdot d(i, j)}
  {\sum_{i \in R}\pi(i) \cdot d(i, L)}
  \leq 2r' \cdot \max_{F \in \mathcal{F}(G)} \frac{\sum_{ij\in E} F(i, j) \cdot d(i, j)}
  {\sum_{i \in R}\pi(i) \cdot d(i, L)} 
  \]
  Finally, note that we can find $\alpha$ using binary search on the range $[\Omega(1/\!\poly(n)), O(\poly(n))]$. 
  Therefore, we only need to invoke \autoref{alg:bidirectional-max-flow} $O(\log n)$ times, leading to a total of $O(\log n)$ maximum flow computations.
\end{proof}

\subsection{Rounding Algorithm for Semidefinite Programming Solution}
\label{sec:ARV-approx}

In this subsection, we prove \autoref{thm:primal-rounding} that the integrality gap of $\primal(G)$ is $O(\sqrt{\log n})$.
The proof is by applying the metric rounding lemma on the two sets provided by the structure theorem of Arora, Rao, and Vazirani (see \autoref{thm:arv-structure}).

We note that by adding triangle inequalities in the reweighted eigenvalues in~\cite{KLT22,LTW23},
essentially the same proof implies $O(\sqrt{\log n})$-approximation algorithms for undirected and directed vertex expansions, and undirected and directed hypergraph expansions (See \autoref{sec:others} for more details). 
These approximation guarantees are all known previously, but with different formulations and proof techniques.
In particular, the SDP relaxation for vertex expansion obtained through our approach is considerably simpler than that obtained by Feige, Hajiaghayi, and Lee \cite{FHL08}.
This demonstrates that our approach of using reweighted eigenvalues and triangle inequalities provides a simple and unifying way to recover all these results.

The proof that $\primal(G)$ is indeed an SDP relaxation of directed edge expansion can be found in \autoref{app:arv}.

\begin{proposition}[Easy Direction]
  \label{prop:arv-easy-direction}
  For any edge-capacitated directed graph $G = (V, E, w)$ with vertex weights $\pi: V \to \R_+$, it holds that $\primal(G) \le 2 \edgeexp(G)$.
\end{proposition}


We will use the structure theorem in~\cite{ARV09} for the proof of $\edgeexp(G) \lesssim \sqrt{\log n} \cdot \primal(G)$.
Since we consider $\pi$-weighted directed edge expansion,
we need the following weighted version of the structure theorem.
The proof of the weighted version is a straightforward reduction to the unweighted version in \autoref{thm:arv-structure} and is deferred to \autoref{app:arv} 
(see~\cite[Algorithm 1]{ACMM05} for a similar weighted structure theorem and reduction).

\begin{lemma}[$\pi$-Weighted Structure Theorem]
  \label{lemma:arv-structure-weighted}
  Let $G = (V, E, w)$ be an edge-capacitated directed graph with vertex weights $\pi: V \to \R_+$ and $\pi(V) = 1$. 
  Let $\{v_i\}_{i=1}^n$ be a set of embedding vectors satisfying $\ell_2^2$ triangle inequalities and $\sum_{i, j \in V} \pi(i) \cdot \pi(j) \cdot \norm{v_i - v_j}^2 = 1$.
  The embedding $\{v_i\}_{i=1}^n$ is said to be well-spread if $\pi\big(B\big(i, 1/\sqrt{10}\big)\big) \le 1/10$ for all $i \in V$.
  If $\{v_i\}_{i=1}^n$ is well-spread, then there exist two subsets $L, R \subseteq V$ with $\pi(L), \pi(R) \ge \Omega(1)$ and
  \[
    d(L, R) := \min_{i \in L, j \in R} \norm{v_i - v_j}^2
    \gtrsim
    1 / \sqrt{\log n}.
  \]
  Moreover, there is a randomized polynomial-time algorithm that finds such sets with high probability.
\end{lemma}

With the $\ell_2^2$ triangle inequalities, 
the function $d(i, j) := \norm{v_i - v_j}^2$ is a metric.
We will apply the metric rounding lemma to find a sparse cut, with the observations that the numerator term $\max_{F \in \mathcal{F}(G)} \sum_{(i, j) \in E} F(i, j) \cdot d(i,j)$ in \autoref{lemma:metric-rounding-lemma} is exactly the inner maximization problem of $\primal(G)$, and the denominator term in \autoref{lemma:metric-rounding-lemma} is large using the two subsets $L,R$ provided by the structure theorem.


\begin{theorem}[Hard Direction] \label{thm:sqrt-logn-rounding}
  Let $G = (V, E, w)$ be an edge-capacitated directed graph with vertex weights $\pi:V\to \R_+$. 
  There is a polynomial-time algorithm which, with high probability, finds a set $S \subseteq V$ with
    $\edgeexp(S) \lesssim \primal(G) \cdot \sqrt{\log n}$.
\end{theorem}

\begin{proof}
  Let $\{v_i\}_{i=1}^n$ be an optimal solution to the $\primal(G)$ program.
  Let $d(i,j) = \norm{v_i-v_j}^2$, which is a metric by the $\ell_2^2$ triangle inequalities in $\primal(G)$.
  By \autoref{lemma:metric-rounding-lemma}, given two subsets $L$ and $R$, there is a subset $S \subseteq V$ with
  \begin{equation}
    \label{eqn:metric-rounding-arv}
    \edgeexp(S) \lesssim
    \frac{r' \cdot \max_{F \in \mathcal{F}(G)} \sum_{(i, j) \in E} F(i, j) \cdot \norm{v_i - v_j}^2}
    {\sum_{i \in R} \pi(i) \cdot d(i, L)}
    = \frac{2r' \cdot \primal(G)}{\sum_{i \in R} \pi(i) \cdot d(i, L)}.
  \end{equation}
  There are two cases to consider: the ``well-spread'' case and the ``large core'' case. 
  The difference in these two cases lies in the different choices of $L$ and $R$ to apply the metric rounding bound in~\eqref{eqn:metric-rounding-arv}.
  In either case, we assume without loss of generality that $\pi(V)=1$.
  Also, by a straightforward calculation that we will show in \autoref{app:arv},
  the two normalization constraints in $\primal(G)$ in \autoref{def:main-sdp} imply the following condition.
  \begin{fact}
    \label{fact:normalization-constraints}
    If
    $\sum_{i \in V} \pi(i) \cdot v_i = \vec{0}$
    and
    $\sum_{i \in V} \pi(i) \cdot \norm{v_i}^2 = 1$, then
    $\sum_{i, j \in V} \pi(i) \cdot \pi(j) \cdot \norm{v_i - v_j}^2 = 2$.
  \end{fact}
  Suppose the vectors $\{v_i\}_{i=1}^n$ are well-spread.
  Since $\sum_{i, j \in V} \pi(i) \cdot \pi(j) \cdot \norm{\frac{1}{\sqrt{2}} v_i - \frac{1}{\sqrt{2}} v_j}^2 = 1$, we can apply \autoref{lemma:arv-structure-weighted} to obtain two subsets $L, R \subseteq V$ with $\pi(L), \pi(R) \ge \Omega(1)$ and $d(L, R) \gtrsim 1 / \sqrt{\log n}$ in randomized polynomial time.
  This implies that the denominator in (\ref{eqn:metric-rounding-arv}) is
  \[
    \sum_{i \in R} \pi(i) \cdot d(i, L) \gtrsim \pi(R) \cdot \frac{1}{\sqrt{\log n}} \gtrsim \frac{1}{\sqrt{\log n}},
  \]
  and thus we get from the metric rounding bound a set $S$ with $\edgeexp(S) \lesssim \sqrt{\log n} \cdot \primal(G)$ as $r' = \max\{1,\pi(R)/\pi(L)\} = O(1)$.

  Otherwise, we are in the large core case, where there is a vertex $i^* \in V$ with $\pi\big(B\big(i^*, 1/\sqrt{10}\big)\big) > 1 / 10$. 
  In this case, we set $L := B\big(i^*, 1/\sqrt{10}\big)$ and $R := V \setminus L$ with $r' = \max\{1,\pi(R)/\pi(L)\} = O(1)$, and use the following lemma to lower bound the denominator in \eqref{eqn:metric-rounding-arv}.
  
  \begin{lemma}[Total Distance to Core]
      \label{lem:large-core-denominator}
      Let $d:V\times V\rightarrow \mathbb{R}_{\ge 0}$ be a semi-metric (i.e.~satisfying all axioms of metric except possibly the triangle inequality). Let $s \ge 1$ so that $d(\cdot, \cdot)$ satisfies an $s$-relaxed triangle inequality: $d(i,j) \leq s\cdot (d(i,k) + d(k,j))$ for all $i,j,k \in V$.
      Let $\pi: V \rightarrow \R_+$ be a weight function with $\pi(V) = 1$ and 
      suppose $d(\cdot, \cdot)$ satisfies $\sum_{i,j \in V} \pi(i) \cdot \pi(j) \cdot d(i,j) = 2$.
      Let $L\subseteq V$ be a subset with diameter $\diam(L) := \max_{i,j \in L} d(i,j)$. Then
      \begin{equation*}
          \sum_{i \notin L} \pi(i) \cdot d(i,L) \geq \frac{1}{s^2} - \frac12 \diam(L)
      \end{equation*}
  \end{lemma}

    Applying \autoref{lem:large-core-denominator} with $s=1$ and $\diam(L)\leq 2 \cdot 1/10 = 1/5$, it follows that
  \begin{align*}
      \sum_{i\in R}\pi(i) \cdot d(i,L)
      = \sum_{i \not\in L} \pi(i) \cdot d(i, L)
      \ge 1 - \frac{1}{10}
      = \frac{9}{10},
  \end{align*}
    and thus we get from the metric rounding bound in \eqref{eqn:metric-rounding-arv} a set $S$ with $\edgeexp(S) \lesssim \primal(G)$.

  The proof of \autoref{lem:large-core-denominator} is in \autoref{app:arv}, which was already done in previous works (\cite{ARV09}, \cite{AK07}) for the uniform case.
\end{proof}

\autoref{thm:primal-rounding} follows immediately from 
\autoref{thm:sqrt-logn-rounding} and \autoref{prop:arv-easy-direction}.

\subsection{Rounding Algorithm for Spectral Solution}
\label{sec:Cheeger}

In this subsection, we provide an alternative proof of the Cheeger-type inequality for directed graphs in~\eqref{eqn:Cheeger-directed} using the metric rounding lemma, where the original proof in~\cite{LTW23} is by a refined ``threshold rounding'' algorithm.
This proof will be used in the proof of \autoref{thm:fast-Cheeger} in \autoref{sec:fast-Cheeger}, as the threshold rounding algorithm in~\cite{LTW23} requires a linear programming duality step which is not clear how to be implemented in almost linear time.

We note that essentially the same proof works for the ordinary Cheeger's inequality~\cite{AM85,Alo86}, as well as the Cheeger-type inequalities for directed vertex expansion and hypergraph edge conductance in~\cite{LTW23} (see \autoref{sec:others}).
This illustrates the max-flow min-cut theorem in the proof of the metric rounding lemma as a unifying method to find sparse cuts in different settings.

Recall from \autoref{sec:LTW23} that $\vec{\phi}(G)$ denotes the directed edge conductance, which is the special case of directed edge expansion in \autoref{def:phi-pi} when $\pi(i) = d_w(i) := w(\delta^+(i)) + w(\delta^-(i))$ is the total degree of $i$.
We will focus on the proof of the ``hard direction'' of \eqref{eqn:Cheeger-directed} that 
\[\vec{\phi}(G) \lesssim \sqrt{\lambda_2^*(G) \cdot \log\big(1/\vec{\phi}(G)\big)}.
\]
Also recall from \autoref{sec:LTW23} that $\lambda_2^*(G)$ can be written as the SDP in \autoref{def:main-sdp} without the triangle inequalities.
In~\cite{LTW23}, the first step of the proof of the hard direction is to relate the ${\lambda_2^{*}}(G)$ program to the following ``one-dimensional $\ell_1$ program'', which was done by using Gaussian projection and applying Cauchy-Schwarz inequality.

\begin{lemma}[One-Dimensional $\ell_1$ Program {\cite[Definition 3.19]{LTW23}}]
\label{lemma:cheeger-l1-program}
Given an edge-capacitated directed graph $G = (V, E, w)$ with vertex weights $d_w: i \mapsto \sum_{e: e \ni i} w(e)$, let
  \begin{align*}
    \eta_e(G) :=
     \min_{v: V \rightarrow \R} \max_{F \in \mathcal{F}(G)} &~~~ \frac12 \sum_{ij \in E} F(i, j) \cdot |v(i) - v(j)|
    \\
    \st
    &~~~
    \sum_{i \in V} d_w(i) \cdot v(i) = 0
    \\
    &~~~ \sum_{i \in V} d_w(i) \cdot |v(i)| = 1.
  \end{align*}
  Then, it holds that
  \[
    \eta_e(G) \lesssim \sqrt{\lambda_2^{*}(G) \cdot  \log \big(1 / \vec{\phi}(G)\big)}.
  \]
\end{lemma}

The second step in~\cite{LTW23} is to use a refined threshold rounding algorithm to prove that $\vec{\phi}(G) \lesssim \eta_e(G)$, thus proving the hard direction.
Here we will use the metric rounding lemma to prove that $\vec{\phi}(G) \lesssim \eta_e(G)$.
The reasons that we can apply the metric rounding lemma to the one-dimensional $\ell_1$ program, but not to the $n$-dimensional $\ell_2^2$ program, are as follows:
(i) $d_1(i, j) := \left| v(i) - v(j) \right|$ is a metric while $d_2(i, j) := \norm{v(i) - v(j)}^2$ needs not be;
(ii) It is natural and straightforward to define sets $L$ and $R$ for a solution to the one-dimensional $\ell_1$ program, but not for the $n$-dimensional $\ell_2^2$ program.
Therefore, we may view the Gaussian projection and Cauchy-Schwarz steps as reducing to $1$-dimension and ``metrifying'' the objective, so that we can apply metric rounding.

\subsubsection{Proof of the Second Step}

We aim to prove that $\vec{\phi}(G) \lesssim \eta_e(G)$ using the metric rounding lemma.
Let $v(1), v(2), \dots, v(n) \in \R$ be an optimal solution to the $\eta_e(G)$ program.
Set $d(i, j) := |v(i) - v(j)|$ which is a metric. 
Let $L := \{i \in V: v(i) \le 0\}$ and $R := \{j \in V: v(j) > 0\}$.
We assume without loss of generality that $r := \mu(R) / \mu(L) \le 1$ so that $r' := \max\{1, r\} = 1$.
  From the definitions of $L$ and $R$ and the constraints on $v(i)$, one can verify that
  \[
    d(i, L) \ge |v(i)|~~\forall i \in R
    \quad \text{ and } \quad 
    \sum_{i \in R} d_w(i) \cdot |v(i)| = \frac{1}{2} \sum_{i \in V} d_w(i) \cdot |v(i)| = \frac{1}{2}.
  \]
  Therefore, applying \autoref{lemma:metric-rounding-lemma}, it follows that
  \begin{eqnarray*}
    \vec{\phi}(G) = \vec{\phi}_{d_w}(G) 
    &\lesssim& 
     \frac{r' \cdot \max_{F \in \mathcal{F}(G)} \sum_{ij \in E} F(i, j) \cdot d(i, j)
    }
    {\sum_{i \in R} d_w(i) \cdot d(i, L)}
    \\
    &\le&
    \frac{
      \max_{F \in \mathcal{F}(G)} \sum_{ij \in E} F(i, j) \cdot |v(i) - v(j)|
    }
    {\sum_{i \in R} d_w(i) \cdot |v(i)|}
    \\
    &=&
    2 \max_{F \in \mathcal{F}(G)} \sum_{ij \in E} F(i, j) \cdot |v(i) - v(j)|
    \\
    &=&
    4 \cdot \eta_e(G).
  \end{eqnarray*}
This completes the proof of the hard direction of \eqref{eqn:Cheeger-directed}.

\section{Almost Linear-Time Primal-Dual \texorpdfstring{$O(\sqrt{\log n})$}{O(sqrt(log n))}-Approximation}
\label{sec:dual-rounding}

The main goal of this section is to prove \autoref{thm:directed-Sherman}.
First, we will derive the dual program of $\primal(G)$ in \autoref{sec:dual}.
Then, in \autoref{sec:mmw-expander-flow}, we describe the primal-dual algorithm using the matrix multiplicative weight update method assuming a black-box algorithm for the oracle exists.
In \autoref{sec:oracle-expander-flow}, we present the geometric results in~\cite{ARV09,AK07,She09} for the design of the oracle, and implement the oracle in the easy ``large core'' case.
Then, in \autoref{sec:well-spread-case}, we implement the oracle in the more difficult ``well spread'' case, in which we use Sherman's chaining theorem to find many paths that violate the triangle inequality.
We conclude with the proofs of \autoref{thm:directed-Sherman} and \autoref{prop:reweighted-phi} in \autoref{sec:concluding-remarks}.

\subsection{Dual Program\texorpdfstring{ of $\primal$}{}} \label{sec:dual}

We construct the dual program of $\primal(G)$ in a similar way as in \autoref{sec:expander-flow-dual} for the dual program of the Goemans-Linial relaxation in~\eqref{eqn:ARV}.

We first write the primal program $\primal(G)$ in \autoref{def:main-sdp} in matrix form.
Let $V$ be the matrix with the $i$-th column being $v_i$ for $1 \leq i \leq n$ and let $X = V^T V$.
Let $L_{i,j}$ be the Laplacian of an undirected edge $ij$.
For a matrix $A$ that is not necessarily symmetric, define the symmetric Laplacian of $A$ as 
$L_{{\rm sym}}(A) = \frac{1}{2} \sum_{i,j} (A(i,j) + A(j,i)) \cdot L_{i,j}$.
As in \autoref{sec:expander-flow-dual}, we express the triangle inequalities redundantly as inequalities along paths in $K_n$, and let $T_p := \sum_{k=1}^{\ell-1} L_{i_k,i_{k+1}} - L_{i_1,i_l}$ for a path $p = (i_1,\ldots,i_\ell)$.
Let $\Pi$ be the diagonal matrix with $\Pi(i, i) = \pi(i)$ for $1 \leq i \leq n$.
Then, check that $\primal(G)$ in \autoref{def:main-sdp} can be written as
\begin{align*}
    \primal(G) \quad = \quad
     \min_{X \succeq 0} 
     \max_{F\in \mathcal{F}(G)}
     & \quad
     \inner{L_{\rm sym}(F)}{X}
    \\
    \st & \quad
    \inner{\Pi\one\one^\top\Pi}{X} = 0 
    \\
    &~~~\inner{\Pi}{X} = 1\\
    &~~~ \inner{T_p}{X} \geq 0\quad\forall p\in \mathcal{P}(K_n).
\end{align*}
To derive the dual of $\primal(G)$, we apply von Neumann's minimax theorem to switch the order of the min and the max, and then take the SDP dual of the inner minimization program to obtain
\begin{align*}
    \max_{F\in \mathcal{F}(G)}     &~~~
    \max_{
      \substack{\lambda, x \in \R
        \\ y_p \geq 0: p \in \mathcal{P}(K_n)}
    } \quad \lambda
    \\
    \st &~~~ \sum_{p}y_pT_p + \lambda \Pi + x \Pi\one\one^\top\Pi \preceq L_{\rm sym}(F).
\end{align*}
The dual constraint can be rewritten as
\[
\lambda I + x \Pi^{\frac12}\one\one^\top \Pi^{\frac12} \preceq \Pi^{-\frac12} \bigg( L_{\rm sym}(F) - \sum_{p}y_pT_p \bigg) \Pi^{-\frac12}.
\]
Note that the vector $\Pi^{\frac12}\one$ is in the null space of the right hand side,
as $\one$ is in the nullspace of any Laplacian matrix.
Therefore, for the dual constraint to hold, an optimal dual solution must set $x = -\lambda / \pi(V)$, so as to make the component $\Pi^{\frac12} \one$ to be zero on the left hand side.
Therefore, the dual program of $\primal(G)$ can be written succinctly as
\begin{equation} \label{eqn:dual}
    \max_{F\in \mathcal{F}(G)}~
    \max_{ y_p \geq 0: p \in \mathcal{P}(K_n) }
\lambda_2 \bigg( 
 \Pi^{-\frac12} \bigg(L_{\rm sym}(F) - \sum_{p}y_pT_p \bigg) \Pi^{-\frac12}
\bigg).
\end{equation}

\subsubsection{Dual Program as Expander Flow}

For our primal-dual algorithm, 
we further rewrite the dual program in \eqref{eqn:dual} to a form that is consistent with the expander flow formulation in \eqref{eqn:expander-flow}, by considering the demand graph of the circulation $F$.

We say $f = \{f_p\}_{p \in {\mathcal P}(G)}$ is a flow path decomposition of $F$ if $F(e) = \sum_{p \ni e} f_p$ for all $e \in E$.
The demand graph $D$ of $f$ is defined such that $D(i,j) = \sum_{p \in \mathcal{P}_G(i,j)} f_p$ for all $i,j\in V$, where $\mathcal{P}_G(i,j)$ denotes the set of directed paths from $i$ to $j$ in $G$.
Note that the demand graph $D$ of a flow path decomposition of a circulation $F$ is Eulerian.
We will use the following formulation of the dual program of $\primal$.

\begin{lemma}[Dual Program of $\primal$] \label{lem:dual-program}
The dual program of $\primal(G)$ can be written as
\begin{align*}
    \max_{
    F \in \mathcal{F}(G)
    }~
    \max_{
    y_p \geq 0: p \in \mathcal{P}(K_n)
    }
    &~~~ \lambda_2 \bigg( 
    \Pi^{-1/2} \bigg(
      L_{\rm sym}(D)-\sum_py_pT_p \bigg)
    \Pi^{-1/2}
    \bigg)
    \\
    \st
    &~~~ \text{$D$ is the demand graph of a flow-path decomposition of $F$}.
\end{align*}
\end{lemma}

Note that a circulation $F \in \mathcal{F}(G)$ can have many different flow path decompositions.
The trivial flow path decomposition is simply to have a path $p=(i,j)$ of length two for each edge $ij$, with the demand graph $D=F$.
Alternatively, we can decompose $F$ into weighted directed cycles and each cycle is expressed as the union of two paths, where each path is assigned a flow value equal to the weight of the cycle. 
In our primal-dual algorithm, we will build a circulation $F$ using a demand graph $D$ of low maximum degree so as to bound the width of the oracle, and this is the reason for the formulation \autoref{lem:dual-program}.

\begin{proofof}{\autoref{lem:dual-program}}
We show that the dual program in the statement is equivalent to that in~\eqref{eqn:dual}.
One direction is easy.
Given a solution to \eqref{eqn:dual}, we can use the trivial flow decomposition of $F$ to obtain a solution to the dual program in the statement.

For the other direction, given a solution to the dual program in the statement,
we consider a flow-path decomposition $f=\{f_p\}_{p \in \mathcal{P}(G)}$ of $\frac12 F$ with demand graph $\frac12 D$.
For any flow path $p\in \mathcal{P}(G)$, 
we write $T_p = L_p - L_{e(p)}$ where $L_p$ is the Laplacian of the undirected path $p$ and $L_{e(p)}$ is the Laplacian of the edge connecting two endpoints of the path. 
As $e(p)$ is simply an edge in the demand graph,
it follows that
\begin{align*}
    \sum_{p}f_pT_p 
  = \sum_{p} f_p (L_p - L_{e(p)})
  = \sum_{ij} \frac12 \big(F(i,j) + F(j,i) - D(i,j) - D(j,i)\big)
  = L_{\rm sym}(F) - L_{\rm sym}(D),
\end{align*}
where the second last equality follows from the definition of the flow-path decomposition and the definition of the demand graph.
Therefore, 
\[L_{\rm sym}(D) - \sum_p y_pT_p 
~=~ L_{\rm sym}(F) - \sum_{p}f_pT_p - \sum_py_pT_p,\]
which is a solution to \eqref{eqn:dual} with the same objective value, where the value of the dual variable for each path $p$ is $f_p+y_p$.
\end{proofof}

\subsubsection{Intuition of the Dual Program}
\label{sec:dual-from-expander-flow}

Since the dual program in \autoref{lem:dual-program} is slightly different from the expander flow formulation in \eqref{eqn:expander-flow} used in all previous works for undirected sparsest cut, we would like to provide some intuition about the term $-\sum_p y_p T_p$ in the objective function and how it will be used to simplify Sherman's algorithm for undirected sparsest cut.

We may interpret each $-T_p$ as a ``shortcut cycle'' $C_p$, where the edges in $p$ have weight $-1$ and the edge connecting the two endpoints have weight $1$.
Since a shortcut cycle has only one positive edge, 
any cut across the cycle has non-positive weight.
Thus, adding a shortcut cycle to a graph does not increase the value of directed edge expansion.
A nice way to understand that the dual program is a lower bound on the directed edge expansion is as follows:
\begin{equation} \label{eqn:certify}
\vec{\phi}_{\pi}(G) 
\gtrsim \phi_{\pi}(D)
\geq \phi_{\pi}\bigg(D + \sum_p y_p C_p\bigg)
\gtrsim \lambda_2 \bigg(\Pi^{-\frac12}\bigg(L_{\rm sym}(D) - \sum_p y_p T_p\bigg) \Pi^{-\frac12} \bigg),
\end{equation}
where the first inequality is by the flow argument because $D$ is the demand graph of a circulation $F \in \mathcal{F}(G)$ (which is Eulerian and so can be considered as an undirected graph), 
the second inequality is by the discussion above that adding shortcut cycles doesn't increase the value of directed edge expansion, 
and the third inequality is by the easy direction of $\primal$ in \autoref{prop:arv-easy-direction}.

Why would adding shortcut cycles help in obtaining a stronger lower bound?
There are graphs where the easy direction of Cheeger's inequality is not tight, such that $\phi \approx \lambda_2^2$ rather than $\phi \approx \lambda_2$.
The prototypical example is a long path $p$, where every edge in the path is short in its spectral embedding, which heavily violates the $\ell_2^2$ triangle inequality.
So, intuitively, given an embedding of the vertices, we would like to add shortcut cycles along the paths that heavily violate $\ell_2^2$ triangle inequalities, so as to increase the objective of this embedding in the hope to improve the lower bound provided by the second eigenvalue, while not decreasing the objective value of sparsest cut of $D$ by much.
Thus, the dual program of $\primal(G)$ can be intuitively understood as finding the best way to add these shortcut cycles to prove the strongest spectral lower bound.
This interpretation is also consistent with the primal program $\primal(G)$ in which we add triangle inequalities to the spectral program.
In our primal-dual algorithm for \autoref{thm:directed-Sherman} that we will present in the next subsection, we will indeed add shortcut cycles for paths that heavily violate the $\ell_2^2$ triangle inequalities in the embedding.

\subsection{Regret Minimization for Approximating Directed Edge Expansion}\label{sec:mmw-expander-flow}


As in the work by Arora and Kale~\cite{AK07} described in \autoref{sec:primal-dual-sparsest} and \autoref{sec:Arora-Kale-directed}, we use the regret bound in \autoref{thm:regret-bound} to design a primal-dual algorithm for approximating directed edge expansion.
The setup is to either certify that the optimal value to $\primal(G)$ is at least $\Omega(1/\kappa)$ by constructing a solution to the dual program in \autoref{lem:dual-program}, or to find a cut of expansion at most $O(\sqrt{\log n}/\kappa)$ for some parameter $\kappa$.
Doing binary search on $\kappa$ will give us a $O(\sqrt{\log n})$-approximation  algorithm.

In each iteration, the algorithm uses the density matrix $X_t$ given by the matrix multiplicative weight update algorithm as a candidate primal solution to $\primal(G)$.
To build a dual solution to \autoref{lem:dual-program}, in each iteration $t$, the oracle tries to either 
\begin{enumerate}
\item find a circulation $f$ with demand graph $D$ such that $\biginner{\Pi^{-\frac12} L_{\rm sym}(D) \Pi^{-\frac12}}{X_t}$ is large (i.e. send a lot of flow between vertices that are far apart in the geometric embedding defined by $X_t$) and $\norm{L_{\rm sym}(D)}$ is small (i.e. the demand graph has small maximum degree), and set the feedback matrix $M_t := \Pi^{-\frac12} L_{\rm sym}(D) \Pi^{-\frac12}$, or
\item find paths $p_1, \ldots, p_k$ and weights $y_1,\ldots,y_k$ such that $-\biginner{\Pi^{-\frac12} \big( \sum_i y_i T_{p_i} \big) \Pi^{-\frac12}}{X_t}$ is large (i.e. paths along which the triangle inequality is violated heavily) and $\norm{\Pi^{-\frac12} \big( \sum_i y_i T_{p_i} \big) \Pi^{-\frac12}}$ is small (i.e. the union of these paths found have small total degree) and set the feedback matrix $M_t := \Pi^{-\frac12} \big( \sum_i y_i T_{p_i} \big) \Pi^{-\frac12}$. 
\end{enumerate}
If the oracle succeeds for $T=O(\rho^2\log n)$ iterations, where $\rho \geq \max_{t\leq T}\norm{M_t}$ , then the regret bound in \autoref{thm:regret-bound} would imply that $\lambda_2( \frac{1}{T} \sum_{i=1}^T M_t )$ is large, and thus we found a solution to the dual program in \autoref{lem:dual-program} with large objective value.
Otherwise, if the oracle fails to find the above objects in some iteration, then the oracle must return a sparse cut $S$. 
Above is the high level description of the algorithm,
while below is the precise description of the algorithm.

\newpage

\begin{algorithm}
\caption{Regret Minimization for Directed Sparsest Cut}
\label{alg:mmw-expander-flow}
\textbf{Input}: An edge-capacitated directed graph $G = (V, E, w)$ with vertex weights $\pi$ such that $\pi(V)=1$; step size $\eta \in (0, 1)$, width bound $\rho \in \R_+$, congestion parameter $\kappa \in \R_+$, and approximation factor $\alpha \in \R_+$.

\vspace{2mm}

\textbf{Output}: Either a sparse cut $S$, or a solution $\overline{M}$ to the dual program in \autoref{lem:dual-program}.

\vspace{2mm}

\textbf{Initialization:} $X_0 = \frac{1}{n-1}(I-\Pi^{\frac12}\one\one^\top \Pi^{\frac12})$.

\vspace{2mm}

\textbf{For} $t = 0$ to $T-1$:
\begin{enumerate}
    \item Given $X_t \succeq 0$ such that $\tr(X_t) = 1$ and $X_t\perp \Pi^{\frac12}\one$, let $Y_t := \Pi^{-\frac12} X_t \Pi^{-\frac12}$ and $v_1, \dots, v_n$ be the Gram decomposition of $Y_t$.
    \item \textbf{(Oracle)} Do one of the following:
    \begin{enumerate}
        \item Find a circulation $f$ on $G$ with congestion $\kappa$ and demand graph $D$ such that $\inner{L_{\rm sym}(D)}{Y_t}\geq 1$ and $ L_{\rm sym}(D) \preceq \rho \cdot \Pi$.
        If this succeeds, set $M_t := \Pi^{-\frac12}L_{\rm sym}(D) \Pi^{-\frac12}$.
        
        \item Find paths $p_1, \dots, p_k$ in $K_n$ and weights $y_1, \dots, y_k\geq 0$ such that $\inner{\sum_i y_iT_{p_i}}{Y_t} \leq -1$ and that $-\rho \cdot \Pi \preceq \sum_i y_i T_{p_i}\preceq \rho \cdot \Pi$.
        If this succeeds, set $M_t := -\Pi^{-\frac12} (\sum_i y_i T_{p_i}) \Pi^{-\frac12}$.
        
        \item If both cases (a) and (b) fail, then we say that Oracle fails. In this case, find a cut $S \subseteq V$ such that $\edgeexp(S) = O(\alpha/\kappa)$. Return $S$ and terminate the algorithm.
    \end{enumerate}
    \item If Oracle succeeds, update $X_{t+1}' := \exp\Big(-\frac{\eta}{\rho}\sum_{i=0}^tM_i\Big)$.
    Let $X_{t+1}$ be obtained from $X_{t+1}'$ by projecting it onto the space orthogonal to $\Pi^{\frac12}\one$ and scaling it to have trace $1$.
\end{enumerate}
Return the average feedback matrix $\overline{M} := \frac{1}{T} \sum_{t=0}^{T-1} M_t$.
\end{algorithm}


We analyze \autoref{alg:mmw-expander-flow} assuming that there is a black-box algorithm for Oracle.

\begin{lemma}[Regret Minimization Algorithm]
\label{lem:MMW-approximation-guarantee}
Suppose there is a black-box algorithm for Oracle.
Set $\eta = \Theta(1/\rho)$.
After $T = \Theta(\rho^2\log{n})$ iterations, \autoref{alg:mmw-expander-flow} either certifies that $\vec{\phi}_\pi(G) \ge \Omega(1/\kappa)$ or finds a cut $S \subseteq V$ with $\vec{\phi}_\pi(S)\leq O(\alpha/\kappa)$.
\end{lemma}

\begin{proof}
  First, suppose Oracle succeeds for $T = \Theta(\rho^2\log{n})$ iterations. 
  By applying the general regret bound~\eqref{eqn:general-regret-bound} in \autoref{thm:regret-bound} restricting to the subspace orthogonal to $\Pi^{\frac12} 1$, it follows that
  \begin{align*}
    \label{eqn:regret-bound-expander-flow}
    \lambda_2(\overline{M})
    &\geq
    \frac{1}{T} \sum_{t=0}^{T-1} \inner{M_t}{X_t} - \eta\rho - \frac{\rho\log{n}}{\eta T}
    =
    \frac{1}{T} \sum_{t=0}^{T-1} \inner{\Pi^{\frac12} M_t \Pi^{\frac12}}{Y_t} - \eta\rho-\frac{\rho\log{n}}{\eta T}
    \geq
    1 - \eta\rho - \frac{\rho\log{n}}{\eta T},
  \end{align*}
  where the last inequality follows from the fact that cases (a) and (b) in Oracle both imply that $\inner{\Pi^{\frac12}M_t\Pi^{\frac12}}{Y_t}\geq 1$.
  By choosing suitable implicit constants in the $\Theta(\cdot)$ for $T$ and $\eta$, 
    \[
      \lambda_2(\overline{M})
      \geq
      1 - \eta\rho - \frac{\rho\log{n}}{\eta T}
      \geq
      1 - \frac{1}{4} - \frac{1}{4}
      \geq \frac12.
    \]
  Note that the average feedback matrix $\overline{M}$ is a Laplacian of the form $\Pi^{-\frac12}(L_{\rm sym}(D)-\sum_py_pT_p)\Pi^{-\frac12}$, where $D$ is the demand graph of a circulation $f$ with congestion $\kappa$ (as $f$ is the average of circulations each with congestion $\kappa$).
  Therefore, by scaling down $f, D$, and all $y_p$ by a factor of $\kappa$, we obtain a solution to the dual program of $\primal(G)$ in \autoref{lem:dual-program} with objective value $\Omega(1/\kappa)$, and this certifies that $\vec{\phi}_\pi(G) \gtrsim 1/\kappa$.
    
  On the other hand, if Oracle fails at some iteration, then it outputs a cut $S$ with $\vec{\phi}_\pi(S) \leq O(\alpha/\kappa)$.
\end{proof}

In \autoref{lem:MMW-approximation-guarantee}, we have set the values of $T$ and $\eta$ in relation to the width bound $\rho$, to obtain the desired approximation guarantee $O(\alpha)$.
The undetermined parameters in the algorithm are $\rho$ and $\alpha$. 
We would like to set them to be as small as possible, so as to minimize both the runtime (as the number of iterations $T$ will be minimized) and the approximation ratio of the algorithm, while the Oracle can still be implemented efficiently.
This is the goal in \autoref{sec:oracle-expander-flow} and \autoref{sec:fast-well-spread}.

\subsection{Geometric Results for Implementation of Oracle}
\label{sec:oracle-expander-flow}


To implement the Oracle in \autoref{alg:mmw-expander-flow},
we need the results proved in~\cite{ARV09,AK07,Kal07,She09} about geometric embeddings.

Let $v_1,\ldots,v_n$ be the Gram decomposition of $Y_t$ in step (1) of \autoref{alg:mmw-expander-flow}.
Note that the trace condition in step (1) implies that $\sum_i \pi(i) \cdot \norm{v_i}^2 = 1$, and the null-space condition in step (1) implies that $\sum_i \pi(i) \cdot v_i = 0$.
It follows from \autoref{fact:normalization-constraints} that $\sum_{i < j} \pi(i) \cdot \pi(j) \cdot \norm{v_i - v_j}^2 = 1$.
As in the SDP rounding result in \autoref{sec:ARV-approx}, 
we consider the following two cases of the geometric embedding.

\begin{proposition}[Dichotomy of Embeddings]
  \label{prop:embedding-cases} 
Let $v_1, \dots, v_n$ be vectors in $\R^n$ satisfying the condition that $\sum_{i<j}\pi(i) \cdot \pi(j) \cdot \norm{v_i - v_j}^2 = 1$. 
One of the following two cases must hold:
\begin{enumerate}[label=(\roman*)]
    \item {\em Large Core}: There exists a vector $v$ such that $\pi(B(v, \frac{1}{2\sqrt{10}}))\geq 1/4$.
    \item {\em Well Spread:} There is a vector $w$ such that if we apply the transformation $u_i := c(v_i - w)$ for $1 \leq i \leq n$ for some constant $c > 0$, then there exists a subset $U$ of vectors with (i) $\pi(U) \gtrsim 1$, (ii) $\norm{u_i}\leq 1$ for all $i \in U$, and (iii) $\sum_{i,j\in U}\pi(i) \cdot \pi(j) \cdot \norm{u_i-u_j}^2\gtrsim 1$.
\end{enumerate}
\end{proposition}

Note that a version of \autoref{prop:embedding-cases} for uniform vertex weights was already proved in \cite{Kal07}. 
The weighted case follows by a simple reduction which we will defer to the appendix.


\subsubsection{Large Core Case}

This is the easy case where we can implement the oracle to either return a circulation in step 2(a) or a sparse cut in step 2(c) of \autoref{alg:mmw-expander-flow}, using a result in \autoref{sec:metric-rounding} for metric rounding proved by the max-flow min-cut theorem.

\begin{lemma}[Oracle in Large Core Case]
\label{lem:large-core-case}
In the large core case in \autoref{prop:embedding-cases}, there is an algorithm that, using two max-flow computations, implements Oracle in \autoref{alg:mmw-expander-flow} so that it either computes a cut $S \subseteq V$ with $\edgeexp(S) \le O(1/\kappa)$ or obtains a circulation $f$ whose demand graph $D$ satisfies $\inner{L_{\rm sym}(D)}{Y_t} \geq 1$ and $L_{\rm sym}(D)\preceq O(1)\cdot \Pi$.
\end{lemma}

\begin{proof}
Let $v_{j}$ be a vector with $\pi\big(B\big(v_j, \frac{1}{2 \sqrt{10}}\big)\big) \ge \frac14$. 
By triangle inequality, for any $v_i \in B\big(v_j, \frac{1}{2\sqrt{10}}\big)$, it holds that $\pi\big(B\big(v_i, \frac{1}{\sqrt{10}}\big)\big) \geq \frac14$. 
So, by random sampling, we can find in $O(n \log n)$ time a vector $v_{i^*}$ such that $\pi\big(B\big(v_{i^*}, \frac{1}{\sqrt{10}}\big)\big) \ge \frac14$ with high probability.

After finding such a vector $v_{i^*}$, we run \autoref{alg:bidirectional-max-flow} (Bidirectional Max-Flow) with $L := B\big(v_{i^*}, \frac{1}{\sqrt{10}}\big)$, $R := \overline{L}$, and $\kappa$ the same as given in \autoref{alg:mmw-expander-flow}.
In order to choose the flow value parameter $\beta$ appropriately, we would first lower bound the total distance to $L$. Applying \autoref{lem:large-core-denominator} with the set $L$ as chosen and the semi-metric $d(i,j) := \norm{v_i-v_j}^2$,
which satisfies the $s$-relaxed triangle inequality for $s=2$,
it follows that
\[
  \sum_{j \in R}\pi(j) \cdot d(j,L)
  =
  \sum_{j \not\in L}\pi(j) \cdot d(j,L) \geq
  \frac14 - \frac{1}{10}
  = \frac{3}{20}.
\]
Apply \autoref{alg:bidirectional-max-flow} with $\beta=20/3$. 
On the one hand, if the algorithm returns a circulation $f$, then its demand graph $D$ satisfies
\begin{equation} \label{eqn:large-core-case}
    \inner{L_{\rm sym}(D)}{Y_t}
    =
    \sum_{i\in L,j\in R} \frac{1}{2} \big(D(i,j)+D(j,i)\big) \norm{v_i-v_j}^2
    \geq
    \sum_{j\in R}\beta \cdot \pi(j) \cdot d(j,L) \geq 1,
\end{equation}
where the first inequality follows from the fact that each vertex $j \in R$ has capacity $\beta \cdot \pi(j)$ and the flow saturates all such capacities.
The capacities also imply that the normalized Laplacian of the demand graph satisfies $\Pi^{-1/2}L_{\rm sym}(D)\Pi^{-1/2} \preceq 2\beta \cdot I$, or equivalently $L_{\rm sym}(D) \preceq 2 \beta \cdot \Pi$.

On the other hand, if the algorithm returns a cut $S \subseteq V$, then by \autoref{lemma:flow-cut-lemma} (unsaturated case) we have
$\edgeexp(S) \leq r \beta/\kappa = 20/\kappa$, since $r := \pi(R)/\pi(L) \leq 3$.
\end{proof}

To summarize, in the large core case, there is an efficient oracle that achieves approximation factor $\alpha = O(1)$ and width bound $\rho = O(1)$.

\subsubsection{Well Spread Case}
\label{sec:well-spread-case}

The well spread case is much more involved, for which we need the correlated chaining theorem of Sherman~\cite{She09}.
In this subsection, we present the background for the correlated chaining theorem, and we defer the implementation of the oracle in the well spread case to the next subsection.


The idea of chaining matchings was the main ingredient that led to the $O(\sqrt{\log{n}})$ approximation result of \cite{ARV09}, and was also used in \cite{AK07, Kal07} to compute expander flows to solve the dual program.
The main idea was to show that if for many directions, there is a large matching between embedding vertices that are well-separated along that direction but close to each other in the overall embedding, then $O(\sqrt{\log{n}})$ such matchings can be chained together to form a path that violates the $\ell_2^2$ triangle inequality.
In \cite{She09}, this was improved so that instead of finding one such violating path, we can find {\em many} such paths {\em efficiently} with good probability through a simple sampling process. 

To handle the arbitrary vertex weights $\pi:V \to \R_+$, we slightly modify Sherman's definitions and results and make use of a version of his main theorem for fractional matchings instead of integral matchings.

\begin{definition}[$\pi$-Fractional Matching] \label{def:fractional-matching}
Let $V$ be a vertex set with weights $\pi: V\rightarrow \mathbb{R}^+$. 
We say that $M$ is a $\pi$-fractional matching if $M$ is a weighted directed subgraph of $K_n$ with edge weights $M(i,j)\in \mathbb{R}_{\ge 0}$ for $i, j \in V$, satisfying the property that each vertex $i \in V$ has either only incoming edges or only outgoing edges and has degree at most $\pi(i)$.
The total weight of $M$ is denoted by $w(M) := \sum_{i,j} M(i,j)$.
\end{definition}

\begin{definition}[Fractional Matching Cover] \label{def:matching-cover}
A $(\sigma, \delta)$-matching cover is a function assigning a $\pi$-fractional matchings $\M_u$ to each vector $u \in \R^n$ satisfying the following properties:
  \begin{enumerate}[label=(\roman*)]
    \item $\forall (i,j)\in \supp(\M_u)$, $\inner{v_j-v_i}{u} \geq \sigma$;
    \item $\M_u(i,j) = \M_{-u}(j,i)$ for all $u \in \R^n$;
    \item $\E_u[w(\M_u)]\geq \delta \cdot \pi(V)$ where $ u \sim \mathcal{N}(0, I)$.
  \end{enumerate}
\end{definition}

We define formally what it means to ``chain together'' fractional matchings.

\begin{definition}[Chained Matchings]  \label{def:chained-matchings}
    Let $\M$ be a fractional matching cover. 
    Given vectors $u_1, \dots, u_\ell \in \R^n$, we define $\M(u_1,\dots, u_\ell)$ to keep track of the paths that result from chaining together matchings $\M_{u_1},\dots, \M_{u_\ell}$.
    Define $\M(u_1, \dots, u_\ell) := (\M_{u_1, \dots, u_\ell},$ $\p_{u_1, \dots, u_\ell} = \{f_{p}, p\}_{p\in \p(K_n)})$, where each $p \in \p_{u_1, \dots, u_\ell}$ is a weighted path of length $\ell+1$ with weight $f_{p}$ and $\M_{u_1,\dots, u_\ell}$ is the graph with $\M_{u_1, \dots, u_\ell}(i, j) = \sum_{p \in \p_{u_1, \dots, u_\ell} \cap \mathcal{P}_{K_n}(i,j)} f_{p}$ being the total weight on paths in $\mathcal{P}_{u_1,\dots u_{\ell-1}}$ going from vertices $i$ to $j$.
    The paths and weights in $\mathcal{P}_{u_1,\ldots,u_l}$ are defined recursively in the following algorithm.
    \begin{framed}
    \textbf{Construction of $\p_{u_1,\dots, u_\ell}$}
    \begin{itemize}
        \item If $\ell = 1$, then $\p_{u_1} = \{\M_{u_1}(i,j),\;(i,j) \mid \M_{u_1}(i,j) > 0\}$. That is, the paths are simply the edges in $\M_{u_1}$ with the corresponding weights.
        \item If $\ell > 1$, then for each $q \in \p_{u_1, \dots, u_{\ell-1}}$ where $q\in \p(i,j)$, run the following loop.
        \begin{enumerate}
            \item While $f_q > 0$ and there exists $j' \in V$ with $\M_{u_\ell}(j,j') > 0$, let $p$ be the path obtained by extending $q$ by $j'$ and add $p$ to $\p_{u_1, \dots, u_\ell}$ with weight $f_p = \min\{\M_{u_\ell}(j,j'), f_q\}$.
            \item Decrement both $f_q$ and $\M_{u_\ell}(j,j')$ by $\min\{\M_{u_\ell}(j,j'), f_q\}$.
        \end{enumerate}
    \end{itemize}
    \end{framed}
\end{definition}

The following simple claim will be used in the runtime analysis of the oracle.

\begin{claim}\label{claim:chaining-runtime} 
Suppose that each matching $\M_{u}$ has at most $m$ edges. Then $\p_{u_1, \dots, u_\ell}$ has at most $m\ell$ paths and can be constructed in $O(m\ell^2)$ time given oracle access to $\M_{u_1}, \dots, \M_{u_\ell}$.
\end{claim}

\begin{proof}
Clearly, the claim holds true for $\ell=1$. 
Now assume by induction that the claim holds for $\ell-1$. 
It suffices to bound the number of times we run the while loop in which we add a new path $p$ to $\p_{u_1, \dots, u_{\ell}}$. 
Since in each iteration of the while loop, we either remove a path from $\p_{u_1, \dots, u_{\ell-1}}$ or an edge from $\M_{u_\ell}$, 
it can run for at most $m(\ell-1) + m = m\ell$ iterations. 
Thus $\p_{u_1, \dots, u_\ell}$ has at most $m\ell$ paths.
\end{proof}

Note that when $\pi$ is uniform, then the definition of $\pi$-fractional matching cover is the same as the matching cover from \cite[Definition 5.2.1]{She09}, in which all edges have weight $0$ or $1$.
Now we present the main theorem that we will use to implement the oracle in \autoref{alg:mmw-expander-flow} in the well spread case.

\begin{theorem}[Sherman's Chaining Theorem]
\label{thm:sherman-main-thm}
For any small enough constant $l$, there is a $k = O(\sqrt{l\log{n}})$ and an efficiently sample-able distribution $\D$ over vectors $(u_1,\dots, u_k) \subseteq \R^d$ 
with the following property:
if $\M$ is a $(\Omega(1),\Omega(1))$-fractional matching cover for the set of embedded vertices $V$, 
then the expected total weight of paths in $\M(u_1,\ldots,u_k)$ between vertices $i, j$ with $\norm{v_i-v_j}^2 \geq l$ is at least $e^{-O(k^2)} \cdot \pi(V)$ when $(u_1, \ldots, u_k)$ is sampled from $\D$.
\end{theorem}

The uniform $\pi$ version of this theorem was proved in \cite[Theorem 5.2.3]{She09}.
The $\pi$-weighted version follows from a simple reduction to the uniform case, which we will defer to \autoref{app:dual-rounding}.

\subsection{Fast Implementation of Oracle for Well-Spread Case} \label{sec:fast-well-spread}

With Sherman's chaining theorem, we are ready to implement the oracle in \autoref{alg:mmw-expander-flow} in the well spread case in this subsection, with approximation ratio $O\Big(\sqrt{\frac{\log n}{\eps}}\Big)$ and width bound $\tilde{O}\Big(\frac{n^\eps}{\eps^{3/2}}\Big)$.

\begin{proposition}[Oracle in Well Spread Case]
\label{prop:well-spread-case}
Let $\epsilon > 0$ be a small enough constant.
In the well-spread case in \autoref{prop:embedding-cases}, there is a randomized implementation of Oracle in \autoref{alg:mmw-expander-flow} that, with high probability, using $\tilde{O}(n^{\epsilon})$ max-flow computations, either outputs a feedback matrix $M_t$ with $\inner{M_t}{X_t}\geq 1$ and $\norm{M_t} \leq \tilde{O}\Big(\frac{n^{\epsilon}}{\eps^{3/2}}\Big)$,
or returns a cut $S \subseteq V$ with $\edgeexp(S) \leq O\Big(\frac{1}{\kappa}\sqrt{\frac{\log{n}}{\epsilon}}\Big)$.
\end{proposition}

\subsubsection{Overview}

The basic subroutine, as in~\cite{AK07,Kal07,She09}, is the Project Max-Flow algorithm (\autoref{alg:project-max-flow}), where we project the vectors along a random direction and set up a bi-directional flow problem between two subsets $L$ and $R$ that are far apart in the projection.
If such a bi-directional flow cannot be sent, then we will show that any min-cut is a sparse cut by \autoref{lemma:flow-cut-lemma}, and so \autoref{alg:mmw-expander-flow} can terminate in step 2(c).
If such a bi-directional can be sent, with the additional property that many flow paths are between vertices that are far apart in the embedding such that 
\[
\inner{L_{\rm sym}(D)}{Y_t} 
= \frac{1}{2} \sum_{i \in L, j \in R} \big(D(i,j) + D(j,i)\big) \cdot \norm{v_i - v_j}^2 \geq 1,
\]
then we will show that the oracle succeeds in finding a circulation in step 2(a) of \autoref{alg:mmw-expander-flow}, and so the algorithm can proceed to the next iteration.

The new observation is that if for many random directions, such a bi-directional flow can be sent but its demand graph does not satisfy $\inner{L_{\rm sym}(D)}{Y_t} \geq 1$, then we can construct a fractional matching cover and use Sherman's chaining theorem to find many paths that violate the triangle inequality heavily.
Thus, the oracle succeeds in finding many violating paths in step 2(b) of \autoref{alg:mmw-expander-flow}, and so the algorithm can proceed to the next iteration.
So, as long as such a bi-directional flow can be sent, then either step 2(a) or 2(b) succeeds in giving a good feedback matrix for the regret minimization algorithm.

This is the main difference with previous algorithms in~\cite{AK07,She09}, where a multi-commodity flow computation is needed to guarantee a condition similar to $\inner{L_{\rm sym}(D)}{Y_t} \geq 1$ for the oracle to succeed. 
We remark that using violating paths as feedback is only possible because of the stronger unmodified dual program in~\autoref{lem:dual-program}, but not in the usual expander flow formulation corresponding to the constrained dual program as in in~\eqref{eqn:expander-flow}.


\subsubsection{Project Max-Flow Algorithm}

In the well spread case in \autoref{prop:embedding-cases}, we will only focus on the vectors in the subset $U$, with $\pi(U) \gtrsim \pi(V)$ and $\norm{v_i} \leq 1$ for $i \in U$ and $\sum_{i,j \in U} \pi(i) \cdot \pi(j) \cdot \norm{v_i-v_j}^2 \gtrsim 1$.


\begin{algorithm}
\caption{Project Max-Flow $(G, u, c, \beta, \kappa)$}
\label{alg:project-max-flow}
\textbf{Input:} An edge-capacitated directed graph $G = (V, E, w)$ with vertex weights $\pi:V \to \R_+$, an embedding $v_1,\ldots,v_{|U|} \in \R^n$ of the vertices in $U$, vector $u \in \R^n$, small constant $c$, 
congestion parameter $\kappa$, and flow value parameter $\beta$. 
\begin{enumerate}
    \item Order the vertices $i \in U$ by the values of $\inner{u}{v_i}$.
    Let $L$ be the $l$ smallest vertices in this ordering, where $l$ is the smallest integer such that $\pi(L)\geq c \cdot \pi(V)$. 
    Let $R$ be the $r$ largest vertices in this ordering, where $r$ is the smallest integer such that $\pi(R) \geq c \cdot \pi(V)$.
    
    \item Compute a bidirectional max-flow using \autoref{alg:bidirectional-max-flow} on $(L,R,\beta,\kappa)$ to obtain either a cut $S \subseteq V$ or a circulation $f$ in $G$ with congestion $\kappa$.
\end{enumerate}
\end{algorithm}


The following lemma shows that with constant probability over the random direction $u$, 
the sets $L$ and $R$ will be well-separated along the direction $u$.

\begin{lemma}[Good Direction]
\label{lem:good-vectors}
Let $v_1,\ldots,v_{|U|}$ be a set of vectors that satisfies (i) $\pi(U) \gtrsim 1$, (ii) $\norm{v_i} \leq 1$ for all $i \in U$ and (iii) $\sum_{i,j \in U} \pi(i) \cdot \pi(j) \cdot \norm{v_i-v_j}^2 \gtrsim 1$.
Then there exist positive constants $\gamma$ and $\sigma$ and $c$ such that if we sample random $u \sim N(0, I)$, then with probability at least $\gamma$ 
the sets $L,R$ in step (1) of \autoref{alg:project-max-flow} satisfy the condition that $\inner{u}{v_i-v_j}\geq \sigma$ for all $i \in L, j \in R$. 
We say that such vectors $u$ are good vectors.
\end{lemma}

The proof is a simple reduction to the uniform $\pi$ case proven in \cite[Lemma14]{Kal07} and \cite[Lemma 5.3.3]{She09}, and so we defer to the Appendix.
We remark that \autoref{lem:good-vectors} is the only place in the proof of \autoref{prop:well-spread-case} that we use the assumption that the vectors are well spread.

For each good vector $u \in \R^n$, 
the sets $L,R$ in step (1) of \autoref{alg:project-max-flow} are disjoint,
and so the bi-directional max-flow in \autoref{alg:bidirectional-max-flow} is well-defined.
Therefore, exactly one the following three cases must happen.
\begin{itemize}
    \item \textbf{A:}
    \autoref{alg:project-max-flow} returns a circulation $f$ with demand graph $D$ such that $\inner{L_{\rm sym}(D)}{Y_t} \ge 1$.

    \item \textbf{B:}
    \autoref{alg:project-max-flow} returns a circulation $f$ with demand graph $D$ such that $\inner{L_{\rm sym}(D)}{Y_t} < 1$.
    
    \item \textbf{C:}
    \autoref{alg:project-max-flow} returns a cut $S \subseteq V$.
\end{itemize}

If we are in case C for some good vector $u$, then we show that the primal-dual \autoref{alg:mmw-expander-flow} can terminate with approximation ratio $O(\beta)$.
(In the proof of \autoref{prop:well-spread-case} that we will present later, we will set $\beta = O(\sqrt{\log n / \eps})$.)

\begin{claim}[Case C] \label{c:C}
If \autoref{alg:project-max-flow} returns a cut $S$ for some good vector $u$,
then $\vec{\phi}_{\pi}(S) = O(\beta / \kappa)$.
\end{claim}
\begin{proof}
Since $u$ is good, the sets $L,R$ are disjoint and $\pi(L), \pi(R) \geq c \cdot \pi(V)$ for some (small) constant $c$.
Then, by \autoref{lemma:flow-cut-lemma} (unsaturated case), \autoref{alg:bidirectional-max-flow} will return a set $S$ with $\vec{\phi}_\pi(S) \leq \beta r' / \kappa \lesssim \beta / \kappa$ as $r' = \max\{1,\pi(R)/\pi(L)\} \leq 1/c$.
\end{proof}

If we are in case A for some good vector $u$, then we show that the oracle succeeds in step 2(a) of \autoref{alg:mmw-expander-flow} with width $\rho = O(\beta)$.

\begin{claim}[Case A] \label{c:A}
If \autoref{alg:project-max-flow} returns a circulation $f$ with demand graph $D$ for some good vector $u$ such that $\inner{L_{\rm sym}(D)}{Y_t} \geq 1$, then the feedback matrix $M_t := \Pi^{-\frac12}L_{\rm sym}(D)\Pi^{-\frac12}$ in step 2(a) of \autoref{alg:mmw-expander-flow} satisfies $\norm{M_t} \lesssim \beta$.
\end{claim}
\begin{proof}
Since $u$ is good, the sets $L,R$ are disjoint and $\pi(L), \pi(R) \geq c \cdot \pi(V)$ for some (small) constant $c$.
In the bi-directional max-flow problem in \autoref{alg:bidirectional-max-flow},
each vertex $i$ in $L \cup R$ has degree at most $r' \cdot \beta \cdot \pi(i) \lesssim \beta \cdot \pi(i)$ as $r' = \max\{1,\pi(R)/\pi(L)\} \leq 1/c$.
This implies that the demand graph satisfies $L_{\rm sym}(D) \preceq O(\beta \cdot \Pi)$, and thus $M_t = \Pi^{-\frac12}L_{\rm sym}(D)\Pi^{-\frac12} \preceq O(\beta I)$. 
\end{proof}




\subsubsection{Finding Many Violating Paths}


If we are in case B for some good vector $u$, then we show how to construct a large matching of flow paths between pairs of vertices with small embedding distance using the following algorithm.

\begin{algorithm}
  \caption{Matching(u)}
  \label{alg:matching-with-direction}
  \textbf{Input:} $s$-$t$ flow $\vec{f}$ and $t$-$s$ flow $\cev{f}$ obtained from Step 2 of \autoref{alg:project-max-flow}, with parameters $(G, u, c, \beta, \kappa)$.
  \begin{enumerate}
    \item Decompose the two flows into at most $m$ flow paths between sets $L$ and $R$.
    Ignore the original direction 
    of the paths and reorient every path from $L$ to $R$. In particular, the paths we get are $(p_r, i_r, j_r, f_{p_r})_{r=1}^k$ where $k\leq 2m$. For each $r\in [k]$, $p_r$ is a path from $i_r \in L$ to $j_r \in R$, with weight $f_{p_r}$.
    
    \item Discard any path $p_r$ with $\inner{v_{j_r} - v_{i_r}}{u} < \sigma$ or $\norm{v_{j_r} - v_{i_r}}^2 > \frac{4}{\beta c}$.
    
    \item Define $\M'_u$ so that $\M'_u(i, j)$ is the sum of the weights of all remaining paths from $i$ to $j$. Return 
    \[\M_u := \frac{1}{\beta \cdot \max\Big\{1,\pi(R)/\pi(L)\Big\}} \M'_u.\]
  \end{enumerate} 
\end{algorithm}

The reason that we ignored the original direction of the paths in \autoref{alg:matching-with-direction} is that we are trying to find paths in $K_n$, rather than in $G$, that violate the triangle inequality. Thus, it is fine if the resulting violating paths from chaining together the matchings do not correspond to paths in $G$. 


\begin{lemma}[Case B] \label{lem:B-matching}
If \autoref{alg:project-max-flow} returns a circulation $f$ with demand graph $D$ for some good vector $u$ such that $\inner{L_{\rm sym}(D)}{Y_t} < 1$, then \autoref{alg:matching-with-direction} returns a fractional matching $\M_u$ with $w(\M_u) \gtrsim c^2$, where each edge $ij$ in $\M_u$ satisfies $\inner{v_j-v_i}{u} \geq \sigma$ and $\norm{v_j - v_i}^2 \lesssim \frac{1}{\beta c}$. Moreover, there is a randomized algorithm to compute $\mathcal{M}_u$ in expected time $O(m\log{n})$.
\end{lemma}

\begin{proof}
Since \autoref{alg:project-max-flow} returns a circulation $f$, 
both flows $\vec{f}$ and $\cev{f}$ from \autoref{alg:bidirectional-max-flow} are saturating.
This implies that the flow value for $\vec{f}$ and $\cev{f}$ is at least $\sum_{j \in R} \beta \cdot \pi(j) = \beta \cdot \pi(R) \geq \beta \cdot c \cdot \pi(V) = \beta \cdot c$,
and thus 
\[\sum_{i \in L, j\in R} \big( D(i,j) + D(j,i) \big) \geq \beta \cdot c\] 
where $D$ is the demand graph of the circulation $f = \frac{1}{2}(\vec{f} + \cev{f})$.

Next, we bound the total weight of the flow paths that we discard in step (2) of \autoref{alg:matching-with-direction}.
Since $u$ is good (see \autoref{lem:good-vectors}), all flow paths are between vectors $i,j$ such that $\inner{v_j-v_i}{u} \geq \sigma$, so no paths will be discarded this way.
Our assumption implies that
\[
  1 > \inner{L_{\rm sym}(D)}{Y_t} = \frac{1}{2}\sum_{i\in L,j\in R} \big(D(i,j) + D(j,i) \big) \cdot \norm{v_i-v_j}^2,
\]
and thus an average flow path is between pairs $i,j$ with $\norm{v_i-v_j}^2 < \frac{2}{\beta c}$.
By Markov's inequality, at least half of the flow of $f$ is on flow paths between $i,j$ with $\norm{v_i - v_j}^2 \leq \frac{4}{\beta c}$.
Therefore, we discard at most half of the flow of $f$ in Step (2) of \autoref{alg:matching-with-direction}, 
and hence the weight of $\M'_u$ is at least $\beta c / 2$. 

By the construction of the bidirectional flow in \autoref{alg:bidirectional-max-flow}, each source and sink vertex has degree at most $\beta \cdot \pi(i) \cdot \max\{1, \pi(R)/\pi(L) \} \leq \beta \cdot \pi(i) / c$. 
So, scaling $\M'_u$ down by this factor gives a fractional matching $\M_u$ as defined in \autoref{def:fractional-matching}, with $w(\M_u)$ at least $c^2 / 2$.

Finally, we bound the runtime of the algorithm. The only non-trivial step is step 1, in which we must decompose a fractional single-commodity flow into integral flow paths. The following result shows that this can be done in nearly linear time on expectation.

\begin{theorem}(\cite[Theorem 5]{LRS13})\label{thm: flow-path-decomposition}
    Given a fractional $s$-$t$ flow $\vec{f}$, there is a randomized algorithm that, in $O(m\log{n})$ expected time, returns a flow path decomposition $(p_r, \vec{f}_{p_r})_{r=1}^k$ where $k\leq m$ and each $p_r$ is a path from $s$ to $t$ with flow value $\vec{f}_{p_r}$ along the path.
\end{theorem}
\end{proof}

It follows that if case B happens often enough, then we can construct a fractional matching cover as defined in \autoref{def:matching-cover}.

\begin{lemma} \label{lem:B-matching-cover}
Suppose that conditioned on $u$ being a good vector, the probability that we are in case $B$ is at least $1/2$. 
Then $\M = \{\M_u\}_{u \in \R^n}$ is a $(\sigma,\delta)$-matching cover with $\sigma,\delta = \Omega(1)$.
\end{lemma}
\begin{proof}
Clearly, conditions (i) and (ii) in \autoref{def:matching-cover} are met. 
As long as $u$ is a good vector, $w(\M_u) \gtrsim c^2$ by \autoref{lem:B-matching}. 
As a random vector is a good vector with probability at least $\gamma$ by \autoref{lem:good-vectors}, we conclude that $\M = \{\M_u\}_{u \in \R^n}$ is a $(\sigma, \delta)$-matching cover with $\sigma = \Omega(1)$ and $\delta = \gamma \cdot c^2 = \Omega(1)$.
\end{proof}

We apply Sherman's chaining theorem on the matching cover to construct many violating paths for step 2(b) in \autoref{alg:mmw-expander-flow}.

\begin{lemma}[Violating Paths] \label{lem:violating-paths}
Given the $(\Omega(1),\Omega(1))$-matching cover $\M$ in \autoref{lem:B-matching-cover}, by setting $\beta = O\Big( \sqrt{\frac{\log n}{\eps}} \Big)$, there is a randomized algorithm using $O(\sqrt{\eps \log n})$ max-flow computations to find paths $p_1, \ldots, p_s$ with weight $f_{p_1}, \ldots, f_{p_s}$, so that the feedback matrix $M_t := -\Pi^{-\frac12} \big( \sum_{r=1}^s f_{p_r} \cdot T_{p_r} \big) \Pi^{-\frac12}$ satisfies $\inner{M_t}{X_t} \geq 1$ and $\norm{M_t} = \tilde{O}(n^\eps / \eps^{3/2})$, with success probability $\Omega(n^{-\eps})$.
\end{lemma}
\begin{proof}
We apply Sherman's \autoref{thm:sherman-main-thm} on $\M$ with $l = \Theta(\epsilon)$ and $k = O(\sqrt{\eps \log n})$ to obtain an efficiently sample-able distribution $\D$ over $(u_1, \ldots, u_k)$ so that the expected total weight of paths in $\M(u_1,\ldots,u_k)$ between $i,j$ with $\norm{v_i-v_j}^2 \geq l = \Theta(\eps)$ is at least $e^{-O(k^2)} \cdot \pi(V) = O(n^{-\eps})$.
Since the total weight is at most $1$, by a reverse application of Markov's inequality, we will find vectors $u_1, \ldots, u_k$ where the weight of such good paths in $\M(u_1,\ldots,u_k)$ is at least $\frac12 n^{-\eps}$, with probability at least $n^{-\eps}$.
With such $u_1, \ldots, u_k$, by \autoref{claim:chaining-runtime}, 
we can find paths $p_1,\dots, p_s$ with $\sum_{r=1}^s f_{p_r} \geq \frac12 n^{-\epsilon}$ in $O(m k^2) = \tilde{O}(m)$ time, such that for $1 \leq r \leq s$ the path $p_r = (v_{i_1},\dots, v_{i_{k+1}})$ satisfies 
\[\sum_{j=1}^{k}\norm{v_{i_j} - v_{i_{j+1}}}^2 \lesssim \frac{k}{\beta \cdot c} \lesssim \frac{\sqrt{\eps \log n}}{\beta \cdot c}
\quad {\rm but} \quad \norm{v_{i_1} - v_{i_{k+1}}}^2\geq l \gtrsim \eps,
\]
where the first inequality is by the property that each edge $ij$ in each fractional matching has $\norm{v_i-v_j}^2 \leq \frac{4}{\beta c}$ in \autoref{lem:B-matching}.
Thus, by choosing $\beta = \Theta\Big(\frac{\sqrt{\eps \log n}}{c \cdot l}\Big) = \Theta\Big(\sqrt{\frac{\log{n}}{\epsilon}}\Big)$ with the appropriate constant, the paths violate triangle inequality so that
\[\inner{T_{p_r}}{Y_t} = 
\sum_{j=1}^{k}\norm{v_{i_j} - v_{i_{j+1}}}^2 - \norm{v_{i_1}-v_{i_{k+1}}}^2 
\lesssim -\eps
\quad \implies \quad
\biginner{\sum_{r=1}^s f_{p_r} T_{p_r}}{Y_t} 
\lesssim -\eps \cdot n^{-\epsilon}
\] 
Setting $y = \Theta(n^\eps / \eps)$ with the appropriate implicit constant and the feedback matrix $M_t := -\Pi^{-\frac12}(y \sum_{r=1}^s f_{p_r} T_{p_r}) \Pi^{-\frac12}$, we ensure that $\inner{M_t}{X_t} = \inner{-y \sum_{r=1}^s f_{p_r}T_{p_r}}{Y_t} \geq 1$.

Finally, we bound $\norm{M_t}$ to bound the width of the oracle.
Note that the edges in these violating paths form a subgraph of the graph $\M_{u_1}\cup\dots \cup \M_{u_k}$.
In each of these graphs, the total degree of vertex $i$ is at most $2 \beta \cdot \pi(i) \leq 2 \beta$, and so the total degree of each vertex $i$ in the union is at most $2\beta \cdot k$. 
Thus, $\norm{M_t} \leq 2y \cdot \beta \cdot k = \Tilde{O}(n^\epsilon / \eps^{3/2})$. 
\end{proof}

\subsubsection{Proof of \autoref{prop:well-spread-case}}

We are ready to put together the results in this subsection to finish the proof of \autoref{prop:well-spread-case}.
Set $\beta = O\Big(\sqrt{\frac{\log n}{\eps}}\Big)$. 
Recall that, by \autoref{lem:good-vectors}, there is a positive constant $\gamma$ such that $u \sim N(0, I)$ is a good vector with probability at least $\gamma$.

Suppose that when conditioned on $u$ being a good vector, we are in case C of \autoref{alg:project-max-flow} with probability at least $\frac14$.
This means that, with probability at least $\frac{\gamma}{4} = \Omega(1)$, a set $S$ with $\vec{\phi}_{\pi}(S) = O\Big(\frac{\beta}{\kappa}\Big) = O\Big(\frac{1}{\kappa} \sqrt{\frac{\log n}{\eps}}\Big)$ will be returned by \autoref{c:C}.
Therefore, after $O(\log n)$ independent samples of $u$, such a sparse cut will be returned with high probability, and so \autoref{alg:mmw-expander-flow} can be terminated.

Similarly, suppose that when conditioned on $u$ being a good vector, we are in case A of \autoref{alg:project-max-flow} with probability at least $1/4$.
This means that, with probability $\Omega(1)$, a feedback matrix $M_t := \Pi^{-\frac12} L_{\rm sym}(D) \Pi^{-\frac12}$ from a circulation $f$ with demand graph $D$ can be returned with $\inner{M_t}{X_t} \geq 1$ and $\norm{M_t} \lesssim \beta \lesssim \sqrt{\frac{\log n}{\eps}}$ by \autoref{c:A}.
Therefore, after $O(\log n)$ independent samples of $u$, such a circulation will be returned with high probability, and so \autoref{alg:mmw-expander-flow} can proceed to the next iteration.

Otherwise, suppose that when conditioned on $u$ being a good vector, we are in case B of \autoref{alg:project-max-flow} with probability at least $1/2$.
By \autoref{lem:violating-paths}, we can use Sherman's result to chain together $O(\sqrt{\eps \log n})$ such flows to find violating paths $p_1, \ldots, p_s$ so that the feedback matrix $M_t := \Pi^{-\frac12} (\sum_r^2 f_{p_r} \cdot T_{p_r}) \Pi^{-\frac12}$ satisfies $\inner{M_t}{X_t} \geq 1$ and $\norm{M_t} = \tilde{O}(n^\eps / \eps^{3/2})$, with probability at least $\Omega(n^{-\eps})$. 
Therefore, after $\tilde{O}(n^{\eps})$ chaining attempts using a total of $\tilde{O}(n^{\eps})$ max-flow computations, such violating paths will be returned with high probability, and \autoref{alg:mmw-expander-flow} can proceed to the next iteration.

These covers all the cases.  
The width and the runtime of the oracle are dominated by the step of finding violating paths.

\subsection{Main Result and Corollary}\label{sec:concluding-remarks}

In this subsection, we prove \autoref{thm:directed-Sherman} and \autoref{prop:reweighted-phi}.

\begin{proofof}{\autoref{thm:directed-Sherman}}
By \autoref{lem:MMW-approximation-guarantee}, if there is an oracle with width $\rho$ and approximation factor $\alpha$, then the regret minimization \autoref{alg:mmw-expander-flow} either certifies that $\vec{\phi}_\pi(G) \geq \Omega(1/\kappa)$ or finds a cut $S \subseteq V$ with $\vec{\phi}_\pi(S) \leq O(\alpha/\kappa)$ for a given $\kappa$ in $O(\rho^2 \log n)$ iterations.
Combining the oracle in the large core case in \autoref{lem:large-core-case} and the oracle in the well spread case in \autoref{prop:well-spread-case}, we obtain an oracle with width $\rho = O(n^\eps / \eps^{3/2})$ and approximation ratio $\alpha = O(\sqrt{\log n / \eps})$.
Therefore, by doing binary search on $\kappa$, we can obtain a $O(\sqrt{\log n / \eps})$-approximation algorithm by running a total of $\Tilde{O}(n^{2\eps})$ iterations of matrix multiplicative weight update in \autoref{alg:mmw-expander-flow}.

Now, we will bound the runtime of each iteration. By \autoref{prop:well-spread-case}, each iteration requires $\tilde{O}(n^\eps)$ max-flow computations. After each max-flow computation, we need to perform a flow-path decomposition either in \autoref{alg:matching-with-direction} or by computing the edges of the demand graph $D_t$, which can be implemented in expected $O(m\log{n})$ time by \autoref{thm: flow-path-decomposition}. In addition, each iteration requires the use of a matrix exponential, whose computation is too long. Thus, instead of computing $Y_t$, we will approximately compute its Gram decomposition using the following lemma, whose proof we will defer to \autoref{app:dual-rounding}:
\begin{lemma}[Matrix Exponential Computation]\label{lem:matrix-exp-computation}
    Let $v_1,\dots v_n$ be the Gram decomposition of the matrix $Y_t$ in step $1$ of \autoref{alg:mmw-expander-flow}. There is a randomized algorithm that, in $\Tilde{O}(\rho m/\delta^2)$ time, computes vectors $\hat{v}_1,\dots \hat{v}_n\in \mathbb{R}^d$ for $d=O(\log{n}/\delta^2)$ such that with probability at least $1-n^{-1}$,
    \begin{align*}
        \norm{\hat{v}_i - \hat{v}_j}^2 \in (1\pm \delta)\norm{v_i - v_j}^2 \pm n^{-\Omega(1)}\quad \forall i,j\in V 
    \end{align*}
\end{lemma}
In particular, \autoref{lem:matrix-exp-computation} implies that 
if $L$ is any Laplacian matrix (possibly with negative edge-weights) satisfying $\inner{L}{\sum_i\hat{v}_i\hat{v}_i^\top} \geq 1$, then $\inner{L}{Y_t}\geq 1-2\delta$. Since all our feedback matrices are always of the form $\Pi^{-1/2}L\Pi^{-1/2}$ for some Laplacian $L$, it suffices to use $\hat{v}_1\dots \hat{v}_n$ as the embedding vectors at step 1 of the algorithm in order to ensure that $\inner{M_t}{X_t}\geq 1-2\delta$ for every iteration $t$ even though we never have to explicitly compute $X_t$.
Since $\rho = \Tilde{O}(n^{\epsilon})$, the overall runtime of each iteration is dominated by the runtime of $\Tilde{O}(n^{\epsilon})$ maxflow computations. 
\end{proofof}

An interesting corollary is about a dual certificate using circulations in \autoref{prop:reweighted-phi}.

\begin{proofof}{\autoref{prop:reweighted-phi}}
Apply \autoref{lem:MMW-approximation-guarantee} with $\kappa \gtrsim \sqrt{\frac{\log n}{\epsilon}} / \vec{\phi}_\pi(G)$ for a suitable implicit constant,
 \autoref{alg:mmw-expander-flow} will always outputs a dual solution with value $\Omega(1/\kappa)$ rather than a cut as there is no cut $S$ with $\edgeexp(S) \le O\Big(\frac{1}{\kappa}\sqrt{\frac{\log{n}}{\epsilon}}\Big) \le \frac{1}{2} \edgeexp(G)$.
Therefore, we can find a circulation $F$ with demand graph $D$, 
and weights $y_p$ over shortcut cycles, such that
    \[
      \phi_\pi(F) \geq
      \lambda_2 \bigg( \Pi^{-\frac12}\bigg(L_{\rm sym}(D) - \sum_py_pT_p\bigg)\Pi^{-\frac12} \bigg)
      \gtrsim \frac{\edgeexp(G)}{\sqrt{\log{n}}},
    \]
where the first inequality is by \eqref{eqn:certify}.
\end{proofof}

\section{Primal-Dual Algorithms for Reweighted Eigenvalues and \texorpdfstring{\\}{}Cut-Matching Game}
\label{sec:primal-dual}

In this section, we show that the regret minimization framework can also be used to compute reweighted eigenvalues in \autoref{sec:mmw-reweighted-eigenvalue} and to derive cut-matching game in \autoref{sec:directed-cut-matching-game}.

\subsection{Reweighted Eigenvalues}
\label{sec:mmw-reweighted-eigenvalue}

In this subsection, we use the regret minimization framework to compute the reweighted eigenvalue defined in~\cite{LTW23}.
The main result is that there is a primal-dual algorithm to compute $\lambda_2^*(G)$ in $O(\log n / \lambda_2^*(G))$ iterations, with each iteration taking almost linear time.
This combined with \autoref{thm:directed-Sherman} will prove \autoref{thm:fast-Cheeger}.

The reweighted eigenvalue was used in~\cite{LTW23} to approximate the directed edge conductance $\vec{\phi}(G)$, which is a special case of the $\pi$-weighted directed edge expansion $\vec{\phi}_{\pi}(G)$ when $\pi(i) = w(\delta^+(i)) + w(\delta^-(i))$, the total degree of vertex $i$.
The result in this subsection only applies to this special case.
To avoid confusion, we use the notation $d_w(i) := w(\delta^+(i)) + w(\delta^-(i))$ to denote the total degree of vertex $i$ instead of using $\pi(i)$, and $D_w := \diag(d_w)$ to denote the diagonal total-degree matrix instead of using $\Pi$.

From \autoref{def:reweighted-eigenvalue}, 
the reweighted eigenvalue is formulated as
\begin{equation} \label{eqn:lambda2*}
    \lambda_2^*(G) := \max_{F\in \mathcal{F}(G)}
    \lambda_2\bigg(
      D_w^{-\frac12} \bigg( L_{\rm sym}(F) \bigg) D_w^{-\frac12}
    \bigg).
\end{equation}
To construct a circulation $F$ that maximizes the objective value, 
we can use the regret minimization framework as in \autoref{sec:Arora-Kale-directed} and \autoref{sec:mmw-expander-flow}.
This framework reduces the above maximization problem to the simpler task of finding a circulation $F_t \in {\mathcal F}(G)$ that maximizes $\inner{F_t}{X_t}$ where $X_t$ is the density matrix in the matrix multiplicative update method in the $t$-th iteration, which can be found using a min-cost flow computation.
Then, the regret bound in \autoref{thm:regret-bound} can be used to prove that the average circulation $\frac{1}{T} \sum_{t=1}^T F_t$ will be an approximate maximizer to $\lambda_2^*(G)$.

Alternatively, using the min-max formulation from \cite{LTW23}[Proposition 3.4]  where 
  \begin{align}
  \label{eqn:phi-d-sdp}
  \begin{split}
    \lambda_2^*(G) =
     \min_{v_1, \ldots, v_n \in \mathbb{R}^n}\max_{F\in \mathcal{F}(G)} 
     &~~~ \sum_{i<j} \frac12\big(F(i,j) + F(j,i)\big) \cdot \norm{v_i - v_j}^2 \\
     \st 
     &~~~\sum_{i=1}^n d_w(i) \cdot v_i = \vec{0}\\
     &~~~\sum_{i=1}^n d_w(i) \cdot \norm{v_i}^2 = 1,
  \end{split}
  \end{align}
we can also interpret the following algorithm as a natural way to play a minimax game between a primal ``embedding'' player and a dual ``circulation'' player.

\begin{algorithm}\caption{Regret Minimization Algorithm for Reweighted Eigenvalue}
  \label{alg:mmw-eigenvalue}
  \textbf{Input:} A directed graph $G = (V,E, w)$ 
   and step size $\eta \in (0,1)$.
 
\vspace{2mm}
 
  \textbf{Initialization:} $X_0 = \frac{1}{n-1}(I-D_w^{\frac12}\one\one^\top D_w^{\frac12})$.

\vspace{2mm}

  For $t = 0$ to $T-1$:
  \begin{enumerate}
    \item Given $X_t \succeq 0$ such that $\tr(X_t) = 1$ and $X_t\perp D_w^{\frac12}\one$, let $Y_t := D_w^{-\frac12} X_t D_w^{-\frac12}$ and $v_1, \dots, v_n$ be the Gram decomposition of $Y_t$. 
    
    \item (Dual Player) Compute circulation $F_t := \arg\max_{F\in \mathcal{F}(G)}\sum_{i<j} \frac12 (F(i,j) + F(j,i)) \cdot \norm{v_i-v_j}^2$ and set the feedback matrix $M_t := D_w^{-\frac12}L_{\rm sym}(F_t)D_w^{-\frac12}$.
    
    \item (Primal Player) Update $X_{t+1}' := \exp\Big(-\frac{\eta}{\rho}\sum_{i=0}^tM_i\Big)$.
    Let $X_{t+1}$ be obtained from $X_{t+1}'$ by projecting it onto the space orthogonal to $D_w^{\frac12}\one$ and scaling it to have trace $1$.

  \end{enumerate}

  Output $\overline{F} = \frac{1}{T}\sum_{t=0}^{T-1} F_t$. 
\end{algorithm}

We bound the number of iterations to obtain a good approximate solution.

\begin{theorem}[Regret Minimization for Reweighted Eigenvalue]
\label{thm:mmw-reweighted-eigenvalues}
    Let $0 < \eta < 1/2$.
    The solution $\overline{F}$ returned by \autoref{alg:mmw-eigenvalue} satisfies
    \[
    \lambda_2\Big(D_w^{-\frac12}L_{\rm sym}(\overline{F})D_w^{-\frac12}\Big)\geq (1-2\eta) \cdot \lambda_2^*(G) 
    {\rm~~after~~} 
    T = \frac{\log{n}}{\eta^2\lambda_2^*(G)} {\rm~~iterations}.
    \]
    Moreover, each iteration can be implemented using one min-cost flow computation. 
\end{theorem}

\begin{proof}
  The main step is to lower bound the inner product $\inner{M_t}{X_t}$ in each iteration.
  The observation is that $v_1, \ldots, v_n$ form a feasible solution to the $\lambda_2^*(G)$ program as stated in \eqref{eqn:phi-d-sdp}.
  To see this, we just need to check that $v_1, \ldots, v_n$ satisfies the constraints in \eqref{eqn:phi-d-sdp}.
  Since $Y_t D_w \one = \vec{0}$ and $Y_t = \sum_i v_iv_i^T$, 
  we have $\sum_i d_w(i) \cdot v_i = \vec{0}$. 
  Also, we have $\tr(X_t) = \sum_i d_w(i) \cdot \norm{v_i}^2 = 1$.
  Therefore, by the definition of $F_t$ in step (2) of \autoref{alg:mmw-eigenvalue} and $\lambda_2^*(G)$ in \eqref{eqn:phi-d-sdp},
  \begin{equation} \label{eqn:key}
    \inner{M_t}{X_t} 
    = \inner{L_{\rm sym}(F_t)}{Y_t}
    =\max_{F\in \mathcal{F}(G)}\sum_{i<j} \frac12 \big( F(i,j) + F(j,i) \big) \cdot \norm{v_i- v_j}^2
    \geq
    \lambda_2^*(G).
  \end{equation}
Note that the width\footnote{This is the reason that this theorem does not hold for general $\pi$.} is $\norm{M_t} \leq 1$ because each vertex $i$ has degree at most $d_w(i)$ in $F_t \in {\mathcal F}(G)$.
As each $M_t \succeq 0$, by applying the regret bound~\eqref{eqn:psd-regret-bound} in \autoref{thm:regret-bound} restricting to the subspace orthogonal to $D_w^{\frac12} \one$, it follows that
  \begin{align}
    \label{eqn:regret-bound-for-lambda2}
    \lambda_2\Big(D_w^{-\frac12}L_{\rm sym}(\overline{F})D_w^{-\frac12}\Big)
     & \geq \frac{1}{T}
    \sum_{t=0}^{T-1} \inner{M_t}{X_t} \cdot (1-\eta) - \frac{\log{n}}{\eta T}
    \geq
    (1-\eta) \cdot \lambda_2^*(G)- \frac{\log{n}}{\eta T}
    \geq 
    (1-2\eta) \cdot \lambda_2^*(G),
  \end{align}
where the last inequality is by our choice of $T$.
Finally, note that the maximization problem in step~(2) of \autoref{alg:mmw-eigenvalue} can be solved using one min-cost flow computation, which can be implemented in $m^{1+o(1)}$ time by ~\cite{CKLPPS22} (see \autoref{app:min-cost-flow}).
\end{proof}


\subsubsection{Fast Algorithm for Cheeger-Type Guarantee} \label{sec:fast-Cheeger}

Note that \autoref{alg:mmw-eigenvalue} is fast when $\lambda_2^*(G)$ is large.
On the other hand, when $\lambda_2^*(G)$ is small, then $\vec{\phi}(G)$ is also small by the directed Cheeger inequality in \eqref{eqn:Cheeger-directed}, and thus the $O(\sqrt{\log n})$-approximation in \autoref{thm:directed-Sherman} is better than the directed Cheeger guarantee.
So, we can combine \autoref{thm:mmw-reweighted-eigenvalues} and \autoref{thm:directed-Sherman} to prove \autoref{thm:fast-Cheeger}.

\begin{proofof}{\autoref{thm:fast-Cheeger}}
First, we apply \autoref{alg:mmw-expander-flow} and \autoref{lem:MMW-approximation-guarantee} with $1/\kappa := 1/\log^{1.5} n$ to either certify $\vec{\phi}(G) \gtrsim 1/\log^{1.5} n$ or to find a set $S$ with $\vec{\phi}(S) \lesssim 1/\log n$.
In the latter case, we know that the set $S_1$ with $\vec{\phi}(S_1) \lesssim \sqrt{\log n} \cdot \vec{\phi}(G)$ returned by \autoref{thm:directed-Sherman} has smaller directed edge conductance than the guarantee by the directed Cheeger inequality in~\eqref{eqn:Cheeger-directed}, and so we are done.

In the former case, we compute a set $S_2$ of directed edge conductance $\vec{\phi}(S_2) \lesssim \sqrt{\phi(G) \log 1/\vec{\phi}(G)}$ and return $S_2$.
Since $\vec{\phi}(G) \geq 1/\log^{1.5} n$ in this case, 
the directed Cheeger inequality in \eqref{eqn:Cheeger-directed} implies that $\lambda_2^*(G) \geq \vec{\phi}^2(G) / \log(1/\vec{\phi}(G)) \geq 1/\log^4 n$.
By setting $\eta=1/4$, we can get a $1/2$-approximation of $\lambda_2^*(G)$ in $O(\log^5 n)$ iterations using \autoref{alg:mmw-eigenvalue}.

We show how to compute $S_2$ from the computations done in \autoref{alg:mmw-eigenvalue} and the flow-based rounding algorithm for the directed Cheeger inequality in \autoref{sec:Cheeger}.
Let $\lambda := \min_{1 \leq t \leq T} \inner{L_{\rm sym}(F_t)}{Y_t}$ and 
$u_1,\ldots,u_n$ be the Gram decomposition of a $Y_t$ that achieves this minimum.
Since $u_1,\ldots,u_n$ is a solution to \eqref{eqn:phi-d-sdp} with objective value $\lambda$,
using the Gaussian projection and the metric rounding step in \autoref{sec:Cheeger},
we can obtain a set $S_2$ with $\vec{\phi}(S_2) \lesssim \sqrt{\lambda \log(1/\vec{\phi}(G))}$.

It remains to argue that $\lambda \lesssim \vec{\phi}(G)$ to prove the approximation guarantee.
By \eqref{eqn:regret-bound-for-lambda2} in the proof of \autoref{thm:mmw-reweighted-eigenvalues}, we have
\[
\lambda_2\Big(D_w^{-\frac12}L_{\rm sym}(\overline{F})D_w^{-\frac12}\Big)
\geq \lambda(1-\eta) - \lambda_2^*(G) \cdot \eta
\geq \lambda(1-2\eta) = \frac12 \lambda,
\]
where the second inequality is because $\lambda \geq \lambda_2^*(G)$ by \eqref{eqn:key} and the last equality is because $\eta=1/4$.
This implies that 
\[
\lambda \leq 2\lambda_2\Big(D_w^{-\frac12}L_{\rm sym}(\overline{F})D_w^{-\frac12}\Big)
\leq 2 \lambda_2^*(G) \leq 4\vec{\phi}(G),
\]
where the second inequality is due to $\overline{F} \in {\mathcal F}(G)$ and \eqref{eqn:lambda2*}, and the last inequality is by the easy direction in~\eqref{eqn:Cheeger-directed}.
This proves that $\vec{\phi}(S_2) \lesssim \sqrt{\phi(G) \log 1/\vec{\phi}(G)}$.

Finally, we bound the time complexity of the algorithm.
Computing $S_1$ takes $\tilde{O}(m^{1+\eps})$ for an arbitrarily small constant $\eps$ using the fast max-flow algorithm in~\cite{CKLPPS22}. In the case that we also need to compute $S_2$,
it takes $O(m^{1+o(1)})$ time to compute a $1/2$-approximation of $\lambda_2^*(G)$, where the bottleneck is the min-cost flow computations in step (2) of \autoref{alg:mmw-eigenvalue}. Note that once again, the matrix exponential can be computed in $\Tilde{O}(m)$ time each iteration by \autoref{lem:matrix-exp-computation}.
Finally, the metric rounding step also takes $O(m^{1+o(1)})$ time, as it also requires $O(\log{n})$ max-flow computations in \autoref{lemma:metric-rounding-lemma}.
\end{proofof}

\subsection{Cut-Matching Game} \label{sec:directed-cut-matching-game}

Louis~\cite{Lou10} considered the following cut-matching game for directed graphs.
In each round, the cut player chooses a bisection $(S,\overline{S})$ of the vertices, and the matching player chooses a directed perfect matching between $(S,\overline{S})$, which is defined as an Eulerian graph where each vertex has indegree and outdegree exactly one.
Louis proved that there is a cut-player strategy such that the union of the directed perfect matchings has edge expansion $\Omega(1)$ in $O(\log^2 n)$ iterations.
Note that the edge expansion in~\cite{Lou10} is the special case of \autoref{def:phi-pi} when $\pi(i)=1$ for all $i \in V$.


In this subsection, we use the matrix multiplicative weight update method in \autoref{alg:mmw-expander-flow} to derive an improved cut-player strategy and prove \autoref{thm:cut-matching-game}.
We also extend the cut-matching game to the more general setting of $\pi$-weighted directed edge expansion in \autoref{def:phi-pi},
for which the bipartition returned by the cut player may not be balanced.
\begin{algorithm}
\caption{Cut Player Strategy}\label{alg:cut-player}
    \begin{enumerate}
         \item Let $D_1,\ldots,D_{t-1}$ be the directed perfect matchings played so far. Let $M_i = \Pi^{-\frac12}L_{\rm sym}(D_i)\Pi^{-\frac12}$. 
Compute $X_t$ from $M_1,\ldots,M_{i-1}$ using step (3) of \autoref{alg:mmw-expander-flow}.
         \item Let $v_1,\ldots,v_n$ be the Gram decomposition of $Y_t:= \Pi^{-\frac12}X_t\Pi^{-\frac12}$ as in step (1) of \autoref{alg:mmw-expander-flow}.
         \item If there is a vertex $i$ with $\pi(i) \geq \frac14 \pi(V)$, then output the bipartition $L = \{i\}$ and $R = V\backslash \{i\}$. 
         \item Otherwise, let $u\sim\mathcal{N}(0,I)$ be a random vector. 
          Let $y = \text{median}\big(\{\inner{u}{v_i}:i\in V\}\big)$ where the median is with respect to $\pi$. Output the bipartition $L = \{i:\inner{u}{v_i}\leq y\}$ and $R = \overline{L}$.
     \end{enumerate}
\end{algorithm}

For general $\pi$-weighted directed edge expansion, the requirement of the matching player is to output a directed fractional perfect matching defined as follows.

\begin{algorithm}
\caption{Matching Player Requirement}\label{alg:matching-player}
Given a bipartition $L,R$ from the cut player, the matching player must play a directed fractional perfect matching $D$, which is defined as a weighted Eulerian subgraph where each $i\in L$ has indegree and outdegree exactly $\pi(R) \pi(i)/\pi(L)$ and each $j\in R$ has indegree and outdegree exactly $\pi(j)$. 
\end{algorithm}

Note that when $\pi$ is uniform, then the cut player will always return a bisection, and the matching player will always return a directed (fractional) perfect matching where each vertex has indegree and outdegree one, and so this is a proper generalization of Louis' cut-matching game.


The plan is to analyze the cut-player strategy using the regret bound in \autoref{thm:regret-bound} as follows.
\begin{align*}
    \vec{\phi}_{\pi}(\overline{D}) 
    \geq \lambda_2\Big(\Pi^{-\frac12} L_{\rm sym}(\overline{D}) \Pi^{-\frac12}\Big) 
    \geq \frac1T\sum_{t=0}^T\inner{L_{\rm sym}(D_t)}{Y_t} \cdot (1-\eta) - \frac{\rho\log{n}}{\eta T} 
    \gtrsim \frac{1}{\log n},
\end{align*}
where $\overline{D} := \frac{1}{T} \sum_{t=1}^T D_t$.
The first inequality is by the easy direction of $\primal$ in \autoref{prop:arv-easy-direction}, the second inequality is by the regret bound in \autoref{thm:regret-bound}, and the third inequality is what we would like to achieve in the following.

The key quantity that we would like to lower bound is $W_t := \inner{L_{\rm sym}(D_t)}{Y_t}$, which is a random variable with respect to the filtration $\F_t$ that is what happened up to round $t$ of the algorithm.
At each round $t$, we would like to lower bound $\E_t[W_t] $ where $ \E_t[\cdot] = \E[\cdot|\Fil_{t-1}]$. To lower bound $\E_t[W_t]$, we will use the following basic property of Gaussians.

\begin{fact}[Gaussian Concentration]
\label{fact:gaussian-concentration}
    Let $X$ be a Gaussian random variable with mean $\mu$ and variance $\sigma^2$. Then,
    \begin{align*}
        \Pr[|X-\mu| > t\sigma ] \leq 2\exp(-t^2/2).
    \end{align*}
\end{fact}

\begin{claim}[Expectation] \label{claim:expectation-Wt}
$\E_t[W_t] \gtrsim \frac{1}{\log{n}}$ for any $t$ in step (4) of \autoref{alg:cut-player}. 
\end{claim}

\begin{proof}
The proof is based on the fact that, with high probability, a random Gaussian projection $v_i\mapsto \inner{u}{v_i}$ will preserve the squared distances between all the vectors within a factor of $\log{n}$. 
Let $x\in \mathbb{R}^n$ be a random vector defined by $x(i) = \inner{v_i}{u}$ where $u \sim \mathcal{N}(0,I)$. 
Then, $\E_t[|x(i) - x(j)|^2] = \E_t|\inner{v_i-v_j}{u}|^2 = \norm{v_i - v_j}^2$,
and by \autoref{fact:gaussian-concentration},
\begin{align*}
     \Pr\Big[\big|x(i)-x(j)\big|^2 \gtrsim \log{n} \cdot \norm{v_i-v_j}^2\Big] \leq n^{-3}.
\end{align*}
Let ${\cal E}$ be the event that $|x(i)-x(j)|^2 \lesssim \log{n} \cdot \norm{v_i-v_j}^2$ for all pairs $i,j\in V$. 
By union bound, $\Pr[{\cal E}]\geq 1-n^{-1}$. 
Therefore,
\begin{align*}
    \E_t[W_t] &\geq \E_t\big[\inner{L_{\rm sym}(D_t)}{Y_t}~\big|~{\cal E}\big] \cdot \Pr[{\cal E}]\\
    &\geq \E_t\Bigg[\sum_{i\in L,j\in R}\frac{1}{2} \big(D(i,j) + D(j,i)\big) \cdot \norm{v_i-v_j}^2~\bigg|~{\cal E} \Bigg] \cdot (1 - n^{-1})\\
    &\gtrsim \frac{1}{\log{n}} \cdot \E_t\Bigg[\sum_{i\in L,j\in R} \big(D(i,j) + D(j,i)\big) \cdot |x(j)-x(j)|^2 \Bigg].
\end{align*}
Let $y$ be the median in step (4) such that $x(i)\leq y \leq x(j)$ for all $i \in L$ and $j \in R$. 
Let $r = \pi(R)/\pi(L)$. 
Since $\max_i\pi(i) \leq \frac14 \pi(V)$ in step (4), it follows that $\frac12 \pi(V) \leq \pi(L)\leq \frac34 \pi(V)$ and thus $\frac13 \leq r \leq 1$. 
Continuing,
\begin{align*}
    & \E_t\Bigg[\sum_{i\in L,j\in R}\big(D(i,j) + D(j,i)\big) \cdot |x(j)-x(j)|^2 \Bigg]  \\
    &\geq \E_t\Bigg[\sum_{i\in L,j\in R}\big(D(i,j) + D(j,i)\big) \cdot \Big( \big(x(i)-y\big)^2 + \big(x(j)-y\big)^2 \Big) \Bigg]  
    \\
    &= 2 \cdot \E_t\Bigg[\sum_{i\in L} r \cdot \pi(i) \cdot \big(x(i)-y\big)^2 + \sum_{j \in R} \pi(j) \cdot \big(x(j)-y\big)^2 \Bigg] \\
    &\geq \frac{2}{3} \cdot \E_t \Bigg[ \sum_i\pi(i) \cdot \big(x(i)^2 - 2y \cdot x(i)\big) \Bigg]\\
    &= \frac{2}{3}\sum_i\pi(i) \cdot \norm{v_i}^2  = \frac{2}{3},
\end{align*}
where the first equality is because $\sum_{j \in R} D(i,j) = r \cdot \pi(i)$ for $i \in L$ and $\sum_{i \in L} D(i,j) = \pi(j)$ by the matching player requirement, the last inequality is because $r \geq \frac13$ in step (4) of \autoref{alg:cut-player},
the second last equality is because $\sum_{i}\pi(i) \cdot x(i) = \inner{u}{\sum_i\pi(i) \cdot v_i} = 0$ and $\E_t[x(i)^2] = \E_t[\inner{u}{v_i}^2] = \E_t[v_i^T(uu^T)v_i] = \norm{v_i}^2$ as $u \sim {\mathcal N}(0,I)$,
and the last equality is because $1=\tr(X_t)=\tr(\Pi^{\frac12}Y_t\Pi^{\frac12})=\sum_{i} \pi(i) \cdot \norm{v_i}^2$.
\end{proof}

To show that with good probability, $\sum_tW_t$ does not deviate much from its conditional expectation, we will apply Azuma's inequality, which we state as follows: 

\begin{theorem}[Azuma's Inequality]
  \label{thm:azuma-ineq}
   Let $X_0, \dots, X_T$ be a Martingale such that $|X_t-X_{t-1}|\leq c_t$ $\forall t\in [T]$. Then we have 
   \[
   \Pr[|X_T-X_0|\geq \delta]
   \leq
   \exp \left( -\frac{\delta^2}{2\sum_{t=1}^Tc_t^2}
   \right).
  \]
\end{theorem}

\begin{claim}[Concentration] \label{claim:mart-concentration}
In step (4) of \autoref{alg:cut-player}, for any constant $\delta > 0$, 
\[\Pr\Bigg[ \sum_{t=0}^T W_t\geq \sum_{t=0}^T \E_t[W_t] - \frac{\delta \cdot T}{\log{n}} \Bigg] \geq 1-\exp\bigg(-\Omega\bigg(\frac{\delta^2 \cdot T}{\log^2{n}}\bigg)\bigg).
\]
\end{claim}
\begin{proof}
Let $Z_t = \sum_{i=1}^t (W_i - \E_i[W_i])$. 
Then $Z_t$ is a martingale with respect to the filtration $\Fil_t$ with $Z_0 = 0$. 
Moreover,
\[|Z_t - Z_{t-1}| 
= \big|W_t - \E_t[W_t]\big| 
\leq 2|\inner{L_{\rm sym}(D_t)}{Y_t}| \leq 2\inner{\Pi}{Y_t} = 2,
\]
where the last inequality is because $L_{\rm sym}(D_t)$ is a Laplacian where the degree of vertex $i$ is at most $\pi(i)$ in step (4) of \autoref{alg:cut-player} and thus $L_{\rm sym}(D_t) \preceq \Pi$, and the last equality is because $1 = \tr(X_t) = \inner{\Pi}{Y_t}$. Using \autoref{thm:azuma-ineq}, we can bound our Martingale $Z_t$ as follows:
\begin{align*}
    \Pr\Bigg[ |Z_T| > \delta \cdot \frac{T}{\log{n}} \Bigg] 
\leq \exp\bigg(-\Omega\bigg(\frac{\delta^2T}{\log^2{n}}\bigg)\bigg).
\end{align*}
Note that $|Z_T| \leq \delta T/\log{n}$ implies that $\sum_{t}W_t \geq \sum_t \E_t[W_t] - \frac{\delta \cdot T}{\log{n}}$.
\end{proof}

We are ready to prove \autoref{thm:cut-matching-game} with these claims.

\subsubsection{Proof of \autoref{thm:cut-matching-game}}

Apply the regret bound in \autoref{thm:regret-bound} with feedback matrices $M_i = \Pi^{-\frac12} L_{\rm sym} \Pi^{-\frac12}$ and let $\overline{D} := \frac{1}{T} \sum_{t=1}^T D_t$, it follows that
\begin{equation} \label{eqn:plan}
    \vec{\phi}_{\pi}(\overline{D}) 
    \geq \lambda_2\Big(\Pi^{-\frac12} L_{\rm sym}(\overline{D}) \Pi^{-\frac12}\Big) 
    \geq \frac1T\sum_{t=0}^T\inner{L_{\rm sym}(D_t)}{Y_t} \cdot (1-\eta) - \frac{\rho\log{n}}{\eta T}, 
\end{equation}
where the first inequality is by the easy direction of the SDP rounding in \autoref{prop:arv-easy-direction}.
The main step is to lower bound $\sum_t W_t = \sum_t \inner{L_{\rm sym}(D_t)}{Y_t}$.

First, we consider the special case in step (3) of \autoref{alg:cut-player}, when there is a vertex $i$ with $\pi(i) \geq \frac14 \pi(V)$.
If this holds then we are in the large core case of \autoref{prop:embedding-cases}. 
We can apply the same argument as \eqref{eqn:large-core-case} in \autoref{lem:large-core-case} to show that $\inner{L_{sym}(D_t)}{Y_t}\geq \Omega(1)$ deterministically. 
Also, as $\pi(R)/\pi(L) \leq 4$, each vertex $i$ has degree at most $4\pi(i)$, 
and thus $M_t = \Pi^{-\frac12}L_{\rm sym}(D_i)\Pi^{-\frac12} \preceq 4I$. 

Otherwise, by \autoref{claim:expectation-Wt} and \autoref{claim:mart-concentration}, 
\[
\sum_{t=1}^T \inner{L_{\rm sym}(D_t)}{Y_t} \geq \Omega\bigg(\frac{T}{\log{n}}\bigg) - \frac{\delta \cdot T}{\log{n}},
{\rm~with~probability~at~least~}
1-\exp\bigg(-\Omega\bigg(\frac{\delta^2T}{\log^2{n}}\bigg)\bigg).
\]
Therefore, by setting $T = O\Big(\frac{\log^2{n}}{\eta^2\delta^2}\Big)$ where $\eta$ and $\delta$ are small enough constants, we have that $\sum_{t=1}^T \inner{L_{\rm sym}(D_t)}{Y_t} \gtrsim T/\log(n)$ with constant probability.
Also, in this case, $\norm{M_t} \leq 1$ as each vertex $i$ has degree at most $\pi(i)$. 

Therefore, plugging in $\eta=\frac14$ and width bound $\rho=4$ to \eqref{eqn:plan}, we conclude that $\vec{\phi}_\pi(\overline{D}) \gtrsim 1/\log n$, which implies \autoref{thm:cut-matching-game} by multiplying $T$ on both sides.

\subsubsection{Approximating Directed Edge Expansion}

As in~\cite{KRV06,OSVV08,Lou10}, a corollary of the cut-matching game is an approximation algorithm for approximating directed edge expansion.

\begin{algorithm}
\caption{Cut-Matching-Game Approximation Algorithm ($G, \kappa$)}\label{alg:cut-matching-game-approx-alg}
Initiate a cut-matching game where the cut player follows \autoref{alg:cut-player}. Each iteration, do the following:
    \begin{enumerate}
        \item Given the cut $L,R$ returned by the cut player and a congestion value $\kappa$, compute a bidirectional max flow on $(L,R,\kappa, \beta=1)$ using \autoref{alg:bidirectional-max-flow}.
         \item If we obtain a cut $S$, output $S$ and terminate. If we obtain saturating flows in both directions, $\vec{f}$ and $\cev{f}$, construct the demand graph of the circulation $f = \frac{1}{2}(\vec{f} + \cev{f})$ as follows:
             \begin{enumerate} 
                \item  Let $\vec{D}$ be the demand graph for $\vec{f}$. That is, for each $i\in L$ and $j\in R$, if there is a flow path $p\in \vec{f}$, then we add an edge $(i,j)$ with weight $\vec{f}_p$.
                \item Construct $\cev{D}$ from $\cev{f}$ the same way.
                \item The matching player plays $D_t = \frac{1}{2}(\vec{D} + \cev{D})$.
            \end{enumerate}
    \end{enumerate}
\end{algorithm}

Note that this algorithm is essentially a special case of \autoref{alg:mmw-expander-flow}, where we implement the Oracle in a similar manner as in the project max flow algorithm (\autoref{alg:project-max-flow}). 
That is, we project our embedding vectors in a random direction and call bi-directional maxflow with $\beta=1$. Then we either output a cut and terminate or update the embedding with the symmetric Laplacian of a circulation. 

\begin{corollary}
  \label{thm:cut-matching-game-approx}
    Given an edge capacitated directed graph $G = (V, E, w)$ and a parameter $\kappa > 0$, there is an algorithm using the cut-matching game in \autoref{thm:cut-matching-game} such that, in $O(\log^2 n)$ iterations, either builds a directed Eulerian subgraph to certify that $\vec{\phi}_\pi(G)\gtrsim \frac{1}{\kappa} \cdot \frac{1}{\log n}$ or outputs a cut $S$ with $\vec{\phi}_\pi(S)\lesssim \frac{1}{\kappa}$. 
Furthermore, each iteration can be computed using $O(1)$ single-commodity flows.
\end{corollary}
\begin{proof}
     If at any point, the algorithm outputs a cut during step 2, then by \autoref{lemma:flow-cut-lemma}, we find a cut of directed edge expansion at most $O(1/\kappa)$ as $\beta= 1$.  
    On the other hand, if for some $T = \Theta(\log^2{n})$ iterations, we always find a saturating flow in step 2, then the cut-matching game would have proceeded for $T$ iterations. 
    Then, by \autoref{thm:cut-matching-game}, the average demand graph $\frac{1}{T}\sum_{t=1}^TD_t$ (which is the average matching played by the matching player) has second eigenvalue at least $\Omega(1/\log{n})$ with constant probability. 
    Thus, we have constructed a circulation with congestion at most $\kappa$ whose demand graph $D$ has $\pi$-weighted edge expansion at least $\Omega(1/\log{n})$, and this implies that $\vec{\phi}_{\pi}(G) \gtrsim \frac{1}{\kappa} \cdot \frac{1}{\log n}$.
\end{proof}

\section{Other Generalizations} \label{sec:others}

As mentioned in the introduction, the reweighted eigenvalue framework captures also vertex expansion and hypergraph edge expansion.
In each case, the framework produces an SDP for which a rounding algorithm with Cheeger-type guarantee exists; see \cite{LTW23}.
In this section, we show that, analogous to the case of directed edge expansion, by adding $\ell_2^2$ triangle inequality constraints to the SDPs for vertex expansion and hypergraph edge expansion, we obtain tighter relaxations which have an integrality gap of $O(\sqrt{\log n})$ to the respective expansion quantities.
Moreover, there is an almost linear-time rounding algorithm for each of these semidefinite programs.


\subsection{Directed Vertex Expansion}
\label{sec:generalization-vertex}
A vertex-capacitated directed graph $G = (V, E, \pi)$ is a graph equipped with a vertex weight/capacity function $\pi: V \rightarrow \R_+$.
Given such a graph, let $S \subset V$ be a nonempty subset of vertices.
The set of out-neighbors of $S$ is defined as $\partial^+(S) := \{v \notin S \mid \exists u \in S \text{ with } uv \in E\}$,
and the directed vertex expansion $\vertexexp(S)$ and $\vertexexp(G)$ are defined as
\[
  \vertexexp(S) := \frac
  {\min\big\{ \pi\big(\partial^+(S)\big), \pi\big(\partial^+(\overline{S})\big) \big\}}
  {\min\big\{\pi(S), \pi(\overline{S})\big\}}
  \quad \text{ and } \quad
  \vertexexp(G) := \min_{\emptyset \neq S \subset V} \vertexexp(S).
\]

Note that these definitions capture undirected vertex expansion as a special case.
Also, let
\[
\mathcal{F}_v(G) := 
\bigg\{
F: E\rightarrow \mathbb{R}_{\ge 0}~
\bigg|~
\sum_{j: ij \in E}F(i,j) = \sum_{j: ji \in E}F(j,i)\leq \pi(i)
\;\; \forall i \in V
\bigg\}
\]

denote the set of feasible vertex-capacitated circulations on $G$.

By adding $\ell_2^2$ triangle inequality constraints to the embedding in \cite[Proposition 3.3]{LTW23},
we arrive at the following program for directed vertex expansion.

\begin{definition}[Vertex Reweighted Eigenvalue with Triangle Inequalities]
  Given a vertex-capacitated directed graph $G = (V, E, \pi)$. The $\lambda_\pi^{\Delta_v}(G)$ program for directed vertex expansion is
  \begin{align*}
  \lambda_\pi^{\Delta_v}(G) :=
     \min_{v_1, \dots, v_n \in \mathbb{R}^n} \max_{F \in \mathcal{F}_v(G)} &~~~ \frac12 \sum_{ij \in E} F(i, j) \cdot \norm{v_i - v_j}^2
    \\
    \st&~~~
    \sum_{i \in V} \pi(i) \cdot v_i = \vec{0}
    \\
    &~~~ \sum_{i \in V} \pi(i) \cdot \norm{v_i}^2 = 1
    \\
    &~~~ \norm{v_i - v_j}^2 + \norm{v_j - v_k}^2 \ge \norm{v_i - v_k}^2 \quad \forall i, j, k \in V.
  \end{align*}
\end{definition}

Note that this is almost identical to $\lambda_\pi^{\Delta}$ for edge expansion in \autoref{def:main-sdp}.
The only difference here is that $F$ is constrained by vertex capacity constraints instead of edge capacity constraints.
An analogous proof to \autoref{prop:arv-easy-direction} shows that $\lambda_\pi^{\Delta_v}$ is indeed a relaxation of $\vertexexp$, and
by combining \autoref{lemma:arv-structure-weighted} with a version of \autoref{lemma:metric-rounding-lemma} for vertex expansion (see \autoref{lemma:metric-rounding-lemma-v} below), we can bound the integrality gap of this SDP relaxation.

\begin{theorem}[Integrality Gap for Vertex Expansion]\label{thm: vertex-arv}
    Let $G$ be a directed graph with vertex weights $\pi: V\rightarrow \mathbb{R}^+$. Then,
    \begin{align*}
        \lambda^{\Delta_v}_\pi(G) \lesssim \vec{\psi}_\pi(G)\lesssim \sqrt{\log{n}}\cdot \lambda^{\Delta_v}_\pi(G).
    \end{align*}
\end{theorem}

In \autoref{sec:dual-rounding}, we gave fast algorithms for approximating directed edge expansion using matrix multiplicative weight update and explander flows. These techniques can be easily adapted to approximating vertex expansion by changing edge-capacitated flows to vertex-capacitated flows.
Moreover, it can be shown that the dual of the $\lambda_\pi^{\Delta_v}(G)$ SDP can be interpreted as finding the best vertex-capacitated expander flow to certify that $G$ has large directed vertex expansion. We can thus obtain the following vertex analogous of \autoref{thm:directed-Sherman}, \autoref{thm:cut-matching-game}, and \autoref{prop:reweighted-phi}.

\begin{theorem}[Fast $O(\sqrt{\log n})$ Approximation to $\vertexexp(G)$]
  \label{thm:directed-Sherman-v}
  For small enough $\epsilon > 0$, there is a randomized algorithm that, given any vertex-capacitated directed graph $G = (V, E, \pi)$,
  uses $\tilde{O}(n^{3 \eps})$ directed max-flow computations to compute a cut $S \subseteq V$, such that $\vertexexp(S) \lesssim \sqrt{\frac{\log n}{\epsilon}}\cdot \lambda_\pi^{\Delta_v}(G)$ with constant probability.
\end{theorem}

\begin{theorem}[Cut Matching Game for Directed Vertex Expansion]
  \label{thm:cut-matching-game-v}
  In the cut-matching game for directed graphs, there is a cut player strategy so that, in $O(\log^2{n})$ iterations, the union of the matchings played by the matching player is an Eulerian graph with vertex expansion $\Omega(\log n)$.
\end{theorem}

\begin{proposition}[Dual Certificate for Vertex Expansion]\label{prop:reweighted-psi} Given a graph $G$ with vertex weights $\pi:V\rightarrow\mathbb{R}_+$, there exists a feasible circulation $F\in \mathcal{F}_v(G)$ such that:
    \begin{align*}
        \phi_\pi(F) \gtrsim \frac{\vec{\psi}_\pi(G)}{\sqrt{\log{n}}}.
    \end{align*}
\end{proposition}

\begin{remark}[Undirected Vertex Expansion]
  We remark that our definition of directed vertex expansion also captures undirected vertex expansion, and that all the results presented above apply to the undirected case.
  Note that the vector program presented in \cite[Section 2.3]{FHL08} can also be rounded to give an $O(\sqrt{\log n})$ approximation of vertex expansion (vertex expansion defined here is within a constant factor of ``minimum ratio vertex cut'' in their paper and is different from the ``vertex expansion'' in their appendix).
  Our program has the advantages that it admits a fast primal-dual rounding algorithm and that it has a considerably simpler form.
\end{remark}

While it is possible to prove \autoref{thm:directed-Sherman-v}, \autoref{thm:cut-matching-game-v}, and \autoref{prop:reweighted-psi} directly by analyzing the $\lambda^{\Delta_v}_\pi$ program, a simpler way to obtain these results is by a reduction from vertex expansion to edge expansion.

\begin{proposition}[Reduction from Directed Vertex Expansion to Directed Edge Expansion]
\label{prop:ve-to-ee}
Let $G = (V, E, \pi)$ be a vertex-capacitated directed graph. Then, there exists an edge-capacitated directed graph $G' = (V', E', w)$ over vertex weights $\pi'$, such that $|V'| = O(|V|)$, $|E'| = O(|V| + |E|)$, and $\edgeexp(G') \sim \vertexexp(G)$.
Moreover, such graph $G'$ can be constructed in linear time, and given any $\emptyset \neq S' \subset V$ we can compute in linear time an $\emptyset \neq S \subset V$ such that $\vertexexp(S) \lesssim \edgeexp(S')$.
\end{proposition}

For brevity, we describe the somewhat standard reduction and leave the verification to the reader.
For each vertex $i$ in $G$, create two copies $i_{in}$ and $i_{out}$ in $G'$, where $\pi'(i_{in}) = \pi(i)$ and $\pi'(i_{out}) = \delta$ for a small positive $\delta \ll \min_i \pi(i)$, then draw an edge from $i_{in}$ to $i_{out}$ with edge weight $\pi(i)$. For each edge $e = ij \in E$, draw an edge from $i_{out}$ to $j_{in}$ with edge weight $M \gg \sum_i \pi(i)$.
Observe that this reduction can be computed in $O(n + m)$ time, and after that we simply apply 
this reduction to obtain \autoref{thm:directed-Sherman-v}, \autoref{thm:cut-matching-game-v}, and \autoref{prop:reweighted-psi} from their edge expansion counterparts.

\subsubsection{Cheeger Rounding for Vertex Expansion}

In this section, we will prove the following generalization of \autoref{thm:fast-Cheeger} for vertex expansion.

\begin{theorem}[Fast Cheeger-type Rounding for Vertex Expansion]\label{thm:fast-cheeger-v}
    Given an graph $G$ with vertex capacities $\pi:V\rightarrow \mathbb{R}_+$, there is an almost-linear time algorithm to obtain a set $S$ such that $\vertexexp(G)\lesssim \sqrt{\vertexexp(S)\cdot \log{\frac{d_{\max}}{\vertexexp(G)}}}$, where $d_{\max}$ is the maximum unweighted degree of $G$.
\end{theorem}
While the previous theorems and proposition can be proved via reduction to $\edgeexp$, the same is not true of \autoref{thm:fast-cheeger-v} because \autoref{thm:fast-Cheeger} only holds for the special case of directed edge conductance rather than the general $\edgeexp$. Instead, we need a version of \autoref{lemma:metric-rounding-lemma} for vertex expansion.

\begin{lemma}[Metric Rounding Lemma for Vertex Expansion]
    \label{lemma:metric-rounding-lemma-v}
    Given a graph $G=(V,E)$,
    let $d(\cdot, \cdot)$ be a metric on $V$, and let
    $\pi: V \rightarrow \mathbb{R}^+$ be an arbitrary weight function over $V$.
    Suppose we are given disjoint vertex subsets $L, R\subseteq V$ as input to the algorithm. 
    Let $r := \pi(R) / \pi(L)$ and $r' := \max\{1,r\}$. Then there is an algorithm using $O(\log n)$ maximum flow computations to output a set $S$ with
    \begin{align*}
        \vertexexp(S) &
        \lesssim
        \frac{
          r' \cdot \max_{F\in \mathcal{F}_v(G)} \sum_{i, j \in V} F(i, j) \cdot d(i, j)
        }{\sum_{i \in R} \pi(i) \cdot d(i, L)}.
    \end{align*}
\end{lemma}

\begin{lemma}[Unsaturated Case, Vertex Version]
    \label{lemma:flow-cut-lemma-v}
    Suppose \autoref{alg:bidirectional-max-flow} with vertex capacities outputs a cut $S$. Then $\vec{\psi}_\pi(S) \leq (\kappa/\beta r' - 1)^{-1}$, where $r' := \max\{1, r\}$.
\end{lemma}

Given these two modifications, the proof of \autoref{lemma:metric-rounding-lemma-v} follows by combining \autoref{lemma:flow-cut-lemma-v} and \autoref{lemma:metric-rounding-saturated} as in the proof of \autoref{lemma:metric-rounding-lemma}.
The rest of the proof of \autoref{thm:fast-cheeger-v} is analogous to that for \autoref{thm:fast-Cheeger} in \autoref{sec:fast-Cheeger}. First, we apply \autoref{thm:directed-Sherman-v}, and if we determine through this algorithm that $\vertexexp(G)$ is small, then the $O(\sqrt{\log{n}})$ approximation dominates the Cheeger bound. If on the other hand, we determine that $\vertexexp(G) = \Omega(\frac{1}{\log^{1.5}{n}})$, then we solve the reweighted eigenvalue program for directed vertex expansion as defined in \cite[Definition 1.2]{LTW23} in $O(\log^{1.5}n)$ iterations of matrix multiplicative weight update method. Then, we apply Cheeger rounding to find a set $S$ in such that $\vertexexp(S)\lesssim\sqrt{\vertexexp(G)\log{\frac{d_{\max}}{\vertexexp(G)}}}$ as guaranteed by \cite{LTW23}. To round the $\ell_1$ program, we use flow rounding by applying \autoref{lemma:metric-rounding-lemma-v} to attain an almost-linear runtime.

\subsection{Directed Hypergraph Expansion}

An edge-capacitated directed hypergraph $H = (V, E, w)$ consists of a set $E$ of weighted directed hyperedges over vertex set $V$.
For each hyperedge $e \in E$, $e = (H_e, T_e)$ where $H_e, T_e\subseteq V$ are the head sets and tail sets in $e$ respectively, and $w(e)$ is its weight.
Given such a hypergraph over vertex weights $\pi: V \rightarrow \R_+$, let $S \subset V$ be a nonempty subset of vertices.
The set of out-neighbours of $S$ is defined as $\delta^+(S) := \{e \in E: T_e \cap S \neq \emptyset \text{ and } H_e \cap \overline{S} \neq \emptyset\}$, and the directed hypergraph expansion $\vec{\phi}_{\pi}(S)$ and $\vec{\phi}_{\pi}(H)$ are defined as
\[
  \vec{\phi}_{\pi}(S) := \frac{\min\big\{w(\delta^+(S)), w(\delta^+(\overline{S}))\big\}}{\min\big\{\pi(S), \pi(\overline{S})\big\}}
  \quad \text{ and } \quad
  \vec{\phi}_{\pi}(H) := \min_{\emptyset \neq S \subset V}\vec{\phi}_{\pi}(S).
\]

Note that this captures expansion in undirected hypergraphs by taking $H_e = T_e$ for each $e \in E$, and also directed expansion in ordinary graphs by constraining $|H_e| = |T_e| = 1$.

We again derive our SDP by adding $\ell_2^2$ triangle inequalities to the reweighted eigenvalue program for directed hypergraphs.
Although the program is not readily available in \cite{LTW23}, its derivation follows the same idea of reducing to the simple case of undirected edge expansion in ordinary graphs, via edge-constrained circulations on the clique graph.

\begin{definition}[Directed Hypergraph Reweighted Eigenvalue with Triangle Inequalities]
  \label{def:main-sdp-h}
Given an edge-capacitated directed hypergraph $H = (V, E, w)$ over vertex weights $\pi: V \rightarrow \R_+$.
Let
\begin{align*}
  \mathcal{F}(H) := \Big\{ F: V \times V \rightarrow \R_{\ge 0}~\Big\vert~
  &
  \exists \{F_e: H_e \times T_e \rightarrow \R_{\ge 0}\}_{e \in E} \text{ s.t. } F(i, j) = \sum_{e: i \in H_e, j \in T_e} F_e(i, j),
  \\
  &
  \sum_{i \in H_e, j \in T_e} F_e(i, j) \le w(e) \; \forall e \in E,
  \\
  & \sum_{i\in V} F(i,j) = \sum_{k \in V} F(j,k) \;\forall j \in V
  \Big\}
\end{align*}

be the set of feasible edge-constrained circulations on $H$.
The $\lambda_{\pi}^{\Delta}(H)$ program for directed hypergraph expansion is
\begin{align*}
    \lambda_{\pi}^{\Delta}(H) :=
     \min_{v_1, \dots, v_n \in \mathbb{R}^n}
     \max_{F \in \mathcal{F}_h(H)}
     &~~~
     \sum_{i<j} \frac12(F(i,j) + F(j,i))\norm{v_i - v_j}^2
     \\
     \st &~~~
     \sum_{i \in V} \pi(i) \cdot v_i = \vec{0}
     \\
     &~~~
     \sum_{i \in V} \pi(i) \cdot \norm{v_i}^2 = 1
     \\
     &~~~
     \norm{v_i - v_j}^2 + \norm{v_j - v_k}^2 \geq \norm{v_i - v_k}^2 \quad \forall i, j, k \in V.
\end{align*}

\end{definition}

The intuition for defining feasible edge-constrained circulations on directed hypergraphs this way is that they correspond to Eulerian reweightings of an underlying ``clique graph'' $K_H$ of the directed hypergraph $H$, where for each edge $(H_e, T_e)$, we add an arc $ij$ from every $i \in H_e$ to $j \in T_e$.
The definition $\primal(H)$ is a natural one for various reasons.
First, it can be shown that $\primal(H)$ is a relaxation of $\edgeexp(H)$.
Second, when $H$ is an undirected hypergraph and $\pi$ is the total weighted degree, i.e.~$H_e = T_e,\;\forall e\in E$, and $\pi(i) = \sum_{e\ni i}w(e)$, then $\primal(H)$ is exactly the reweighted eigenvalue program for undirected hypergraphs as defined in \cite[Section 5.1]{LTW23} but with $\ell_2^2$ triangle inequalities. Third,
just as our program for directed graphs is a relaxation of the SDP in \cite{ACMM05}, this program is a relaxation of the SDP in \cite{CS18}.
Note also that $\lambda_\pi^{\Delta}$ for ordinary graphs may be considered a special case of \autoref{def:main-sdp-h}.

Again, our main results for edge expansion extend to hypergraph expansion.

\begin{theorem}[Integrality Gap for Hypergraph Expansion]\label{thm: hypergraph-arv}
    Let $H = (V, E, w)$ be an edge-capacitated directed hypergraph with vertex weights $\pi: V\rightarrow \mathbb{R}^+$. Then,
    \begin{align*}
        \lambda^{\Delta}_\pi(H) \lesssim \edgeexp(H)\lesssim \sqrt{\log{n}}\cdot \lambda^{\Delta}_\pi(H)
    \end{align*}
\end{theorem}

\begin{theorem}[Fast $O(\sqrt{\log n})$ Approximation to $\edgeexp(H)$]
  \label{thm:directed-Sherman-h}
  For small enough $\epsilon > 0$, there is a randomized algorithm that, given any edge-capacitated directed hypergraph $H = (V, E, w)$ over vertex measure $\pi: V \rightarrow \R_+$,
  uses $\tilde{O}(n^{3 \eps})$ directed max-flow computations to compute a cut $S \subseteq V$, such that $\edgeexp(S) \lesssim \sqrt{\frac{\log n}{\epsilon}}\cdot \lambda_\pi^{\Delta}(H)$ with constant probability.
\end{theorem}

\begin{proposition}[Dual Certificate for Hypergraph Expansion]
\label{prop:reweighted-phi-h}
Given a hypergraph $H = (V, E, w)$ with vertex weights $\pi:V\rightarrow\mathbb{R}_+$, there exists a feasible circulation $F \in \mathcal{F}_h(H)$ such that
    \begin{align*}
        \phi_\pi(F) \gtrsim \frac{\edgeexp(H)}{\sqrt{\log{n}}}.
    \end{align*}
\end{proposition}

We can also define a cut-matching game for directed hypergraphs, where the matching player is required to return an Eulerian subgraph of the clique graph $K_H$ satisfying the indegree and outdegree constraints as in \autoref{alg:matching-player}.

\begin{theorem}[Cut Matching Game for Directed Hypergraph Expansion]
  \label{thm:cut-matching-game-h}
  In the cut-matching game for directed hypergraphs, there is a cut player strategy so that, in $O(\log^2{n})$ iterations, the union of the matchings played by the matching player is a feasible circulation on $H$ with hypergraph expansion $\Omega(\log n)$.
\end{theorem}

The key for obtaining these results is to relate hypergraph expansion of $H$ to the edge expansion of an ordinary derived graph $G_H$ as in \cite{CS18}, which we present here for completeness.

\begin{definition}[Derived Graph of Directed Hypergraphs {\cite[Fact 1.1]{CS18}}]
  Let $H = (V, E, w)$ be an edge-capacitated directed hypergraph over vertex weights $\pi: V \rightarrow \R_+$.
  The derived graph $G_H = (V', E', w')$ over vertex weights $\pi': V' \rightarrow \R_+$ is defined as follows:
  \begin{itemize}
    \item $V' := V \cup \{i_{e}^{in}: e \in E\} \cup \{i_{e}^{out}: e \in E\}$,
    \item $E' := \{(j, i_e^{in}): j \in H_e, e \in E\} \cup \{(i_e^{in}, i_e^{out}): e \in E\} \cup \{(i_e^{out}, k): k \in T_e, e \in E\}$,
    \item $w'(j, i_{e}^{in}) = w'(i_e^{out}, k) = \infty$ and $w'(i_e^{in}, i_e^{out}) = w(e)$
    for all $e \in E$, $(j, k) \in H_e \times T_e$,
    \item $\pi'(i) = \pi(i)$ for all $i \in V$, and $\pi'(i_e^{in}) = \pi'(i_e^{out}) = 0$ for all $e \in E$.
  \end{itemize}
\end{definition}

From \cite[Fact 1]{CS18}, there is a correspondence between subsets $S \subseteq V$ and $S' \subseteq V'$ so that $\edgeexp(S) \sim \vec{\phi}_{\pi'}(S')$. Therefore, if we perform a black-box reduction from $\edgeexp(H)$ to $\vec{\phi}_{\pi'}(G_H)$, we obtain \autoref{thm:directed-Sherman-h}, \autoref{thm:cut-matching-game-h}, and \autoref{prop:reweighted-phi-h}, although the approximation guarantees using this approach degrade to $O(\sqrt{\log (n+m)})$ or $O(\log (n+m))$ (since $|V'| = \Theta(n+m)$), which are worse when $m = \omega(\poly(n))$.

To obtain these results in full, one needs to derive hypergraph analogues of \autoref{lemma:metric-rounding-lemma}, and of the algorithms in \autoref{sec:dual-rounding} and \autoref{sec:directed-cut-matching-game}.
To this end, the key modification is to replace the bidirectional max-flow algorithm in \autoref{alg:bidirectional-max-flow} by its hypergraph counterpart, and we may leave the other components essentially unchanged.
We will need to define flows on hypergraphs $H$ and obtain a hypergraph version of max-flow min-cut theorem,
and this is achieved by considering flows and cuts on the derived graph $G_H$:
\begin{enumerate}
    \item Given $L,R$, a partition of $V$, we add vertices $\{s,t\}$ to $G_H$ with $s$ connected to $L$ and $t$ connected to $R$ as in \autoref{alg:bidirectional-max-flow} and compute maximum $s$-$t$ and $t$-$s$ flows. 
    \item Each flow path is of the form $(s, j_1, i_{e_1}^{in}, i_{e_1}^{out}, j_2, i_{e_2}^{in}, i_{e_2}^{out}, \dots, j_\ell, t)$, where $(j_t, j_{t+1}) \in H_{e_t} \times T_{e_t}$ for all $1 \le t \le \ell - 1$.
    It corresponds to the flow path $(s, j_1, j_2, \dots, j_\ell, t)$ in the respective $s$-$t$ flow problem in the clique graph $K_H$.
    One can then check that bidirectional flows on $G_H$ correspond to Eulerian reweightings on $K_H$, i.e.~feasible circulations on $H$.
    \item The max-flow min-cut theorem yields an $s$-$t$ cut in $G_H$. Since $w'(j, i_e^{in})$ and $w'(i_e^{out}, k)$ are large, the cut edges will only be in one of the following types: $(s, j)$, $(j', t)$, or $(i_e^{in}, i_e^{out})$ (where $j, j' \in V$). Thus, we derive a hypergraph version of \autoref{lemma:flow-cut-lemma}, whose proof follows closely that of the original version.
    Consequently, we obtain a hypergraph version of \autoref{lemma:metric-rounding-lemma}.
\end{enumerate}
Thus, the overall idea for generalizing our arguments for directed graphs to directed hypergraphs is to use the derived graph $G_H$ to compute bi-directional flows. Then we can either find a directed sparse cut or many feasible circulations in $\mathcal{F}(H)$, whose average can be used to certify that $\edgeexp(H)$ is large through the matrix multiplicative weight update method.

Finally, we will give the following generalization of \autoref{thm:fast-Cheeger} to \textit{undirected} hypergraph conductance, improving on the runtime of the algorithm in \cite[Section 5]{LTW23}.
Recall that for undirected hypergraphs, $H_e = T_e$ for all $e \in E$.

\begin{theorem}[Hypergraph Fast Cheeger-type Rounding]\label{thm:fast-cheeger-h}
    Given an edge-capacitated undirected hypergraph $H = (V, E, w)$ with vertex weights $\pi(i) = \sum_{e: i \in H_e} w(e)$, there is an almost-linear time algorithm to obtain a set $S \subseteq V$ such that 
    $\phi_\pi(S)\lesssim \sqrt{\phi_\pi(G)\cdot \log{r}}$, where $r := \max_e |H_e|$ is the maximum edge size of $H$.
\end{theorem}
The proof of \autoref{thm:fast-cheeger-h} is analogous to that for vertex expansion, except that this time, it suffices to use threshold rounding as in \cite{LTW23}, which can be done in linear time.
 
We remark that by using the directed hypergraph metric rounding lemma outlined above, as well as a version of \autoref{lemma:flow-cut-lemma} for directed hypergraphs, a fast Cheeger-type rounding algorithm exists for directed hypergraphs, with the guarantee that
\[
  \edgeexp(S) \lesssim \sqrt{\edgeexp(G) \cdot \log \frac{r}{\edgeexp(G)}}.
\]

This would necessitate a Cheeger inequality for directed hypergraphs, which is not available in \cite{LTW23} but follows readily from their technique.

\section{Summary}
\label{sec:summary}
In this paper, we have given a unifying approach for generalizing all the major approximation algorithms for undirected edge expansion
to other settings, including directed edge expansion,
directed vertex expansion and directed hypergraph expansion.
These algorithms may be summarized in a one-sentence formula:
use flows to implement a matrix multiplicative weight update algorithm for solving a reweighted eigenvalue program or playing a cut-matching game.
This formula either recovers or improves all relevant past results.

On the practical side, it is worth noting that the algorithms presented in this paper are almost-linear time.
While we have theoretical guarantee on their runtimes and approximation ratios, we are curious about whether they may be implemented to find good sparse cuts in large graphs quickly.
Such implementation would bring these algorithms into the practical realm; in particular, fast spectral algorithms for computing hypergraph sparse cuts would be useful in certain machine learning applications, and fast algorithms for finding reweightings could be useful in graphical neural networks for hypergraphs and directed graphs.

We believe our approach leaves room for further research into graph partitioning problems. Since multi-way graph partitioning has found many applications in clustering and classification, one interesting open question is to design fast approximation algorithms for multi-way graph partitioning and generalize it to the vertex, directed graph, and hypergraph settings. In \cite{Yos19}, Yoshida recovered Cheeger-type inequalities for partitioning problems on all submodular functions, which is more general than directed hypergraphs. Another open question is to obtain fast approximation algorithms for partitioning problems on more general classes of submodular functions, possibly using flows and reweighted eigenvalues.

\bibliographystyle{plain}

\newpage

\begin{appendix}

\section{Missing Proofs of \autoref{sec:rounding-algorithms}}
\label{app:arv}

\begin{proof}[Proof of \autoref{prop:arv-easy-direction}]
  Let $\emptyset \neq S \subset V$. We construct an SDP solution to show that $\xi(G) \le 2 \edgeexp(S)$.
  Consider the vector solution
  \[
    v_i := \begin{cases}
      (a, 0, \dots, 0), & \text{ if } i \in S, \\
      (b, 0, \dots, 0), & \text{ otherwise},
    \end{cases}
  \]
  where $a, b \in \R$ satisfies $a \cdot \pi(S) + b \cdot \pi(\overline{S}) = 0$ and $a^2 \cdot \pi(S) + b^2 \cdot \pi(\overline{S}) = 1$. Note that such $(a, b)$ must exist.
  A routine check reveals that all the constraints on $v_i$ are satisfied.
  It remains to show that
  \[
    \frac{1}{2} \sum_{ij \in E} F(i, j) \norm{v_i - v_j}^2 \le 2 \edgeexp(S) \qquad \forall F \in \mathcal{F}(G).
  \]

  Solving for $a$ and $b$, we see that $(a-b)^2 = \pi(V) / \pi(S) \pi(\overline{S})$.
  Then,
  \begin{eqnarray*}
    \frac{1}{2} \sum_{ij \in E} F(i, j) \norm{v_i - v_j}^2
    & = &
    \frac{1}{2} \left[ \sum_{i \in S, j \in \overline{S}} + \sum_{i \in \overline{S}, j \in S} \right] F(i, j) (a - b)^2
    \\
    & = &
    \left( F(S, \overline{S}) + F(\overline{S}, S) \right) \cdot \frac{\pi(V)}{2 \pi(S) \pi(\overline{S})}
    \\
    &\le&
    \frac{F(S, \overline{S}) + F(\overline{S}, S)}{\min\{\pi(S), \pi(\overline{S})\}}
    \\
    &\le&
    \frac{2 \min\{\delta^+(S), \delta^+(\overline{S})\}}{\min\{\pi(S), \pi(\overline{S})\}}
    = 2\edgeexp(S),
  \end{eqnarray*}

  where the last inequality uses the fact that $F \in \mathcal{F}(G)$ is an Eulerian reweighting, so that $F(S, \overline{S}) = F(\overline{S}, S) \le \min\{\delta^+(S), \delta^+(\overline{S})\}$.
  This finishes the proof that $\xi(G) \le 2 \edgeexp(G)$.
\end{proof}

\begin{proof}
[Proof of \autoref{lemma:arv-structure-weighted}]
The algorithm for the unweighted version in \autoref{thm:arv-structure} proceeds as follows. Let $\sigma > 0$ be a suitable absolute constant.
First, choose a random direction $u \sim \mathbb{S}^{n-1} \subseteq \R^n$, and order the vertices $i$ by $\inner{v_i}{u}$.
Second, if the median value is $M$, set $L$ to be the set of vertices $i$ such that $\inner{v_i}{u} \ge M + \sigma / \sqrt{n}$, and set $R$ to be the set of vertices $i$ such that $\inner{v_i}{u} < M$.
Third, while there are pairs $(i, j) \in L \times R$ such that $\norm{v_i - v_j}^2 < \Delta = \Theta(1 / \sqrt{\log n})$, remove $i$ from $L$ and $j$ from $R$.
If $|L| \ge \Omega(n)$ and $|R| \ge \Omega(n)$, then the procedure successfully finds two large subsets that are at least $\Delta \ge \Omega(1 / \sqrt{\log n})$ $\ell_2^2$-distance apart.
Refer to \cite{ARV09} for complete details.

To prove the $\pi$-weighted version in \autoref{lemma:arv-structure-weighted}, we do a reduction to the unweighted case.
Recall the assumption that $\pi(V) = 1$. Let $K \in \N$ such that $K \cdot \min_{i \in V} \pi(i) \ge 1/2$, and let $\pi'(i) := \ceil{K \pi(i)}$ for $i \in V$. We may further assume that $\min_{i \in V} \pi(i) \ge \Omega(1 / \poly(n))$ (vertices with smaller measure may be ignored), so that $K \le O(\poly(n))$.
Create $\pi'(i)$ copies of $v_i$ and feed the embedding to the unweighted algorithm. Note that the embedding consists of $\Theta(K)$ vectors.
In the end of the unweighted algorithm, w.h.p. the output sets $L$ and $R$ will have size $\Theta(K)$ each, and they will be at least $\Omega(1/\sqrt{\log n})$ $\ell_2^2$-distance apart.

Note that if one copy of $v_i$ is in either of the output set, we may include all copies of $v_i$ in that output set, without affecting the distance between $L$ and $R$. Then, the $\pi$-measure of vertices in $L$ will be at least
\[
  \sum_{i \in L} \pi(i)
  \ge \sum_{i \in L} \frac{\pi'(i)}{2K}
  = \frac{|L|}{2K} \ge \Omega(1);
\]
same for $R$. We have proved that w.h.p. $\pi(L), \pi(R) \ge \Omega(1)$.
The runtime is polynomial in the number of vectors which is $\Theta(K)$, and hence polynomial in $n$.

To get rid of the $K$-dependence in the runtime, we may modify the unweighted algorithm as follows:
In the second step, compute the weighted median. In the third step, instead of removing both vertices $i$ and $j$, subtract $\min(\pi(i), \pi(j))$ from both $\pi(i)$ and $\pi(j)$, and remove the vertex whose $\pi$-measure drops to zero. \end{proof}

\begin{proof}[Proof of \autoref{lem:large-core-denominator}]
Direct calculation gives 
  \begin{eqnarray*}
    2
    &=&
    \sum_{i, j \in V} \pi(i) \cdot \pi(j) \cdot d(i, j)
    \\
    &\le&
    s^2\sum_{i, j \in V} \pi(i) \cdot \pi(j) \Big[ d(i, L) + {\rm diam}(L) + d(j, L) \Big]
    \\
    &=&
    s^2 \cdot \Big(\pi(V)^2 \cdot {\rm diam}(L) + 2 \pi(V) \cdot \sum_{j \in V} \pi(j) \cdot d(j, L)\Big)
    \\
    &=&
    s^2 \cdot \Big({\rm diam}(L) + 2 \cdot \sum_{j \in R} \pi(j) \cdot d(j, L)\Big),
  \end{eqnarray*}
  where the inequality comes from applying the $s$-relaxed triangle inequality twice and the last equality uses $\pi(V) = 1$. Rearranging gives the desired result.
\end{proof}

\section{Missing proofs of \autoref{sec:dual-rounding}}
\label{app:dual-rounding}

\begin{proof}[Proof of \autoref{prop:embedding-cases}]
    We will make use of \cite[Lemma 5]{Kal07}, which is restated as follows.
    \begin{lemma}[{\cite[Lemma 5]{Kal07}}]
        Given a set of vectors $v_1,\dots v_n$ such that $\sum_{i,j}\norm{v_i-v_j}^2 >4n^2/5$. Then, one of the two cases hold:
        \begin{itemize}
            \item There is a node $i$ such that $\Big|B\Big(i,\frac{1}{2\sqrt{10}}\Big)\Big| > \frac{n}{4}$. 
            \item There is a set of nodes $S\subseteq V$ and an $i_0\in S$ such that $\forall i\in S$, $\norm{v_i -v_{i_0}}^2 = O(1)$ and $\sum_{i,j\in S}\norm{v_i-v_j}^2 = \Omega(n^2)$.
        \end{itemize}
    \end{lemma}
    Note that this immediately implies \autoref{prop:embedding-cases} in the case where $\pi$ is uniform. For general $\pi$, suppose we re-scale $\pi$ so that for each $i$, $\pi(i)$ is an integer. Then, we have $\sum_{i,j}\pi(i)\pi(j)\norm{v_i-v_j}^2 = \pi(V)^2$. We define a new set of vertices $V'$ with $|V'| = \pi(V)$ with embedding vectors $w:V'\rightarrow \mathbb{R}^n$. In particular, we replace each $i\in V$ with $\pi(i)$ vertices in $V'$ each embedded at the point $v_i$.  Then, we have $\sum_{i',j'\in V'}\norm{w_{i'}-w_{j'}}^2 = \sum_{i,j\in V}\pi(i)\pi(j)\norm{v_i-v_j}^2 = \pi(V)^2$. Now, by applying \cite[Lemma 5]{Kal07}, we must have one of the following two cases:
    \begin{itemize}
        \item There is a vertex $i'\in V'$ such that $\Big|\Big\{j':\norm{w_{j'}-w_{i'}}\leq \frac{1}{2\sqrt{10}}\Big\}\Big| > \frac{1}{4} \pi(V)$, which means there is a vertex $i\in V$ such that $\pi\Big(B\Big(i,\frac{1}{2\sqrt{10}}\Big)\Big) > \frac{1}{4} \pi(V)$.
        \item There is a set of nodes $S'\subseteq V'$ and an $i'_0\in S$ such that $\forall i'\in S'$, $\norm{w_{i'} -w_{i'_0}}^2 = O(1)$ and $\sum_{i',j'\in S'}\norm{w_{i'}-w_{j'}}^2 = \Omega(\pi(V)^2)$. This means that there is a set of nodes $S\subseteq V$ and an $i_0\in S$ such that $\forall i\in S$, $\norm{v_i -v_{i_0}}^2 = O(1)$ and $\sum_{i,j\in S}\pi(i) \cdot \pi(j)\norm{v_i-v_j}^2 = \Omega(\pi(V)^2)$.
    \end{itemize}
\end{proof}

\begin{proof}[Proof of \autoref{lem:good-vectors}]
    We will prove the theorem via a simple reduction to \cite[Lemma 14]{Kal07}, which we will state as follows:
    \begin{lemma}[{\cite[Lemma 14]{Kal07}}]
        Suppose there are vectors $v_1,\dots, v_n$ such that $\norm{v_i}^2\leq 1$ for each $i$ and $\sum_{i,j}\norm{v_i-v_j}^2 \geq an^2$ for some constant $a$. Let $c \leq a/256$. Then with probability $8c$ over $u$, There exist sets $L$, $R$, of size at least $2cn$ such that for each $i\in L$ and $j\in R$, it holds that $\inner{v_j-v_j}{u}\geq \sigma$.
    \end{lemma}
    Now suppose we re-scale $\pi$ so that for each $i$, $\pi(i)$ is an integer. Then we have $\sum_{i,j}\pi(i)\pi(j)\norm{v_i-v_j}^2 \geq a \cdot \pi(V)^2$ for some constant $a$. We define a new set of vertices $V'$ with $|V'| = \pi(V)$ with embedding vectors $w:V'\rightarrow \mathbb{R}^n$. In particular, we replace each $i\in V$ with $\pi(i)$ vertices in $V'$ each embedded at the point $v_i$.  Then we have $\sum_{i',j'\in V'}\norm{w_{i'}-w_{j'}}^2 = \sum_{i,j\in V}\pi(i)\pi(j)\norm{v_i-v_j}^2 \geq a \cdot \pi(V)^2$. Then, applying \cite[Lemma 14]{Kal07}, we have that with probability at least $8c$ over $u\sim N(0,I)$, two sets $L',R'\subseteq V'$, each of size at least $2c \cdot \pi(V)$ such that for all $i\in L',j\in R'$, $\inner{v_j-v_i}{u}\geq \sigma$. This implies that there exist $L, R\subseteq V$ such that $\pi(L),\pi(R)\geq 2c\pi(V)$. 
\end{proof}

\subsection{Sherman's Main Theorem}

First, we will formally define the distribution used in \autoref{thm:sherman-main-thm}. In the statement of the theorem, the distribution is over $k$ vectors $u_1,\dots, u_k$ for some $k = O(\sqrt{\log{n}})$. The explicit distribution is over a shuffling of independent Gaussian vectors and correlated Gaussian vectors as defined in Section 5.4.1 of \cite{She09}. 

\begin{definition}(Correlated Sequence of Gaussian Vectors)
    Let $\mathcal{N}_{\rho}^k$ be a distrbution over vectors $u_1,\dots, u_k \in \mathbb{R}^d$ such that each $u_i$ has distribution $\mathcal{N}(0,I_d)$, and $u_{i+1}$ is $\rho$-correlated with $u_i$. In particular, if we define the matrix $U\in \mathbb{R}^{k\times d}$ with $U_{i,j} = u_i(j)$, then each column of the matrix, $(u_1(i),...u_k(i))$, is a $k$-dimensional multivariate normal distribution with covariance matrix $\Sigma_{a,b} = \rho^{|a-b|}$, and the $d$ columns are mutually independent.
\end{definition}

To prove \autoref{thm:sherman-main-thm}, we first note that it is invariant under scaling of $\pi$, which means we again assume without loss that $\pi(i)$ is an integer for each $i$.
Once again, we apply the reduction in which we have a set of $\pi(V)$ vertices, call it $V'$, and for each $i\in V$, we embed $\pi(i)$ vertices in $V'$ at the point $v_i$. Given the $\pi$-fractional matching cover $\M$, we can define a matching cover $\M'$ over $V'$ as follows. Given a vector $u$, for each $i,j\in V\times V$, we add $\M_u(i,j)$ edges between the corresponding vertices at $v_i$ and $v_j$ in $V'$. Clearly, $\M'$ is a $(\sigma,\delta)$ matching cover $V'$. 
The idea of Sherman's original argument was to show that if $k$ matchings chained together does not give many paths between vertices $i,j$ such that $\norm{v_i-v_j}^2 \geq l$, then there is a vertex $i$ such that with constant probability over random vectors $u\sim\mathcal{N}(0,I)$, we have $\inner{v_j-v_i}{u}\geq \Omega(k\sigma)$ for some $j$ such that $\norm{v_i-v_j}^2\leq l$. 
Then he shows that for large enough $k$ (in particular $k \approx \sqrt{l\log{n}}$), the probability of the later event happening cannot be $\Omega(1)$ by union bounding over all $i,j$ pairs.
In our case, even though we have $\pi(V)$ points, there are still only $n$ distinct positions so it still suffices to union bound over $n$ points instead of $\pi(V)$ points. For completeness, we will give the details of the argument in the rest of the section. 

Sherman defines a \textit{uniform $(\sigma,\delta)$-matching cover} \cite[Definition 5.4.1]{She09} as a $(\sigma,\delta)$-matching cover in which each vertex has at least a $\delta$ probability of having out-degree $1$ in the matching. By iteratively pruning vertices $V'$ whose probability of being matched is less than $\delta/4$, we obtain $X\subseteq V'$ of size at least $\pi(V)/4$ such that $\M'$ is a $(\sigma, \delta/4)$-uniform matching cover over $X$.
\begin{definition}
Given a distribution $\D$ over vectors $u_1,\dots, u_k$, and a matching-cover over the vertices $X$, a vertex $i$ is $(\sigma, \delta,\gamma,l)-$covered, by $\D$ if with probability at least $\delta$ over a random Gaussian $u\sim\mathcal{N}(0,I)$, 
\begin{align}
    \Pr_{u_2,\dots, u_k}\Big[\exists j \in B(i, l) :\;(i,j)\in \M'(\mathcal{D}),\;\inner{v_j-v_i}{u}\geq \sigma \mid u_1 = u \Big]\geq \gamma. \label{eqn:conditional}
\end{align}
\end{definition}

Now, we will use the following lemma from \cite{She09}.

\begin{lemma}[{\cite[Lemma 5.4.8]{She09}}]
\label{lem:app-main-lemma}
Let $\M'$ be a $(\sigma,\delta)$-uniform matching cover of $X$ where $\delta \leq 1/16$. Let $l\leq \sigma/2^7\sqrt{\log(1/\delta)}$ and $k\geq 1$. Then one of the following must occur:
\begin{enumerate}
    \item There are distributions $\D^0,\dots, \D^k$ such that for every $b\leq k$, at least $\delta^{6b}|X|$ vertices are $(b\sigma/4, \delta^8, \delta^{56bk},\sqrt{l})$-covered in $\M'(\D^{b})$
    \item There is a efficiently sample-able distribution $\D^*$ such that at least $\delta^{6k}|X|$ vertices $i$ have at least $\delta^{59k^2}$ probability of having an out-going edge to some $j \in i \backslash B(i,l)$ in $\M'(\D^*)$. Furthermore, $\D^*$ is a shuffling of $\mathcal{N}_{1-1/k}^{k'}$ with $\mathcal{N}_{0}^{k''}$ for some $k'\leq k$ and $k''\leq 6k$.
\end{enumerate}
\end{lemma}

To show that case 1 cannot hold for too many rounds, we will use the following lemma.

\begin{lemma}\label{lem:app-gaussian-property}
    Let $\M'$ be any matching cover for $X$ and $\gamma > 0$. There are no vertices $i\in X$ that are $(\sqrt{2l\log{(n/\delta})}, \delta, \gamma, \sqrt{l})$-covered by $\mathcal{M}(\mathcal{D})$.
\end{lemma}
\begin{proof}
Let $i\in X$ be arbitrary, and let $j\in B(i,l)$. This means $\norm{v_j - v_i}^2\leq l$. Then, by \autoref{fact:gaussian-concentration},  
$$\Pr_u\Big[\inner{v_j-v_i}{u} \geq \sqrt{2l\log{(n/\delta)}}\Big] \leq \exp(-\log{(n/\delta)}).$$
Since there are at most $n$ distinct embedding positions in $X$, the probability over $u_1$ that there is any $j\in B(i,l)$ such that $\inner{v_j-v_i}{u_1} \geq l\sqrt{2\log{(n/\delta)}}$ is at most $(n-1)\delta/n$. In this case, the conditional probability in \eqref{eqn:conditional} must be $0$, which is less than $\gamma$, for greater than $1-\delta$ fraction of $u_1$.
\end{proof}
\begin{proof}[Proof of \autoref{thm:sherman-main-thm}]
For any constant $l$, if $k \geq C\sqrt{l\log{n}}$ for some constant $C$, then case 1 of \autoref{lem:app-main-lemma} would imply that there is some $x \in X$ that contradicts \autoref{lem:app-gaussian-property}. Thus, we must be in case 2 for some $k < C\sqrt{l\log{n}}$. This means that for some distribution $\D^*$, the expected number of edges in $\M'(\D^*)$ between vertices $i,j\in X$ such that $\norm{v_i-v_j}^2 \geq l$ is at least $e^{-O(k^2)}|X|$. Finally, we note that $w$ paths between vertices embedded at $v_i$ and $v_j$ in $\M'(u_1,\dots, u_k)$ correspond to a path of weight $w$ between $i$ and $j$ in $\M$. Moreover, the algorithm for constructing $\p_{u_1,\dots, u_k}$ in $\M(u_1,\dots, u_k)$ is equivalent to the natrual algorithm chaining together $0/1$-matchings in $\M'(u_1,\dots, u_k)$. Thus, the expected total weight of paths in $\p_{u_1,\dots, u_k}$ between vertices $i,j$ at distance at least $l$ apart is at least $e^{O(-k^2)}\pi(V)$.

Finally, we note that while our distribution $\mathcal{D}^*$ is a shuffling of $\mathcal{N}_{1-1/k}^{k'}$ and $\mathcal{N}_0^{k''}$, we have not yet given a way to find the correct ordering. However, since $k' + k'' = O(\sqrt{l\log{n}})$, the total number of sequences is at most $O(\sqrt{l\log{n}}!) = n^{o(1)}$. Thus, if we pick a random shuffling, the probability that it will be the correct ordering is $n^{-o(1)}$. Thus, after taking into account the randomness over shuffling orders, the expected total weight of paths between vertices $i,j$ at distance at least $L$ apart is at least $n^{-o(1)}e^{O(-k^2)} \cdot \pi(V) = e^{-O(k^2)} \cdot \pi(V)$ since $k^2 = \Theta(l\log{n})$.
\end{proof}

\subsection{Matrix Exponential}

In this section, we give details on how to implement the matrix exponential step in Algorithms \ref{alg:mmw-eigenvalue}, \ref{alg:mmw-expander-flow} and \ref{alg:cut-player} and prove \autoref{lem:matrix-exp-computation}. Given feedback matrices $M_1,\dots, M_t$, such that $\norm{M_i}\leq \rho$ for each $i\in [t]$, we would like to approximately compute the Gram decomposition of the matrix
$$Y_t = \frac{\Pi^{-1/2}\exp(-\frac\eta\rho\sum_{i=1}^tM_i)\Pi^{-1/2} - \Pi^{1/2}\one\one^\top \Pi^{1/2}}{\tr(\exp(-\frac\eta\rho\sum_{i=1}^tM_i))-1}.$$

Note in particular that if we take $A = \frac{\eta}{\rho}\sum_{i=1}^tM_i$, then it suffices to compute the rows of the matrix $\exp(-\frac12A)\Pi^{-1/2}$ projected into the space orthogonal to $\Pi^{1/2}\one$. Since computing the matrix exponential exactly is costly, we will instead compute a low-dimensional approximation of its rows by multiplying the matrix with random vectors and applying the Johnson-Lindenstrauss lemma. Thus, the problem of computing the embedding vectors reduces to the problem of computing $\exp(S) \cdot u$ for some vector $u$ and symmetric matrix $S$. For our purpose, it suffices to compute the first terms in the Taylor expansion of the matrix exponential. 

\begin{lemma}[{\cite[Lemma 23]{Kal07}}]
    \label{lem:tayler-approx}
    Given a symmetric matrix $S$ and a unit vector $u$, let $v = \sum_{i=0}^k\frac{1}{i!}S^iu$. If $k \geq \max(e^2\norm{S}, \ln\frac{1}{\tau})$, then $v$ satisfies
    \[
    \norm{\exp(S) \cdot u - v} \leq \norm{\exp(S)}\tau
    \]
    Moreover, the time to compute $v$ is $O(km)$ where $m$ is the number of non-zero entries in $S$.
\end{lemma}

Given the previous result, we can give the algorithm for computing the matrix exponential based on \cite[Section 4.7]{Kal07}. However, there are two modifications we must make. First, to take into account the $\pi$ vertex weights, we let $\pi_{\min} = \min_i\pi(i)$, and we will show that it suffices to take $\tau = \pi_{\min}/\poly(n)$. Second, we must take into account the projection into the subspace orthogonal to $\Pi^{1/2}\one$, which is the exactly the nullspace of $A$. We will call this subspace $\mathcal{U}_{\pi}^\perp$, and for a matrix $M$, we will define the matrix $M|_{\mathcal{U}_{\pi}^\perp}$ as the matrix whose columns are those of $M$ projected into $\mathcal{U}_{\pi}^\perp$. We can then modify  \autoref{lem:tayler-approx} as follows. 

\begin{lemma}\label{lem:modified-taylor-approx}
   Let $u$ be a unit vector in $\mathcal{U}_{\pi}^\perp$, and let $v$ be the vector obtained from applying the Taylor approximation in \autoref{lem:tayler-approx} for $k = \max(e^2\norm{A},\ln{\frac1\tau})$ iterations to the matrix $\exp(-\frac12A)u$. In $\Tilde{O}(\rho m)$ time, we can ensure that
    \begin{align*}
        \norm{\exp\Big(-\frac12A\Big)\Big|_{\mathcal{U}_\pi^\perp} \cdot u - v} \leq \norm{\exp\Big(-\frac12A\Big)\Big|_{\mathcal{U}_\pi^\perp}} \cdot \tau.
    \end{align*}
\end{lemma}
\begin{proof}
    Let $P\in \mathbb{R}^{n\times n-1}$ be a unitary matrix mapping $\mathcal{U}_{\pi}^\perp$ to $\mathbb{R}^{n-1}$. Note that this means $P^\top P = I_{n-1}$ and for any $x\in \mathcal{U}_\pi^{\perp}$, we have $PP^\top x = x$. Since $u\in \mathcal{U}_\pi^\perp$, we have $\exp(-\frac12A)u = \exp(-\frac12A)|_{\mathcal{U}_\pi^\perp}u$. Moreover, the Taylor approximation $v$ is also in $\mathcal{U}_\pi^\perp$, which means
    \begin{align*}
    \norm{\exp\Big(-\frac12A\Big)\Big|_{\mathcal{U}_\pi^\perp}u - v} &= \norm{P^\top\exp\Big(-\frac12A\Big)\Big|_{\mathcal{U}_\pi^\perp}u - P^\top v}\\
    &= \norm{P^\top\exp\Big(-\frac12A\Big)\Big|_{\mathcal{U}_\pi^\perp}PP^\top u - P^\top v}\\
    &= \norm{P^\top\exp\Big(-\frac12A\Big)PP^\top u - P^\top v}\\
    &= \norm{\exp\Big(-\frac12P^\top AP\Big)P^\top u - P^\top v},
    \end{align*}
    where the last equality follows from the fact that $\text{Range}(A) \in \mathcal{U}_\pi^\perp$, which means $(P^\top A P)^i = P^\top A^i P$ for any $i$. Finally, we check that
    \begin{align*}
         P^\top v = P^\top\sum_{i=1}^k  \frac{(-1)^i}{2^ii!}A^{i}PP^\top u =\sum_{i=1}^k \frac{(-1)^i}{2^ii!}(P^\top AP)^{i}P^\top u.
    \end{align*}
    Thus, we can apply \autoref{lem:tayler-approx} with $S = -\frac12 P^\top AP$ and the input unit vector being $P^\top u$ to obtain
    \begin{align*}
        \norm{\exp\Big(-\frac12P^\top AP\Big)P^\top u - P^\top v} \leq \norm{\exp\Big(-\frac12P^\top AP\Big)} \cdot \tau = \norm{\exp\Big(-\frac12A\Big)\Big|_{\mathcal{U}_\pi^\perp}} \cdot \tau
    \end{align*}
    Finally, to bound the runtime, we see that it suffices to take $k = \max\{e^2\norm{P^\top AP}, \ln{(1/\tau)}\}$. We can bound the matrix norm by $\norm{A}$ since $P$ is unitary. Since each $M_t$ has spectral norm at most $\rho$, we have $\norm{A} \leq \eta T$. In Algorithms \ref{alg:mmw-eigenvalue} and \ref{alg:cut-player}, we have $\eta = O(1)$ and $T = O(\log^3{n})$ and in Algorithm \ref{alg:mmw-expander-flow}, we have, $\eta = 1/\rho$ and $T = O(\rho^2\log{n})$. Thus, in the worst case, we have $\norm{A}\leq \Tilde{O}(\rho)$.
\end{proof}
Now, we are ready to give our algorithm for approximating the matrix exponential.

\newpage 

\begin{algorithm}[h]\caption{Matrix Exponential}\label{alg:matrix-exponential}
\textbf{Input:} a symmetric matrix: $A = \frac{\eta}{\rho}\sum_{i=1}^tM_i$, embedding dimension $d$, and accuracy parameter $\tau$.
\begin{enumerate}
    \item Let $U$ be a $n\times d$ matrix whose $d$ columns form an orthogonal basis of a random $d = O(\log{n})$-dimensional subspace orthogonal to the vector $\Pi^{1/2}\one$ with each column vector having length $\sqrt{n/d}$. Let $U_1$ be another random matrix defined similarly but whose columns are orthogonal to the vector $\one$ instead.
    \item Pick $k\geq \Omega(\max(\rho, \log{(1/\tau)}))$. Compute $Z_\pi = \sum_{i=0}^k\frac{(-1)^i}{2^ii!}A^i\Pi^{-1/2}U_1 $ and $Z = \sum_{i=0}^k\frac{(-1)^i}{2^ii!}A^iU$.
    \item Let $\hat{v}_1,\dots \hat{v}_n$ be the rows of the matrix $Z_{\pi}/\sqrt{\tr(ZZ^\top)}$. Return these as the approximate embedding vectors. 
\end{enumerate}
\end{algorithm}

For the sake of analysis, we will define the following matrices: let $W_{\pi}:= \exp(-\frac12A)\Pi^{-1/2}U_1 = \exp(-\frac12A)|_{\mathcal{U}_\pi^\perp}\Pi^{-1/2}U_1$ and let $W: = \exp(-\frac12A)| U = \exp(-\frac12A)|_{\mathcal{U}_\pi^\perp} U$. Note the second inequalities follow from the fact that since the columns of $U$ and $\Pi^{-1/2}U_1$ are in $\mathcal{U}_\pi^\perp$.
By applying the Johnson Lindenstrauss lemma, we can show that with good probability, the rows of the matrix $W_\pi$ and $W$ are good approximations to those of the matrix exponential.
\begin{lemma}[Johnson-Lindenstrauss Lemma~\cite{JL84}]\label{lem:JL}
    Let $x^\pi_1,\dots x^\pi_n$ and $x_1,\dots x_n$ be the rows of the matrices $\exp(-\frac12 A)|_{\mathcal{U}_{\pi}^\perp}\Pi^{-1/2}$ and   $\exp(-\frac12 A)|_{\mathcal{U}_\pi}^\perp$ respectively. Let $r^\pi_1,\dots r^\pi_n$ and $r_1,\dots r_n$ be the rows of the matrices $W_\pi$ and $W$ respectively. For some $d = O(\frac{1}{\delta^2}\log{n})$, with probability $1-n^{-1}$,
    \begin{align*}
        \norm{r_i^\pi - r_j^\pi}^2 &\in \norm{x_i^\pi - x_j^\pi}^2(1\pm\delta)\quad\forall i,j\in V \quad {\rm and} \quad
        \norm{r_i}^2 \in \norm{x_i}^2(1\pm\delta) \quad\forall i\in V.
    \end{align*}
\end{lemma}

 The following lemma shows that our approximations of $W$ and $W_\pi$ are also good for some $\tau = 1/\poly(\frac{n}{\pi_{\min}})$.

\begin{lemma}\label{lem:exp-approx}
     Let $Y' := W_{\pi}W_{\pi}^\top/\tr(WW^\top)$ and $Y'' = Z_{\pi}Z_{\pi}^\top/\tr(ZZ^\top)$. Suppose $W$ satisfies $\tr(WW^\top) \in \tr(\exp(-A)|_{\mathcal{U}_{\pi}^\perp})(1\pm \delta)$ for $\delta \leq 1/5$. Then for small enough $\tau \in 1/\poly(\frac{n}{\pi_{\min}})$,  
    \[\norm{Y'' - Y'} \leq O(\pi_{\min}^{-1}n^{3/2})\cdot \tau.\]
\end{lemma}
\begin{proof}[Proof Sketch]
    The proof of this lemma follows very closely to the proof of \cite[Lemma 25]{Kal07} so we will only sketch out the details. First, we define the error matrices $E_\pi = W_\pi-Z_\pi$ and $E = W-Z$. Let $w^\pi_1,\dots w^\pi_d$ and $z^\pi_1\dots z^\pi_d$ be the columns of the matrices $W_\pi$ and $Z_\pi$ respectively. By \autoref{lem:modified-taylor-approx}, 
    \[\norm{E_\pi}^2 \leq \norm{E_\pi}_F^2 = \sum_{i=1}^d\norm{w_i^\pi - z_i^\pi}^2 \leq d \cdot \pi_{\min}^{-1}\norm{\exp\Big(-\frac12A\Big)\big|_{\mathcal{U}_{\pi}^\perp}}^2\tau^2 \leq d \cdot \pi_{\min}^{-1}\norm{\exp(-A)|_{\mathcal{U}_{\pi}^\perp}}\tau^2,\]
    where the second inequality follows from the fact that if $u$ is a unit vector orthogonal to $\one$, then $\Pi^{-1/2}u$ is a vector of length at most $\pi_{\min}^{-1/2}$ orthogonal to $\Pi^{1/2}\one$. 
    Similar calculations show that $\norm{E}^2\leq d \cdot \norm{\exp(-A)|_{\mathcal{U}_{\pi}^\perp}}\tau^2$. 
    Using the $E$ and $E_\pi$, matrices, we can bound the following matrix distances
    \begin{align}
        \norm{W_\pi W_\pi^{\top} - Z_{\pi}Z_{\pi}^{\top}} = \norm{E_\pi E_\pi^\top + E_\pi V_\pi^\top + V_\pi E_\pi^\top}\leq 3d \cdot \pi_{\min}^{-1}\norm{\exp(-A)|_{\mathcal{U}_{\pi}^\perp}}\tau. \label{eqn: norm-distance-bound}
    \end{align}
    And similar calculations show that
    \begin{align}
        |\tr(WW^\top - ZZ^\top)| = |\tr(E E^\top + E V^\top + V E^\top)| \leq 3d^{3/2}\norm{\exp(-A)|_{\mathcal{U}_{\pi}^\perp}}\tau. \label{eqn: trace-distance-bound}
    \end{align}
    Thus,
    \begin{align*}
        \norm{Y'-Y''} &\leq \norm{\frac{W_\pi W_\pi^\top}{\tr(WW^\top)} - \frac{W_\pi W_\pi^\top}{\tr(VV^\top)}} + \norm{\frac{W_\pi W_\pi^\top}{\tr(VV^\top)} - \frac{V_\pi V_\pi^\top}{\tr(VV^\top)}}\\
        &\leq \frac{\norm{W_\pi W_\pi^\top}}{\tr(WW^\top)}\cdot \frac{|\tr(WW^\top - \tr(ZZ^\top))|}{\tr(ZZ^\top)} + \frac{\norm{W_\pi W_\pi^\top - Z_\pi Z_\pi^\top}}{\tr(ZZ^\top)}\\
        &\leq \pi_{\min}^{-1}\cdot  \frac{3d^{3/2}\norm{\exp(-A)|_{\mathcal{U}_{\pi}^\perp}}\tau}{\tr(ZZ^\top)} + \frac{3d \cdot \pi_{\min}^{-1}\norm{\exp(-A)|_{\mathcal{U}_{\pi}^\perp}}\tau}{\tr(ZZ^\top)}\\
        &\leq \frac{6\pi_{\min}^{-1}n^{3/2}\norm{\exp(-A)|_{\mathcal{U}_{\pi}^\perp}}\tau}{(1-\delta - 3n^{3/2}\tau)\norm{\exp(-A)|_{\mathcal{U}_{\pi}^\perp}}\tau}\\
        &\leq 8n^{3/2} \cdot \pi_{\min}^{-1} \cdot \tau,
    \end{align*}
where the third inequality follows from bounds in (\ref{eqn: norm-distance-bound}), (\ref{eqn: trace-distance-bound}) and $W_\pi W_\pi^\top \preceq \pi_{\min}^{-1} WW^\top$, and the fourth inequality follows from $\tr(WW^\top) \in (1\pm \delta)\norm{\exp(-A)|_{\mathcal{U}_{\pi}^\perp}}$ and (\ref{eqn: trace-distance-bound}).
\end{proof}

Finally, we are ready to prove \autoref{lem:matrix-exp-computation}.

\begin{proof}[Proof of \autoref{lem:matrix-exp-computation}]
     Suppose we pick $\tau=\pi_{\min}/n^c$ for some constant $c$. By \autoref{lem:JL}, with probability at least $1-n^{-1}$, for all $i,j\in V$
     \begin{align*}
         \norm{v_i - v_j}^2 = \frac{\norm{x^\pi_i - x^\pi_j}^2}{\sum_i\norm{x}^2} \in \frac{\norm{r^\pi_i - r^\pi_j}^2(1\pm \delta)}{\sum_i\norm{r_i}^2(1\pm \delta)} = \inner{L_{i,j}}{Y'} \cdot (1\pm 2\delta)
     \end{align*}
     and in particular, $W$ satisfies $\tr(WW^\top) = \sum_i\norm{r}_i^2\in (1\pm \delta)\sum_i\norm{x_i}^2 = \tr(\exp(-A)|_{\mathcal{U}_{\pi}^\perp})(1\pm \delta)$. Thus, \autoref{lem:exp-approx} implies that
     \begin{align*}
         |\inner{L_{i,j}}{Y'} - \inner{L_{i,j}}{Y''}| \leq 2\norm{Y-Y''} \leq 1/n^{-c+3/2}. 
     \end{align*}
     Since $\inner{L_{i,j}}{Y''} = \norm{\hat{v}_i - \hat{v}_j}^2$, it follows that $\norm{\hat{v}_i - \hat{v}_j}^2 \in \norm{v_i - v_j}^2(1\pm 2\delta) \pm n^{-\Omega(1)}$. 
     Finally, \autoref{lem:modified-taylor-approx} implies that the runtime of \autoref{alg:matrix-exponential} is $\Tilde{O}(\rho m)$. 
\end{proof}


\subsection{Fast Computation of Maximum Circulation}\label{app:min-cost-flow}

Recall that our main program in \autoref{def:main-sdp} is to minimize the maximum edge-constrained circulation over all feasible embeddings $v_1, \dots, v_n$.
We remark that the inner maximization problem
\[
  \max_{F \in \mathcal{F}(G)} \sum_{i < j} \frac{1}{2} (F(i, j) + F(j, i)) \norm{v_i - v_j}^2
\]
is a special case of the minimum-cost flow in \cite{CKLPPS22}, each edge $e = ij$ having lower edge capacity $0$, upper edge capacity $w(e)$, and cost $-\norm{v_i - v_j}^2$ (so that it becomes a maximization problem), and each vertex $i$ having demand $d(i) = 0$.
Therefore, this problem can be computed in $O(m^{1 + o(1)})$ time.

\section{Missing proofs of \autoref{sec:others}}

\label{app:others}
\begin{proof}[Proof of \autoref{lemma:flow-cut-lemma-v}]
    Suppose $\vec{f}$ is the non-saturating flow.
  We obtain a cut which consists of edges incident to either $s$ or $t$ and vertices of $G$, whose removal would make it impossible to go from $s$ to $t$.
  Let $S \subseteq V$ be the set of vertices reachable from $s$ after removing the cut edges and vertices.
  Let $V_s \subseteq L$ and $V_t \subseteq R$ be defined similarly to the edge-capacitated case. 

  Let $\nu \cdot \beta \cdot \pi(R)$ be the max-flow value of $\vec{f}$ with $\nu < 1$. 
  Note that
  \[
    \kappa \cdot \pi(\partial^+(S)) = \beta (\nu \cdot \pi(R) - r \cdot \pi(V_s) - \pi(V_t)),
  \]
  because $\{si \mid i \in V_s\} \cup \partial_G^+(S) \cup \{jt \mid j \in V_t\}$ is the minimum $s$-$t$ cut obtained, with total weight equal to $r \cdot \beta \cdot \pi(V_s) + \beta \cdot \pi(V_t) + \kappa \cdot \pi(\partial^+(S)) = \nu \cdot \beta \cdot \pi(R)$ by our construction of $\vec{G}$.
  Also, since $L \setminus (V_s \cup \partial^+(S)) \subseteq S$ and $R \setminus (V_t \cup \partial^+(S)) \subseteq V-S$, it follows that 
  \[
      \pi(S) \geq \pi(L) - \pi(V_s) - \pi(\partial^+(S))
      \quad \text{ and } \quad
      \pi(V-S) \geq \pi(R) - \pi(V_t) - \pi(\partial^+(S)).
  \]
  Therefore, 
  \begin{align*}
     \frac{1}{\vertexexp(S)} = \frac{\min\{\pi(S),\pi(V-S)\}}
     {\pi(\partial^+(S))}
     & \geq
       \frac{\min\{\pi(R) - \pi(V_t) - \pi(\partial^+(S)),
             ~\pi(L)- \pi(V_s) - \pi(\partial^+(S))\}}
       {\pi(\partial^+(S))}
     \\
     &=
     \frac{\kappa}{\beta} \cdot
     \frac{\min \{\pi(R) - \pi(V_t),~\pi(L) - \pi(V_s)\}}
       {\nu \cdot \pi(R) - r \cdot \pi(V_s) - \pi(V_t)}
     - 1
     \\
     &\geq
     \frac{\kappa}{\beta} \cdot
     \min \left\{
     \frac{\pi(R) - \pi(V_t)}{\pi(R) - \pi(V_t)},~
     \frac{\pi(L) - \pi(V_s))}{r \cdot \pi(L) - r \cdot \pi(V_s)}
     \right\}
     - 1 
     \\
     &=
     \frac{\kappa}{\beta \cdot r'} - 1,
  \end{align*}
  where the last inequality is because $\nu < 1$ and $\pi(R) = r \cdot \pi(L)$.
  Rearranging gives the desired conclusion.
\end{proof}

\end{appendix}

\end{document}